\documentclass[12pt]{article}
\usepackage{amsmath,amssymb,framed,mathrsfs}
\usepackage{diagrams}
\usepackage[colorlinks]{hyperref}

\usepackage[all]{xy}

\newcommand{\M}{{\ \overline{\!\!\mathcal{M}\!}}\, }
\newcommand{\rem}[1]{}

\makeatletter

\@addtoreset{figure}{section}
\def\thefigure{\thesection.\@arabic\c@figure}
\def\fps@figure{h, t}
\@addtoreset{table}{bsection}
\def\thetable{\thesection.\@arabic\c@table}
\def\fps@table{h, t}
\@addtoreset{equation}{section}

\makeatother

\newcommand{\todo}[1]{\vspace{5 mm}\par \noindent
\framebox{\begin{minipage}[c]{0.95 \textwidth}
\tt #1 \end{minipage}}\vspace{5 mm}\par}

\begin{document}

\newtheorem{theorem}{Theorem}[section]
\newtheorem{definition}[theorem]{Definition}
\newtheorem{lemma}[theorem]{Lemma}
\newtheorem{remark}[theorem]{Remark}
\newtheorem{proposition}[theorem]{Proposition}
\newtheorem{corollary}[theorem]{Corollary}
\newtheorem{example}[theorem]{Example}

\def\below#1#2{\mathrel{\mathop{#1}\limits_{#2}}}

\title{Reduction theory for symmetry breaking\\with applications to nematic systems}
\author{Fran\c{c}ois Gay-Balmaz${}^1$ and Cesare Tronci${}^2$\\
\footnotesize\it ${}^1$ Control and Dynamical Systems, California Institute of Technology\\
\footnotesize\it ${}^2$ Section de Math\'ematiques, \'Ecole Polytechnique
F\'ed\'erale de Lausanne. Switzerland}


\date{ }
\maketitle

\makeatother
\maketitle




\begin{abstract}
We formulate Euler-Poincar\'e and Lagrange-Poincar\'e equations for systems with broken
symmetry. We specialize the general theory to present explicit equations of motion for nematic systems, ranging from single nematic molecules to biaxial liquid crystals. The geometric construction applies to order parameter spaces consisting of either unsigned unit vectors (directors) or symmetric matrices (alignment tensors). On the Hamiltonian side, we provide the corresponding Poisson brackets in both Lie-Poisson and Hamilton-Poincar\'e formulations. The explicit form of the helicity invariant for uniaxial nematics is also presented, together with a whole class of invariant quantities (Casimirs) for two dimensional incompressible flows.
\end{abstract}



\medskip

\section{Introduction}

\subsection{A simple example: the heavy top}

Symmetry breaking phenomena are widely common in several physical
contexts, from classical mechanics to particle physics. The simplest
example is probably the heavy top dynamics, that is the motion of a rigid body with a fixed point in a gravitational field. This motion
takes place on the
special Euclidean group $SE(3)$, which carries a semidirect
product structure $SE(3)=SO(3)\,\circledS\,\Bbb{R}^3$ involving the
special orthogonal group $SO(3)$. Semidirect product Lie group
structures of this kind may be understood as ``signatures'' of
broken symmetries in physical systems.
For the case of the heavy top, the physical
description involves a basic configuration space
which is $SO(3)$, i.e. the same as in rigid body dynamics. Although
one can simply write Hamilton's equations on the cotangent bundle $T^*SO(3)$, one
realizes that the heavy top dynamics is very different from the
simple rigid body case.
Indeed, the presence of gravity in the
system produces a Hamiltonian which is {\it not} $SO(3)$-invariant,
contrarily to what happens for the rigid body. Thus, besides the
body angular momentum, one also considers the direction of gravity
(in the body reference frame) as a dynamical variable in $S^2$. Upon
identifying $S^2$ with unit vectors in $\Bbb{R}^3$, one is leaded to
enlarge the Lie group $SO(3)$ thereby considering the semidirect
product $SO(3)\,\circledS\,\Bbb{R}^3$. Such an approach has a
relevant infinite-dimensional analogue which yields the theory of
compressible fluid flows. The main references for geometric
mechanics on semidirect product Lie groups are
\cite{MaRaWe1984,HoMaRa1998}. One of the targets of the present paper is
to generalize these semidirect product structures in the context of reduction by symmetry, thereby extending to the case when the vector space $V$ in the semidirect product $G\,\circledS\,V$ is replaced by a manifold $M$, so that
the semidirect-product group structure breaks into the Cartesian product
$G\times M$. This introduction reviews the high points of the Euler-Poincar\'e
reduction theory and summarizes the results obtained in the paper, after a brief discussion of the geometric methods that will be used.

\subsection{Euler-Poincar\'e approach to heavy tops}\label{subsec_rigid_body}

This section briefly reviews the concrete mathematical background for the
dynamics of the heavy top. This example will be still
considered in the remainder of the paper for direct comparison.

Although it is defined on the configuration space $SO(3)$, the heavy top dynamics exhibits rotational symmetry breaking
through the appearance of a potential term, depending on the direction of gravity $\mathbf{e}_3\in S^2$. The heavy top Lagrangian $L_{\mathbf{e}_3}:TSO(3)\rightarrow\mathbb{R}$
is written as
\begin{equation}\label{HT-lagr}
L_{\mathbf{e}_3}(\chi,\dot{\chi})=\frac{1}{2}\int_\mathcal{B}\rho(A)|\dot{\chi}A|^2d^3A-Mg\ell\,
\mathbf{e}_3\cdot\chi\boldsymbol{\zeta}
\end{equation}
where $\mathcal{B}\subset\Bbb{R}^3$ is the reference configuration of the body, $\rho(A)$ is the mass density, $M$ is the total mass, $g$ is the acceleration of gravity, and
$\boldsymbol{\zeta}$ is the unit vector along the segment of length
$\ell$ that connects the fixed point of the body with its center of
mass at $t=0$.
Evidently, the $SO(3)$-symmetry is broken:
$L_{\mathbf{e}_3}(\chi,\dot{\chi})\neq
L_{\mathbf{e}_3}(\chi^{-1}\dot{\chi})$. Rather, the system is
invariant only with respect to $SO(2)$ (rotations around the
vertical axis), since this is the isotropy subgroup of
$\mathbf{e}_3$. At this point, one denotes
$\mathbf{e}_3=\boldsymbol{\Gamma}_0\in\mathbb{R}^3$ and defines
$L:TSO(3)\times \mathbb{R}^3\rightarrow\mathbb{R}$ by
$L(\chi,\dot{\chi},\boldsymbol{\Gamma}_0):=L_{\boldsymbol{\Gamma}_0}(\chi,\dot{\chi})$.
where $\boldsymbol{\Gamma}_0$ is interpreted as a new variable. We now let $SO(3)$ act on the extended space $TSO(3)\times\mathbb{R}^3$ by the left action
$(\chi,\dot\chi, \boldsymbol{\Gamma}_0)\mapsto(\psi\chi,\psi\dot\chi,\psi\boldsymbol{\Gamma}_0)$,
with the result that the function $L$ is now $SO(3)$-invariant since
\[
Mg\ell\,
\psi\boldsymbol{\Gamma}_0\cdot\psi\chi\boldsymbol{\zeta}=Mg\ell\,\boldsymbol{\Gamma}_0\cdot\chi\boldsymbol{\zeta},\quad\text{for all $\psi\in SO(3)$}.
\]
\rem{ 
F: Ok for you if we use the above text instead of the following  sentence:\\
It is clear that, upon defining
$\boldsymbol{\Gamma}:=\chi^{-1}\,\boldsymbol{\Gamma}_0$ (the
direction of gravity as seen from the body), the potential energy $M
g\ell\,\Phi(\chi,\mathbf{e}_3):=M g\ell\,\boldsymbol{\Gamma}_0\cdot
\chi\boldsymbol{\zeta}$ is rewritten according to
\[
\Phi(\chi,\boldsymbol{\Gamma}_0)=\mathbf{e}_3\cdot
\chi\boldsymbol{\zeta}=\chi^{-1}\boldsymbol{\Gamma}_0\cdot
\chi^{-1}\chi\boldsymbol{\zeta}=\boldsymbol{\Gamma}\cdot\boldsymbol{\zeta}=\Phi(\chi^{-1}\chi,\chi^{-1}\boldsymbol{\Gamma}_0)=\phi(\chi^{-1}\boldsymbol{\Gamma}_0)=:\phi(\boldsymbol{\Gamma})
\]
}   
Thus, by $SO(3)$-invariance, $L$ induces the reduced Lagrangian
$l:\mathfrak{so}(3)\times \mathbb{R}^3\rightarrow\mathbb{R}$ given
by
\[
l(\boldsymbol{\Omega},\boldsymbol{\Gamma})=\frac{1}{2}\mathbb{I}\boldsymbol{\Omega}\!\cdot\!\boldsymbol{\Omega}-Mg\ell
\boldsymbol{\Gamma}\!\cdot\!\boldsymbol{\zeta}.
\]
where $\boldsymbol{\hat{\Omega}}=\chi^{-1}\dot\chi$ (so that `hat'
denotes the usual isomorphism $\mathfrak{so}(3)\simeq\Bbb{R}^3$) and
$\boldsymbol{\Gamma}=\chi^{-1}\boldsymbol{\Gamma}_0$. The moment of inertia tensor $\Bbb{I}$ is defined by (cf.
\cite{MaRa1999})
\[
\Bbb{I}=-\int_{\mathcal{B}}\rho(A)\left(AA^T-\left|A\right|^2{\bf I}\right)d^3A\,.
\]
Again, the
reduced Lagrangian produces the Euler-Poincar\'e variational
principle
\[
\delta\!\int_{t_0}^{t_1}l(\boldsymbol\Omega,\boldsymbol{\Gamma})dt=0
\]
thereby yielding the equations
\[
\frac{d}{d t}\frac{\delta
l}{\delta\boldsymbol\Omega}+\boldsymbol\Omega\times\frac{\delta
l}{\delta\boldsymbol\Omega}=-\boldsymbol{\Gamma}\times\frac{\delta
l}{\delta\boldsymbol{\Gamma}} \qquad \frac{d\boldsymbol{\Gamma}}{d
t}=\boldsymbol{\Gamma}\times\boldsymbol{\Omega}
\]
where the variations are evaluated as $\delta
\boldsymbol\Omega=\delta
(\chi^{-1}\dot{\chi})=\dot{\boldsymbol\Sigma}+\boldsymbol\Omega\times\boldsymbol\Sigma$
(with $\boldsymbol\Sigma:=\chi^{-1}\delta \chi$) and
$\delta\boldsymbol{\Gamma}=\boldsymbol{\Omega}\times\boldsymbol{\Gamma}$.
This construction is referred to as Euler-Poincar\'e approach for
semidirect products (cf. \cite{HoMaRa1998}). In this particular
case, the configuration Lie group is the well known special
Euclidean group $SE(3):=SO(3)\,\circledS\,\Bbb{R}^3$.  Such a
formulation is only possible upon identifying points on the sphere
 (e.g. the direction of gravity) with unit vectors in $\Bbb{R}^3$ and by
 noticing that the modulus of the latter is preserved by the motion.
 Although
 in many situations this process leads to no contradiction,  this is not always the case in condensed matter physics and it is useful
to consider  parameters (called `order parameters')
belonging to a manifold $M$ rather than to a vector space $V$. For example, the heavy
top would require a Lagrangian $L:TSO(3)\times S^2\to\Bbb{R}$. The next section gives an overview of the general setting in the context of condensed matter
physics.

\subsection{Broken symmetries in condensed matter: nematic molecules}

An enlightening example of broken symmetry appears in the theory of
nematic liquid crystals \cite{Chandra1992,deGennes1993}. In the simplest case of uniaxial molecules, such systems are
continuum systems composed of oriented particles, i.e.  particles
that are endowed with a special orientation identified with an
unsigned unit vector ${\bf n} \sim -{\bf n}$,  called {\it
director}. The presence of such a special direction in the system
plays exactly the same role as in heavy top dynamics, as it is shown
by the equations of motion for the single nematic particle:
\begin{equation}
\frac{d}{dt}\frac{\delta l}{\delta
\boldsymbol{\nu}}=\boldsymbol{\nu}\times\frac{\delta l}{\delta
\boldsymbol{\nu}}+{\bf n}\times\frac{\delta l}{\delta \bf n}
\,,\qquad\quad \dot{\bf n}=\boldsymbol\nu\times {\bf n} \,.
\end{equation}
A direct comparison with heavy top dynamics shows the strict
similarity between nematic particles and heavy tops. However, a
director carries no sign and thus it does not belong to $S^2$, but
rather it takes values in $S^2/\Bbb{Z}_2$, i.e. the unit sphere
$S^2$ with diametrically opposite points identified, also known as
the real projective plane $\Bbb{R}P^2$. Therefore the occurrence of
the director breaks the full $SO(3)$ rotational symmetry of the
particle, whose rotational motion is now invariant only under
a subgroup $\mathcal{P}\subset SO(3)$, i.e. the isotropy subgroup of
the director itself. For nematic particles, the isotropy subgroup of
unsigned unit vectors is $\mathcal{P}=O(2)$. (More precisely
$\mathcal{P}=D_\infty$, where the group $D_\infty$ consists of rotations about the
molecular axis and $180^\circ$-rotations about a normal to the
molecular axis). In more general condensed matter applications, if
$\mathcal{O}$ is the broken symmetry (the order parameter group)
acting \textit{transitively} on the order parameter space $M$, then
the latter is isomorphic to the coset space $\mathcal{O/P}$, where
$\mathcal{P}$ is the isotropy subgroup $\mathcal{P=O}_{n_0}$ of a
reference point $n_0\in M$. In the case of nematics, the order
parameter space is $\mathcal{O/P}=SO(3)/O(2) \simeq\mathbb{R}P^2$.
Thus, after starting with the configuration space $\mathcal{O}$
($SO(3)$ for nematics), the symmetry breaking requires to consider
also dynamics on the order parameter space $M\simeq\mathcal{O/P}$.
Notice that, this description refers only to the rotational motion
of complex fluid particles and we have forgotten about its motion in
physical space. However, the latter can be included {a posteriori}
by considering Cartesian product configuration manifolds such as
$Q\times\mathcal{O}$, where $Q$ denotes physical space and we assume
that $\mathcal{O}$ does not act on $Q$. In the first part of this paper we shall
neglect motion in physical space in order to keep the treatment
sufficiently compact.

Nematic particles differ from the heavy top in that the order
parameter space can be a manifold, rather than a vector space. Thus,
the ordinary theory of heavy top dynamics needs to be extended for
condensed matter applications. In particular, even if the order
parameter coset space $\mathcal{O/P}$ is naturally associated to
symmetry breaking, it is not clear how this coset space arises from
a fundamental approach in terms of reduction by symmetry. For
example, the emergence of a broken symmetry often leads to consider
$\mathcal{O/P}$ (rather than $\mathcal{O}$) as the configuration
manifold. In the context of nematic particles, the Ericksen-Leslie
equations (cf. \cite{Le1979}) are simply the Euler-Lagrange
equations on the projective plane $\Bbb{R}P^2\simeq SO(3)/O(2)$. One
of the tasks of this work is to show how this new configuration
space emerges naturally from reduction theory. In more generality,

\bigskip
\noindent{\it this paper aims to  provide a unified systematic
framework for systems with broken symmetry by presenting the
corresponding Euler-Poincar\'e  and Lagrange-Poincar\'e approaches
to formulate their dynamics. }

\medskip
\rem{ 
 These two approaches turn out
to be equivalent whenever the $\mathcal{O}$-action on any manifold
$M$ is transitive, so that $M\simeq\mathcal{O/P}$. If this is not
the case, then one still concludes that the coset space
$\mathcal{O/P}$ is the configuration manifold in the
Lagrange-Poincar\'e setting, provided the $\mathcal{P}$-action is
free and proper (a condition often satisfied in physics). \todo{F: I
am not sure to understand the two above sentences. EP and Lgr-P are
always equivalent. The action of $\mathcal{P}$ on $\mathcal{O}$ is
always proper.}
}     
We shall mainly concentrate on the symmetry properties of finite
dimensional systems, while the last part of the paper extends the
treatment to consider the infinite dimensional cases. In terms of
complex fluid systems, this means that we shall focus on the symmetry
properties of the fluid single-particle dynamics. The question of
how the underlying geometric structure is preserved in passing from
the microscopic single-particle approach to the macroscopic
continuum description will be the subject of future work.


\paragraph{Notation.} We have used the notation $\mathcal{O}$ for the order parameter group of a system with broken symmetry and have denoted by $M$ the order parameter space, on which $\mathcal{O}$ act transitively. From now on we will use the notation $\mathcal{O}$ only when the group acts transitively on $M$. For general actions, the Lie group will be denoted by $G$.

\subsection{Geometric setting for symmetry breaking}\label{subsec_math_background}

This section introduces some of the mathematical theory of systems with
broken symmetry and anticipates how the coset space $\mathcal{O/P}$
emerges in a more rigorous framework.

Let $\mathcal{O}$ be a Lie group (the order parameter group),
acting \textit{transitively} on the \textit{left} of a manifold
$M$ (the order parameter space). Choose an element $n_0\in M$, and
consider the isotropy subgroup $\mathcal{P}:=\mathcal{O}_{n_0}$.
We have the isomorphism $\mathcal{O}/\mathcal{P}\rightarrow M,\quad
[\chi]=\chi\,\mathcal{P}\mapsto \chi n_0$,
where $\mathcal{P}$ acts on $\mathcal{O}$ by right multiplication.
The dynamics is described by a $\mathcal{P}$-invariant Lagrangian
$L_{n_0}:T\mathcal{O}\rightarrow \mathbb{R}$,
which produces the Euler-Lagrange equations on $T\mathcal{O}$.
 Now we notice that, from the transitivity of the
action, any  $\mathcal{P}$-invariant function $L_{n_0}$ determines a
unique function
\[
L:T\mathcal{O}\times
M\rightarrow\mathbb{R},\quad L(v_\chi,\varphi n_0):=L_{n_0}(v_\chi\varphi)\,,\quad\text{with}\quad \varphi\in \mathcal{O}
\]
that is  $\mathcal{O}$-invariant under the right action of
$\psi\in\mathcal{O}$ given by $(v_\chi,n)\mapsto (v_\chi\psi,\psi^{-1}n)$.
When $L_{n_0}$ is hyperregular, one can obtain the Hamiltonian
formulation associated to a $\mathcal{P}$-invariant Hamiltonian
$H_{n_0}:T^*\mathcal{O}\rightarrow\mathbb{R}$ by Legendre
transformation. By proceeding analogously, one defines an
 $\mathcal{O}$-invariant Hamiltonian
 $H(\alpha_\chi,\varphi n_0)=H_{n_0}(\alpha_\chi\varphi)$ on $T^*\mathcal{O}\times
 M$ (with the obvious notation $\alpha_\chi\in T^*\mathcal{O}$).

As explained in the previous section, this paper uses symmetry reduction theory to present the equations of motion on the corresponding reduced space.
In particular, given a arbitrary \textit{left} action $n\mapsto gn$ of a Lie group $G$ on a manifold $M$ and a $G$-invariant function $L:TG\times M\rightarrow\mathbb{R}$ under the \textit{right} action $(v_h,n)\mapsto (v_hg,g^{-1}n)$,

\medskip\noindent
{\it
we develop two approaches:\\
}

\noindent $(1)$ \textbf{Euler-Poincar\'e approach to symmetry breaking} (Section \ref{new_EP}).
We make use of the diffeomorphism
\[
(TG\times M)/G\rightarrow \mathfrak{g}\times M,\quad [v_g,n]\mapsto \left(v_gg^{-1},g n\right),
\]
so that the function $L$ induces a reduced Lagrangian
$l:\mathfrak{g}\times M\rightarrow \mathbb{R}$, which is defined by
$l(v_gg^{-1},g n)= L(v_g,n)$.
We obtain the equation of motion on the reduced space $\mathfrak{g}\times M$.
This construction
generalizes the well known Euler-Poincar\'e theory for semidirect
products. We explore the Hamiltonian side and determine the
associated noncanonical Poisson brackets.
In this context, we obtain
a restriction of the Lie-Poisson bracket of \cite{KrMa1987}, who
considered the case when the manifold $M$ is Poisson. Indeed, the
Legendre transform of the Euler-Poincar\'e equations yields the
bracket in \cite{KrMa1987} in the special case when $M$ is endowed
with the trivial Poisson structure.
As from the discussion above, a
particularly interesting situation is the case of a transitive
action of $G=\mathcal{O}$ on $M$, so that $M$ is isomorphic to the
coset space $\mathcal{O/P}$,  $\mathcal{P}$ being the isotropy
subgroup of $\mathcal{O}$ for a fixed element $n_0\in M$. This
result
justifies the usual emergence of coset spaces for symmetry breaking in terms of Euler-Poincar\'e reduction.\\

\noindent $(2)$ \textbf{Lagrange-Poincar\'e approach to symmetry
breaking} (Section \ref{sec_Lagr_red}). We shall derive
Lagrange-Poincar\'e equations by applying standard Lagrangian
reduction to the ordinary Lagrangian
$L_{n_0}:TG\rightarrow\mathbb{R}$, defined by
$L_{n_0}(v_g):=L(v_g,n_0)$ for a \textit{fixed reference element}
$n_0\in M$. Since $L_{n_0}$ is invariant under the isotropy group
$G_{n_0}$ of $n_0$, this process involves the quotient $TG/G_{n_0}$.
For simplicity, we suppose here that $G=\mathcal{O}$ acts
transitively on $M$ and denote by $\mathcal{P}$ the isotropy group
of $n_0$ (the case of a general action is treated in Section
\ref{sec_LPA}). In order to explicitly write the reduced equations,
one needs a connection on the right principal bundle
$\mathcal{O}\rightarrow \mathcal{O}/\mathcal{P}\simeq M$. Using this
connection, the reduced tangent bundle $T\mathcal{O/P}$ can be
identified with the vector bundle
$TM\oplus_M\widetilde{\mathfrak{p}}$ over $M\simeq\mathcal{O/P}$,
where $\oplus$ denotes Whitney sum, $\widetilde{\mathfrak{p}}$
denotes the adjoint bundle
$(\mathcal{O}\times\mathfrak{p})/\mathcal{P}$ and gothic fonts
denote Lie algebras of corresponding Lie groups, as usual. This is a
general construction \cite{CeMaRa2001}, which is here applied to the
coset bundle $\mathcal{O}\rightarrow\mathcal{O}/\mathcal{P}$.
\medskip

\subsection{Summary of main results}
After slightly extending the ordinary Euler-Poincar\'e theory to account
for order parameter manifolds, this construction is applied to recover the
equations of motion for nematic particles, thereafter formulating new Euler-Poincar\'e
equations for biaxial nematic particles and $V$-shaped molecules.
As a further step, the Euler-Poincar\'e equations for ordinary nematic and biaxial particles are also expressed in terms of the corresponding alignment tensors, e.g. \makebox{${\sf Q}=1/2\left({\bf nn}^T-1/3\,{\bf I}\right)$} for nematics.

The second important result concerns the application of the
Lagrange-Poincar\'e reduction method to the coset bundle $G\to
G/G_{n_0}$. Indeed, this method is shown to be extremely powerful
when combined with the definition of a mechanical connection on
$G\to G/G_0$. The result of this combination is the formulation of
the nematic particle motion on the configuration space
$M=SO(3)/O(2)$, that allows to relate systematically the variables
$({\bf n},\dot{\bf n})\in TM$ with the corresponding
Euler-Poincar\'e variables $(\boldsymbol\nu,{\bf n})\in
\mathfrak{so}(3)\times M$. This step requires particular care
because of the appearance of the extra constant $r={\bf
n}\cdot\boldsymbol\nu$, belonging to the commutative Lie algebra
$\mathfrak{o}(2)$. For example, setting $r=0$ neglects rotations
about $\bf n$, consistently with the rod-like nature of nematic
particles, while this is not sensible for heavy top dynamics, which
turns out to be also explained by exactly the same procedure (upon
replacing $O(2)$ with $SO(2)$). It is worth mentioning that the
possibility of a non-zero quantity $r={\bf n}\cdot\boldsymbol\nu$
for the heavy top does not represent a limit of the theory: rather
it represents a strong point of this approach that enables to split
Euler-Lagrange (EL) dynamics on $T(G/G_{n_0})=TS^2$ from
Euler-Poincar\'e dynamics on the Lie algebra
$\mathfrak{g}_{n_0}=\mathfrak{so}(2)$ (i.e. trivial motion, in this
specific case). As a final result of this method, we obtain the
Ericksen-Leslie equation as a covariant EL equation on $TS^2$ and we
extend its validity to the heavy top case, including a non-zero
$r\in\mathfrak{so}(2)$.

In the last part of this paper, the finite-dimensional treatment is
extended to the infinite-dimensional fluid theory of liquid
crystals. This is performed in two reduction stages: the first
corresponding to the broken symmetry in the micromotion of nematic
particles and the second corresponding to the fluid relabeling
symmetry. On the Lagrange-Poincar\'e side, this requires the use of
a recent reduction procedure (known as `metamorphosis'), first
appeared in  the study of the shape evolution in image dynamics
\cite{HoTrYo2008}.
After establishing the direct correspondence
between the dynamics obtained by this approach and the
Euler-Poincar\'e reduced system, we perform the Legendre transform and apply a well known theorem
\cite{KrMa1987} to produce two equivalent Poisson brackets for the same set of equations. Moreover, we present the explicit form of the helicity invariant for nematic liquid crystals, as well as a whole class of Casimirs functionals for two dimensional incompressible flows.

\section{Euler-Poincar\'e equations and Lie-Poisson brackets}\label{new_EP}

\subsection{The Euler-Poincar\'e
approach}\label{subsec_lagr}

Let $G$ be a Lie group acting on the \textit{left\/} on a manifold $M$. We denote by $n\mapsto gn$ the action of $g\in G$ on $n\in M$. Then $G$ acts naturally on the \textit{right\/} on $TG\times M$ via the free action
\[
(v_g,n)\mapsto (v_gh,h^{-1}n)\,,
\]
where the action on the first factor is given by tangent lift of right translation on $G$, that is we denote
$v_gh:=TR_h(v_g)$,
where $TR_h$ is the tangent map to the right translation $R_h$ by the element $h\in G$.
Similarly, we will denote by $\alpha_gh$ the cotangent lifted action of right translation by an element $h\in G$, that is, we have
$\alpha_gh:=T^*R_{h^{-1}}(\alpha_g)$
where evidently $\alpha_g\in T_g ^*G$. The quotient space $(TG\times
M)/G$ can be identified with $\mathfrak{g}\times M$ via the
diffeomorphism
\begin{equation}\label{diffeo_EP}
[v_g,n]\mapsto (v_gg^{-1},gn).
\end{equation}
\begin{itemize}
\item Assume that we have a function $L:TG\times M\rightarrow\mathbb{R}$
which is \textit{right\/} $G$-invariant.
\item In particular, if $n_0\in M$, define the Lagrangian
\begin{equation}\label{def_L_n_0}
L_{n_0}:TG\rightarrow\mathbb{R},\quad L_{n_0}(v_g):=L(v_g,n_0).
\end{equation}
Then $L_{n_0}$
is \textit{right\/} invariant under the lift to $TG$ of the right action of $G_{n_0}$ on
$G$, where $G_{n_0}$ is the isotropy group of $n_0$.
\item Right $G$-invariance of $L$ allows us to define $l:\mathfrak{g}\times
M\rightarrow\mathbb{R}$ by
\begin{equation}\label{def_l}
l(v_gg^{-1},gn)=L(v_g,n).
\end{equation}
\item For a curve $g_t\in G$, let $\xi_t:=\dot{g}_tg_t^{-1}$ and
define the curve $n_t\in M$ as the unique solution of the following differential equation with time dependent coefficients
\[
\dot{n}_t=(\xi_t)_M(n_t),
\]
with initial condition $n_0$. Here $(\xi_t)_M\in\mathfrak{X}(M)$ denotes the infinitesimal generator associated to the time dependent Lie algebra element $\xi_t\in\mathfrak{g}$. The solution of this differential equation can be written as
$n_t=g_tn_0$.
\end{itemize}

\begin{theorem}\label{gen_EP} With the preceding notations, the following are
equivalent:
\begin{itemize}
\item[\bf{(i)}] With $n_0\in M$ held fixed, Hamilton's variational principle
\begin{equation}\label{Hamilton_principle}
\delta\int_{t_1}^{t_2}L_{n_0}(g,\dot{g})dt=0,
\end{equation}
holds, for variations $\delta g$ of $g$ vanishing at the endpoints.
\item[\bf{(ii)}] $g$ satisfies the Euler-Lagrange equations for $L_{n_0}$ on
$G$.
\item[\bf{(iii)}] The constrained variational principle
\begin{equation}\label{Euler-Poincare_principle}
\delta\int_{t_1}^{t_2}l(\xi,n)dt=0,
\end{equation}
holds on $\mathfrak{g}\times M$, upon using variations of the form
\[
\delta\xi=\frac{\partial\eta}{\partial t}-[\xi,\eta],\quad \delta n=\eta_M(n),
\]
where $\eta\in\mathfrak{g}$ vanishes at the endpoints.
\item[\bf{(iv)}] The Euler-Poincar\'e equations hold on $\mathfrak{g}\times M$:
\begin{equation}\label{EP}
\frac{\partial}{\partial t}\frac{\delta
l}{\delta\xi}+\operatorname{ad}^*_\xi\frac{\delta l}{\delta\xi}=\mathbf{J}\left(\frac{\delta
l}{\delta n}\right),
\end{equation}
where $\mathbf{J}:T^*M\rightarrow\mathfrak{g}^*$, given by $\langle\mathbf{J}(\alpha_m),\xi\rangle=\langle\alpha_m,\xi_M(m)\rangle$, is the momentum map associated to the cotangent lifted action of $G$ on $T^*M$.
\end{itemize}
\end{theorem}
The proof is an immediate generalization to that given in \cite{HoMaRa1998}, in the case where $M$ is a vector space on which $G$ acts by a representation.

\medskip

\begin{remark}\normalfont  The Euler-Poincar\'e equations \eqref{EP} can be rewritten in the form
\begin{equation}\label{cons_law}
\frac{\partial }{\partial t}\left[\operatorname{Ad}_g^*\frac{\delta l}{\delta\xi}\right]=\mathbf{J}\left(g^{-1}\frac{\delta
l}{\delta n}\right).
\end{equation}
Note that we have $g^{-1}\frac{\delta
l}{\delta n}\in T^*_{n_0}M$, since $n_0=g^{-1}n$.
\end{remark}

\begin{remark}[Relation to previous Euler-Poincar\'e formulations]\normalfont In the special case when $M$ is a vector space acted on by ordinary representations, we recover the formulation in \cite{HoMaRa1998}, useful for heavy tops and compressible fluids. Also, when the vector space $M$ is acted on by affine representations, the above picture recovers the results in \cite{GBRa2008}, useful for complex fluids.
\end{remark}
\begin{remark} [Terminology]\normalfont
Throughout the paper, we will refer to equations \eqref{EP} as \textit{Euler-Poincar\'e equations} or \textit{Euler-Poincar\'e equations for symmetry breaking} since they generalize the (\textit{pure}) \textit{Euler-Poincar\'e equations}
\[
\frac{\partial}{\partial t}\frac{\delta
l}{\delta\xi}+\operatorname{ad}^*_\xi\frac{\delta l}{\delta\xi}=0
\]
on a Lie algebra (see \cite{MaRa1999}).
\end{remark}

\begin{remark}\label{i_n_0}
\normalfont
Since the Lagrangian $L_{n_0}:TG\rightarrow \mathbb{R}$ is $G_{n_0}$-invariant, it induces a reduced Lagrangian $\ell_{n_0}:TG/G_{n_0}\rightarrow\mathbb{R}$. Denoting by $i_{n_0}:TG\rightarrow TG\times M$ the injection $v_g\mapsto (v_g,n_0)$, we can write $L_{n_0}=L\circ i_{n_0}$. By passing to the quotient spaces, the injection $i_{n_0}$ induces an injection
\[
\bar i_{n_0}:TG/G_{n_0}\rightarrow\mathfrak{g}\times M,\quad [v_g]_{G_{n_0}}\mapsto (v_gg^{-1},gn_0)
\]
whose image is $\mathfrak{g}\times \operatorname{Orb}(n_0)$, where $\operatorname{Orb}(n_0)\subset M$ denotes the orbit of $n_0$. We have the relation $\ell_{n_0}=l\circ\bar i_{n_0}$ between the two reduced Lagrangians $\ell_{n_0}$ and $l$.
\end{remark}

\begin{remark}[Including translational motion in physical space]\normalfont
As we remarked in the introduction, the present treatment neglects
translational motion in physical space $Q$. It is important to
emphasize that this simplification is completely irrelevant. Indeed,
translational motion may always be included a posteriori, upon
extending the Lagrangian $L: TG\times M\to\Bbb{R}$ to be defined on
the Cartesian product $TQ\times\left(TG\times M\right)$. Thus, by
assuming that $G$ acts trivially on $Q$, the new Lagrangian
$L_{n_0}(q,\dot{q},g,\dot{g})$ will produce the extra Euler-Lagrange
equation on $TQ$, which is then accompanied by the Euler-Poincar\'e
equation on $\mathfrak{g}\times M$. In conclusion, the reduced
Lagrangian $l:TQ\times\left(\mathfrak{g}\times M\right)$ produces
the equations
\begin{equation*}
\frac{\partial}{\partial t}\frac{\delta l}{\delta
\dot{q}}-\frac{\delta l}{\delta q}=0\,, \qquad
\frac{\partial}{\partial t}\frac{\delta
l}{\delta\xi}+\operatorname{ad}^*_\xi\frac{\delta
l}{\delta\xi}=\mathbf{J}\left(\frac{\delta l}{\delta n}\right),
\end{equation*}
where one has to remember the relation $\dot{n}=\xi_M(n)$.
\end{remark}

\subsection{The Lie-Poisson equations}\label{Ham_approach}

Consider a function $H: T ^\ast G \times M\rightarrow \mathbb{R}$ \textit{right\/}
invariant under the $G$-action on $T ^\ast G \times M$ given by $(\alpha_g,n)\mapsto (\alpha_gh,h^{-1}n)$. In particular,
the
function $H_{n_0}: T^*G\rightarrow\mathbb{R}$ defined by $H_{n_0}(\alpha_g)= H(\alpha_g,n_0)$ is invariant under the induced action
of the isotropy subgroup $G_{n_0}: = \{g \in G \mid g n_0 = n_0\}$
for
any $n_0 \in M$. The reduced Hamiltonian $h:\mathfrak{g}^*\times M\rightarrow\mathbb{R}$ is defined by $h(\alpha_gg^{-1},gn):=H(\alpha_g,n)$.

\begin{theorem}\label{Hamilton_side}
Fix an element $n_0\in M$. For $\alpha\in T^*_gG$ and
$\mu:=\alpha g^{-1}\in\mathfrak{g}^*$, the following are
equivalent:
\begin{itemize}
\item[\bf{i}] The curve $\alpha$ satisfies Hamilton's equations for
$H_{n_0}$ on $T^*G$.
\item[\bf{ii}] The curve $(\mu,n)\in \mathfrak{g}^*\times M$ is a solution of the Lie-Poisson equations
\begin{equation}\label{red_Ham}
\left\{
\begin{array}{l}
\vspace{0.2cm}\displaystyle\dot\mu=-\operatorname{ad}^*_{\frac{\delta
h}{\delta \mu}}\mu-\mathbf{J}\left(\frac{\delta h}{\delta n}\right)\\
\displaystyle\dot n=\left(\frac{\delta h}{\delta
\mu}\right)_M(n),\quad n(0)=n_0.
\end{array}
\right.
\end{equation}
These equations are Hamiltonian relative to the Poisson bracket
\begin{equation}\label{red_Poisson}
\{f,g\}(\mu,n)=\left\langle\mu,\left[\frac{\delta
f}{\delta\mu},\frac{\delta g}{\delta\mu}\right]\right\rangle+\left\langle
\mathbf{J}\left(\frac{\delta f}{\delta n}\right),\frac{\delta g}{\delta\mu}\right\rangle-\left\langle
\mathbf{J}\left(\frac{\delta g}{\delta n}\right),\frac{\delta f}{\delta\mu}\right\rangle
\end{equation}
on $\mathfrak{g}^*\times M$.
\end{itemize}
As on the Lagrangian side, the evolution of the variable $n$ is given by
$n=gn_0$.
\end{theorem}
\textbf{Proof.} Canonical Hamilton's equations for $H_{n_0}$ on $T^*G$ are equivalent to Hamilton's equation for $H$ on $T^*G\times M$, endowed with the direct sum of the canonical Poisson bracket on $T^*G$ and the zero Poisson bracket on $M$, with initial value $n_0$. Since $H$ is $G$ invariant, one can apply Poisson reduction to obtain the reduced Hamiltonian equations on the quotient manifold $(T^*G\times M)/G$ under the right action $(\alpha_g,n)\mapsto (\alpha_gh, h^{-1}n)$. We identify the quotient manifold $(T^*G\times M)/G$ with $\mathfrak{g}^*\times M$, via the diffeomorphism $[\alpha_g,n]\mapsto (\alpha_gg^{-1},gn)$. Using Proposition 2.1 in \cite{KrMa1987}, in the particular case where the Poisson structure is trivial on $M$, we obtain the reduced bracket \eqref{red_Poisson}. One then observes that the Hamilton's equations associated to this bracket are given by \eqref{red_Ham}.

\paragraph{Legendre transformation.} The preceding theorem is compatible with Theorem \ref{gen_EP}. Indeed, we can start with a Lagrangian $L_{n_0}:TG\rightarrow\mathbb{R}$ as in
\S\ref{subsec_lagr}, that is, we have a function $L:TG\times
M\rightarrow\mathbb{R}$ which is right $G$-invariant under the action
$(v_g,n)\mapsto
(v_gh,h^{-1}n)$, such that
$L_{n_0}(v_g)=L(v_g,n_0)$. Then $L_{n_0}$ is right invariant under the lift to
$TG$ of the right action of the isotropy group $G_{n_0}$ on $G$. Suppose that the Legendre transformation $\mathbb{F}L_{n_0}$ is
invertible and form the corresponding Hamiltonian
$H_{n_0}=E_{n_0}\circ\mathbb{F}L_{n_0}^{-1}$, where $E_{n_0}$ is the energy of
$L _{n_0}$, see \cite{MaRa1999}. Then the function $H: T^\ast G \times M
\rightarrow \mathbb{R}$ so defined is $G$-invariant and one can apply this
theorem. At the level of the reduced space, to a reduced Lagrangian
$l:\mathfrak{g}\times M\rightarrow\mathbb{R}$ we associate the reduced
Hamiltonian $h:\mathfrak{g}^*\times M\rightarrow\mathbb{R}$ given by
\[
h(\mu,n):=\langle\mu,\xi\rangle-l(\xi,n),\quad\mu=\frac{\delta l}{\delta\xi}.
\]
Since
\[
\frac{\delta h}{\delta\mu}=\xi\quad\text{and}\quad\frac{\delta h}{\delta
n}=-\frac{\delta l}{\delta n},
\]
we see that the Hamilton's equations for $h$ on $\mathfrak{g}\times M$ are
equivalent to the Euler-Poincar\'e equation \eqref{EP} for $l$ together
with the kinematic equation
\[
\dot n=\xi_M(n).
\]
Again, when $M$ is a vector space acted on by a linear or affine representation, the above Lie-Poisson reduction recovers the results in \cite{MaRaWe1984} and \cite{GBRa2008}, respectively.

\medskip

\section{Transitive actions: the dynamics of nematic
molecules}

This section considers the interesting case of a Lie group $G$
acting transitively from the \textit{left}  on a manifold $M$,
called \textit{order parameter space}. In this particular case we
use the notation $G=\mathcal{O}$, since we think of $G$ as the
\textit{order parameter group} or the \textit{broken symmetry}.

This is a typical situation in condensed matter physics, which
describes exactly the situation presented in the Introduction. The
particular choice of the group $\mathcal{O}$ and the manifold $M$
depends on the system under consideration. For example, the common
case $\mathcal{O}=SO(3)$ and $M=\Bbb{R}P^2\simeq S^2/\Bbb{Z}_2$
yields ordinary nematic particles, while $\mathcal{O}=SO(3)$ and
$M=SO(3)/D_2$ yields biaxial nematics. Here $D_2$ is the dihedral
group generated by $180^\circ$-rotations and reflections. Many other
choices are certainly possible and we address the reader to
\cite{Mermin1979,Mi1980} for reviews on these topics.

Before proceeding, we emphasize that the present approach focuses on
the rotational dynamics of a fixed nematic particle (e.g. ordinary
or biaxial) in an external potential. Although this situation is
physically improbable, its extension to more realistic
configurations is straightforward. For example, translational motion
in physical space $Q$ can be included as discussed previously, while
the dynamics of $N$ interacting particles requires extending the
present treatment to the space $Q^N\times(\mathcal{O\times O/P})^N$.
The continuum limit of a nematic lattice and the full hydrodynamic
model are analyzed in section \ref{continuum_OP}.

As we have shown in the Introduction, if we fix a reference point
$n_0\in M$ and consider the isotropy group
$\mathcal{P}=\mathcal{O}_{n_0}=\{\psi\in\mathcal{O}\mid \psi
n_0=n_0\}$, then the orbit map
\[
\mathcal{O}\rightarrow M,\quad \chi\mapsto \chi n_0
\]
induces a diffeomorphism
\[
\Psi^{n_0}:\mathcal{O}/\mathcal{P}\rightarrow M,\quad
[\chi]=\chi\mathcal{P}\mapsto \chi n_0\,.
\]
The reference point $n_0$ corresponds to the coset $[e] =
e\mathcal{P} = \mathcal{P}$ of $\mathcal{O}/\mathcal{P}$. Since the
action is transitive, given a reference point $n_0$, a
$\mathcal{P}$-invariant Lagrangian
$L_{n_0}:T\mathcal{O}\rightarrow\mathbb{R}$ completely determines
the right invariant function $L:T\mathcal{O}\times
(\mathcal{O/P})\rightarrow\mathbb{R}$. Indeed, it suffices to define
$L(v_\chi,n):=L_{n_0}(v_\chi\psi)$,
where $\psi\in\mathcal{O}$ is such that $n=\psi n_0$, by
transitivity of the action. In particular, $L_{n_0}$ determines the
Lagrangians $L_{m_0}$ for all other reference points $m_0\in M$. Note also that the injection $\bar i_{n_0}$ is here a
diffeormorphism since $\operatorname{Orb}(n_0)=M$, that is, we have
the diffeomorphism
\begin{equation}\label{lisboa}
\bar i_{n_0}:T\mathcal{O}/\mathcal{P}\rightarrow \mathfrak{o}\times
M,\quad [v_\chi]\mapsto (v_\chi\chi^{-1},\chi n_0).
\end{equation}

\subsection{Dynamics of uniaxial nematic molecules}

Theorem \ref{gen_EP} applies to nematic particles without changes. Thus, in this case $\mathcal{O}=SO(3)$,
$\mathcal{O/P}=\Bbb{R}P^2$ and we have the following Euler-Poincar\'e
equations arising from the reduced Lagrangian
$l:\mathfrak{so}(3)\times \Bbb{R}P^2\to\Bbb{R}$:
\begin{equation}\label{EP_nematics}
\frac{d}{dt}\frac{\delta l}{\delta
\boldsymbol{\nu}}=\boldsymbol{\nu}\times\frac{\delta l}{\delta
\boldsymbol{\nu}}+{\bf n}\times\frac{\delta l}{\delta \bf n}
\,,\qquad\quad \dot{\bf n}=\boldsymbol\nu\times {\bf n}
\end{equation}
where we have confused the space $\Bbb{R}P^2$ of unsigned unit
vectors with ordinary unit vectors in $\Bbb{R}^3$, thereby
emphasizing the strict relation between rotational motion of nematic
particles and the heavy top dynamics. Indeed, it is easier to
consider the Lagrangian $l:\mathfrak{so}(3)\times
(S^{2\!}/\Bbb{Z}_2) \to\Bbb{R}$ as a function
$l:\mathfrak{so}(3)\times S^2\to\Bbb{R}$ possessing the
$\Bbb{Z}_2$-symmetry
$l(\boldsymbol{\nu},\mathbf{v})=l(\boldsymbol{\nu},-\mathbf{v})$,
for all $\boldsymbol{\nu}\in\mathfrak{so}(3)$ and $\mathbf{v}\in
S^2$. This is how equations \eqref{EP_nematics} are written: one
chooses a representative vector for $\mathbf{n}$ to compute the
equations (i.e. a unit vector $\mathbf{v}\in S^2$ such that
$[\mathbf{v}]=\mathbf{n}$), and then checks that the result does not
depend on the chosen representative.

The Lagrangian for the nematic particle is
$L_{\mathbf{n}_0}:TSO(3)\rightarrow \mathbb{R}$
\[
L_{\mathbf{n}_0}(\chi,\dot{\chi})=\frac{1}{2}j\left|\dot{\chi}\right|^2-\Phi_{\mathbf{n}_0}(\chi),
\]
where $j>0$ is the moment of inertia and $|\cdot|$ denotes the norm
associated to the $SO(3)$-invariant Riemannian metric $\langle
\dot\chi_1,\dot{\chi}_2\rangle=\operatorname{Tr}(\dot\chi_1^T\dot\chi_2)$
on $SO(3)$.
We notice that the inertia tensor $\mathbb{I}=j\,\mathbf{I}$ leads to the schematic description of a nematic molecule in terms of a spherical rigid body endowed with a special direction $\mathbf{n}_0$.
The unreduced potential energy is
$\Phi_{\mathbf{n}_0}:SO(3)\rightarrow\mathbb{R}$. An example of such
an external potential for a nematic particle is given by the
quadratic expression:
$\Phi_{\mathbf{n}_0}(\chi)=\lambda\left|\mathbf{n}_0\cdot\chi^{-1}{\bf
k}\right|^{2\!}/2$, where $\lambda\in\Bbb{R}$ is a constant parameter and
${\bf k}$ plays the role of an external force field, e.g. an external magnetic
(or electric) field. (Another
possibility could be
$\Phi_{\mathbf{n}_0}(\chi)=\lambda\left|\mathbf{n}_0\times\left(\chi^{-1}{\bf
k}\right)\right|^{2\!}/2$). Besides the quadratic form of the potential,
the important difference from the heavy-top case is that we now
identify $\mathbf{n}_0$ with $-\mathbf{n}_0$ and the same for the
fixed vector ${\bf k}$, which thus can point upwards or downwards
without distinction. The presence of the potential
$\Phi_{\mathbf{n}_0}$ breaks the symmetry, since $L_{\mathbf{n}_0}$
is only $O(2)$-invariant (or better $D_\infty$-invariant, as
mentioned in the Introduction) under right translation. If
$\mathbf{n}_0\in S^2/\Bbb{Z}_2$ is considered as an arbitrary
director and we define $L:TSO(3)\times (S^2/\Bbb{Z}_2)\to\Bbb{R}$ by
$L(\chi,\dot{\chi},\mathbf{n}_0):=L_{\mathbf{n}_0}(\chi,\dot{\chi})$,
then $L$ is $SO(3)$-invariant under the right action
$(\chi,\dot{\chi},\mathbf{n}_0)\mapsto
(\chi\psi,\dot{\chi}\psi,\psi^{-1}\mathbf{n}_0)$. The argument  is
the same in the heavy-top case, as shown by the following
calculation:
\[
\Phi(\chi,\mathbf{n}_0)=\frac\lambda2\left|\mathbf{n}_0\cdot\chi^{-1}{\bf
k}\right|^2 =\frac\lambda2\left|\chi\mathbf{n}_0\cdot{\bf k}\right|^2
=\Phi(\chi\chi^{-1},\chi\mathbf{n}_0)=:\phi(\chi\mathbf{n}_0)
\]
where we have introduced the reduced potential
$\phi:S^2/\Bbb{Z}_2\to\Bbb{R}$. Thus, by $SO(3)$-invariance, $L$
induces the reduced Lagrangian $l$ governing the rotational dynamics
of a single nematic particle:
\begin{equation}\label{Nem-EP-Lagrangian}
l(\boldsymbol{\nu},
\mathbf{n})=\frac12\,j\,|\boldsymbol{\nu}|^2-\phi(\mathbf{n})
=\frac12\,j\,|\boldsymbol{\nu}|^2-\frac\lambda2\left|\mathbf{n}\cdot{\bf
k}\right|^2
\end{equation}
where $\boldsymbol{\nu}\in\mathfrak{so}(3)$, and $ \mathbf{n}\in
S^2/\Bbb{Z}_2$. In this case, the Euler-Poincar\'e equations for
nematics reads
\begin{equation}\label{equation_nematics_EP}
j\dot{\boldsymbol{\nu}}=\nabla\phi(\mathbf{n})\times\mathbf{n}=\lambda\left|\mathbf{n}\cdot\mathbf{k}\right|\mathbf{k}\times\mathbf{n},\quad \dot{\mathbf{n}}=\boldsymbol{\nu}\times\mathbf{n}
\end{equation}
by \eqref{EP_nematics}. Moreover, the Legendre transform
$\boldsymbol{\mu}=\delta l/\delta\boldsymbol\nu$ produces the
reduced quadratic Hamiltonian on
$\mathfrak{so}(3)^*\times\mathbb{R}P^2$:
\begin{equation}\label{Nem-EP-Hailtonian}
h(\boldsymbol{\mu},{\bf n})=\frac1{2j}|\boldsymbol{\mu}|^2+\frac\lambda2\left|\mathbf{n}\cdot{\bf
k}\right|^2
\end{equation}
The corresponding Lie-Poisson equations are
\[
\frac{d}{dt}\boldsymbol{\mu}=\frac{\delta h}{\delta
\boldsymbol{\mu}}\times\boldsymbol{\mu}+\frac{\delta h}{\delta \bf
n}\times{\bf n} \,,\qquad\quad \dot{\bf n}=\frac{\delta
h}{\delta \boldsymbol{\mu}}\times{\bf n}
\]
while the Lie-Poisson bracket of functions
$f,g\in\mathcal{F}\!\left(\mathfrak{so}(3)^*\times\mathbb{R}P^2\right)$
is
\begin{equation}\label{Poisson_nematic_particle}
\{f,g\}(\boldsymbol{\mu},\mathbf{n})=\boldsymbol{\mu}\!\cdot\!\left(\frac{\delta
f}{\delta \boldsymbol{\mu}}\times \frac{\delta g}{\delta
\boldsymbol{\mu}}\right)+\mathbf{n}\!\cdot\!\left(\frac{\delta
f}{\delta \mathbf{n}}\times \frac{\delta g}{\delta
\boldsymbol{\mu}}-\frac{\delta g}{\delta \mathbf{n}}\times
\frac{\delta f}{\delta \boldsymbol{\mu}}\right).
\end{equation}

A special case of physical interest in microscopic theories of
liquid crystals is the case of an ensemble of $N$ interacting
nematic particles moving in the physical space $Q$. In this case,
the unreduced Lagrangian
$L_{\mathbf{n}^{(0)}_1,...,\mathbf{n}_N^{(0)}}:TQ^N\times TSO(3)^N\rightarrow\mathbb{R}$ is given by
\[
L_{\mathbf{n}^{(0)}_1,...,\mathbf{n}_N^{(0)}}(\chi_1,...,\chi_N)=\frac12\sum_{i=1}^N\|\dot{q}_i\|^2+\frac{1}{2}\,j\sum_{i=1}^N|\dot\chi_i|^2-\sum_{i\neq
k}\Phi_{\mathbf{n}^{(0)}_1,...,\mathbf{n}_N^{(0)}}(q_i,q_k,\chi_i,\chi_k),
\]
where the first norm is associated to a Riemannian metric on $Q$, while $\mathcal{O}=SO(3)^N$ is endowed with
the direct product group structure. The order parameter space is now
$M=(S^2/\Bbb{Z}_2)^N$. As before, Euler-Poincar\'e reduction yields
the reduced Lagrangian $l:TQ^N\times\left(\mathfrak{so}(3)\times
S^2/\Bbb{Z}_2\right)^{\!N}\to\Bbb{R}$, which is written as
\begin{equation}\label{nematic-EPLagrangian}
l(q_i,\boldsymbol{\nu}_i,\mathbf{n}_i)=\frac12\sum_{i=1}^N\|\dot{q}_i\|^2+\frac12\,j\sum_{i=1}^N\,|\boldsymbol{\nu}_i|^2-\sum_{i\neq
k}\phi(q_i,q_k,\mathbf{n}_i,\mathbf{n}_k)
\end{equation}
and produces Euler-Lagrange equations on $TQ^N$ and Euler-Poincar\'e
equations on
$\left(\mathfrak{so}(3)\times\mathbb{R}P^2\right)^{\!N}$. This
approach is adopted in the physics literature to formulate
microscopic approaches for nematic liquid crystals
\cite{Nemtsov1975,LeTo1977,BePr1987}. In this framework, when
$Q=\Bbb{R}^n$, the reduced potential is usually expressed as (cf.
\cite{LeTo1977})
\[
\phi(q_i,q_k,\mathbf{n}_i,\mathbf{n}_k)=V(q_i-q_k)+W(q_i-q_k)\,
\left|{\bf n}_i\cdot{\bf n}_k\right|^2.
\]
where the angular factor $\left|{\bf n}_i\cdot{\bf n}_k\right|^2$ is
reminiscent of the well known Maier-Saupe potential, in alternative
to Onsager's expression $\left|{\bf n}_i\times{\bf n}_k\right|^2$.
In what follows, however, we shall pursue the question of how
rotational motion is affected by symmetry breaking, without
considering translational motion in physical space, since the latter
can be always considered a posteriori, once the reduction has been
carried out on the rotational part of the Lagrangian.

\rem{ 
\todo{F: Are the Lagrangian for other kind of particles (like
biaxial nematics or smectics) always of the above form: kinetic -
potential?\\$\,$\\

If yes, I will write the text below on p.9 in the paragraph
``Transitive actions, order parameter and nematic particles'':

The Lagrangians $L_{n_0}:T\mathcal{O}\rightarrow\mathbb{R}$ are
usually of the form
\[
L_{n_0}(\dot\chi)=\frac{1}{2}|\dot\chi|^2-\Phi(\chi n_0),
\]
where the norm is taken relative to a $\mathcal{O}$-invariant metric
and $\Phi:M\rightarrow\mathbb{R}$ is a potential function. The
reduced Lagrangian reads $l:\mathfrak{o}\times
M\rightarrow\mathbb{R}$, $l(\nu,n)=\frac{1}{2}|\nu|^2-\Phi(n)$. The
expression of $L_{n_0}$ clearly explain the breaking symmetry
phenomenon: the presence of the potential $\Phi$ (i.e. the presence
of the parameter $n_0$) breaks the symmetry from $\mathcal{O}$ to
$\mathcal{P}$.

C: To my knowledge, it is not clear what the Hamiltonian/Lagranian
is for generic order parameter dynamics. I will have a look into
this.

}

} 


\begin{remark}[The heavy top revisited]\normalfont
The same Euler-Poincar\'e
approach also applies to the simpler case when the order parameter
manifold is $M=S^2$. In the case of symmetry breaking, the
Lagrangian $L:TSO(3)\to\Bbb{R}$ is not $SO(3)$-invariant. This is
exactly the heavy-top case (see equation \eqref{HT-lagr}), which
involves a Lagrangian $L$ that is only \makebox{$SO(2)$-invariant}
(rotations with respect to the vertical axis). The transitivity of
the \makebox{$SO(2)$-action} on $SO(3)$ recovers the well known
isomorphism $S^2=SO(3)/SO(2)$, thereby producing a reduced
Lagrangian $l:\mathfrak{so}(3)\times (SO(3)/SO(2))\to\Bbb{R}$. This
approach recovers the well known treatment of heavy top dynamics
(cf. e.g. \cite{HoMaRa1998}), avoiding the necessity of identify
points in $S^2$ with vectors in $\Bbb{R}^3$.
\end{remark}

\subsection{Alignment tensor dynamics for uniaxial nematics}

Although the order parameter of a nematic liquid crystal often considered as a director, such a quantity is also difficult to work with analytically. In fact, the Landau-de Gennes theory typically involves the components of a symmetric matrix ${\sf Q}$, usually known as alignment tensor.
In matrix notation, this tensor is given by (cf. e.g. \cite{Chandra1992})
\begin{equation}
{\sf Q}=\frac12\left(\mathbf{n}\mathbf{n}^T-\frac13\,\mathbf{I}\right)
\end{equation}
where $\mathbf{n}\mathbf{n}^T$ is also denoted as the dyadic form $\mathbf{n}\mathbf{n}$. In order to write the Poisson bracket \eqref{Poisson_nematic_particle} in terms of the $\sf Q$, we first need to compute functional derivatives. This is done by simply imposing the equality $\delta f(\boldsymbol\sigma,\mathbf{n})=\delta f(\boldsymbol\sigma,{\sf Q})$, which yields the relation
\begin{equation*}
\frac{\delta f}{\delta \mathbf{n}}=\mathbf{n}^T\frac{\delta
f}{\delta \sf Q}
\end{equation*}
At this point, it suffices to express the following term in terms of the alignment tensor:
\begin{align*}
\mathbf{n}\cdot\frac{\delta f}{\delta
\boldsymbol{\mu}}\times\frac{\delta g}{\delta
\mathbf{n}}
&\ =
-\operatorname{Tr}\!\left(\mathbf{n}^T\frac{\delta g}{\delta
{\sf Q}}\frac{\delta f}{\delta \hat{\boldsymbol{\mu}}}\mathbf{n}\right)
=-\operatorname{Tr}\!\left(\left(2{\sf Q}+\frac13\,\mathbf{I}\right)\frac{\delta
g}{\delta {\sf Q}}\frac{\delta f}{\delta \hat{\boldsymbol{\mu}}}\right)
\\
&\ =-2\operatorname{Tr}\!\left({\sf Q}\,\frac{\delta g}{\delta
{\sf Q}}\frac{\delta f}{\delta \hat{\boldsymbol{\mu}}}\right)
=-\operatorname{Tr}\!\left(\left[{\sf Q},\,\frac{\delta g}{\delta
{\sf Q}}\right]\frac{\delta f}{\delta \hat{\boldsymbol{\mu}}}\right)
 =\operatorname{Tr}\!\left({\sf Q}\left[\frac{\delta f}{\delta
\hat{\boldsymbol{\mu}}}\,,\frac{\delta g}{\delta {\sf Q}}\right]\right)
\end{align*}
where we have introduced the antisymmetric hat matrix
$\hat{\mu}_{ij}=\varepsilon_{ijk}{\mu}_k$ and we have used the fact
that ${\sf Q}$ is symmetric. Thus, the bracket \eqref{Poisson_nematic_particle} becomes
\begin{equation}\label{Poisson_ordpartensor}
\{f,g\}(\boldsymbol{\mu},\mathbf{n})=\boldsymbol{\mu}\!\cdot\!\left(\frac{\delta
f}{\delta \boldsymbol{\mu}}\times \frac{\delta g}{\delta
\boldsymbol{\mu}}\right)+\operatorname{Tr}\!\left(\!{\sf Q}\left(\left[\frac{\delta
f}{\delta \hat{\boldsymbol{\mu}}}\,,\frac{\delta g}{\delta
{\sf Q}}\right]-\left[\frac{\delta g}{\delta \hat{\boldsymbol{\mu}}}\,,\frac{\delta
f}{\delta {\sf Q}}\right]\right)\right)\,,
\end{equation}
which is the Lie-Poisson structure associated to the semidirect product $SO(3)\,\circledS\,\operatorname{Sym}(3)$, relative to the representation ${\sf Q}\mapsto\chi{\sf Q}\chi^{-1}$. Consequently, we have shown that the map
$\left(\boldsymbol{\mu},\mathbf{n}\right)\mapsto\left(\boldsymbol{\mu},
{\sf Q}\right)\in\mathfrak{so}^*(3)\times\operatorname{Sym}(3)$ is Poisson relative to the Lie-Poisson brackets \eqref{Poisson_nematic_particle} and \eqref{Poisson_ordpartensor}. This result should not come as surprise, since the map $\mathbf{n}\mapsto
{\sf Q}$ is equivariant with respect to the $SO(3)$ actions $\mathbf{n}\mapsto\chi\mathbf{n}$ and ${\sf Q}\mapsto\chi{\sf Q}\chi^{-1}$.

The resulting Lie-Poisson equations on
$\mathfrak{so}^*(3)\times\operatorname{Sym}(3)$ are
\[
\frac{d}{dt}\boldsymbol{\mu}=\frac{\delta h}{\delta
\boldsymbol{\mu}}\times\boldsymbol{\mu}+\!\overrightarrow{\,\left[{\sf Q},\frac{\delta
h}{\delta {\sf Q}}\right]\,} \,,\qquad\quad \dot{{\sf Q}}=\left[\frac{\delta
h}{\delta \hat{\boldsymbol{\mu}}},{\sf Q}\right]
\]
with the notation
$\overrightarrow{A\,}_{\!i}=\varepsilon_{ijk}A_{jk}$ and the
following Hamiltonian (up to an irrelevant constant) arising from its previous expressions
\eqref{Nem-EP-Hailtonian}:
\begin{equation}\label{Q-EP-Hailtonian}
h(\boldsymbol{\mu},Q)=\frac1{2j}|\boldsymbol{\mu}|^2+\frac\lambda2\,\mathbf{k}^T
Q\,\mathbf{k}\,.
\end{equation}
Notice that more general potential terms of the type $\mathbf{k}_1^T
Q\,\mathbf{k}_2$ are also allowed, which correspond to $\left(\mathbf{n}\cdot{\bf
k}_1\right)\left(\mathbf{n}\cdot{\bf
k}_2\right)$ in \eqref{Nem-EP-Hailtonian}. However, in the Landau-de Gennes theory, the potential energy $\phi({\sf Q})$ involves more complicated expansions of the form
\[
\phi({\sf Q})=\sum_{s,n\geq1} a_{sn} \left(\operatorname{Tr}\!\left({\sf Q}^n\right)\right)^{s}
\]
where the coefficients $a_{sn}$ are physical constants, see \cite{deGennes1993} .

At this point, it is important to notice that the Euler-Poincar\'e and Lie-Poisson equations involving the alignment tensor ${\sf Q}$ can also be derived directly by reduction of the Lagrangian $L_{\sf Q_0}(\chi,\dot{\chi})$ or the Hamiltonian $H_{\sf Q_0}(\chi,\Psi)$ (with $(\chi,\Psi)\in T^*SO(3)$), for a given ${\sf Q_0}\in\operatorname{Sym}(3)$. This process is independent of the definition of the director variable $\bf n$.

\subsection{Dynamics of biaxial nematic molecules}
The same treatment applies to the case of biaxial nematic particles.
In this context, we can identify the order parameter space  with the
manifold $M$ of ordered couples $n=({\bf n}_1,\, {\bf n}_2)$ of
mutually orthogonal, unsigned unit vectors ${\bf n}_1$, ${\bf n}_2$
(cf. e.g. \cite{Levy1985}). This choice is consistent with the
microstructure of a biaxial molecule, which can thus be envisioned
as a particle carrying two orthogonal director variables determining
a special rotational state. The group $SO(3)$ acts transitively on
the ordered couples in $M$ by matrix multiplication on each
director: $(\mathbf{n}_1,\mathbf{n}_2)\mapsto
(\chi\mathbf{n}_1,\chi\mathbf{n}_2)$. In order to determine the
nature of symmetry breaking for biaxial particles, we can fix a
reference couple $n_0\in M$: for example, one considers
$\left((1,0,0),(0,1,0)\right)$. Then, the associated isotropy
subgroup $SO(3)_{n_0}\subset SO(3)$ is readily seen to be the
dihedral group
\[
D_2=\left\{\operatorname{diag}(1,1,1), \operatorname{diag}(-1,-1,1), \operatorname{diag}(-1,1,-1), \operatorname{diag}(1,-1,-1)\right\}=SO(3)_{n_0}
\,.
\]
 Therefore we can write $M= SO(3)/D_2$ and we can express the Euler-Poincar\'e equations for a biaxial nematic particle as
\begin{equation}\label{biaxial}
\frac{d}{dt}\frac{\delta l}{\delta
\boldsymbol{\nu}}=\boldsymbol{\nu}\times\frac{\delta l}{\delta
\boldsymbol{\nu}}+{\bf n}_i\times\frac{\delta l}{\delta {\bf n}_i}
\,,\qquad\quad \dot{\bf n}_i={\bf n}_i\times\boldsymbol\nu
\end{equation}
where we assume summation over repeated indexes in the first
equation. A straightforward calculation may verify that the
orthogonality condition ${\bf n}_1\cdot{\bf n}_2=0$ is
consistently preserved at all times. A simple form of Lagrangian for a single biaxial nematic particle is given by
\[
l(\boldsymbol{\nu},
\mathbf{n}_1,{\bf n}_2)
=\frac12\,j\,|\boldsymbol{\nu}|^2-\frac12
\Big(\lambda_1\left|\mathbf{n}_1\cdot{\bf k}_1\right|^2+
\lambda_2\left|\mathbf{n}_2\cdot{\bf
k}_2\right|^2+\lambda_3\left|\mathbf{n}_1\times\mathbf{n}_2\cdot{\bf
k}_2\right|^2\Big)
\]
where ${\bf k}_1$, ${\bf k}_2$ and ${\bf k}_3$ again play the role
of an external force field. A simple Legendre transform of equations
\eqref{biaxial} yields the following Poisson bracket for biaxial
particles:
\begin{multline}\label{Poisson_biaxial_particle}
\{f,g\}(\boldsymbol{\mu},\mathbf{n}_1,\mathbf{n}_2)=\boldsymbol{\mu}\!\cdot\!\left(\frac{\delta
f}{\delta \boldsymbol{\mu}}\times \frac{\delta g}{\delta
\boldsymbol{\mu}}\right)+\mathbf{n}_1\!\cdot\!\left(\frac{\delta
f}{\delta \mathbf{n}_1}\times \frac{\delta g}{\delta
\boldsymbol{\mu}}-\frac{\delta g}{\delta \mathbf{n}_1}\times
\frac{\delta f}{\delta \boldsymbol{\mu}}\right)
\\
+\mathbf{n}_2\!\cdot\!\left(\frac{\delta f}{\delta
\mathbf{n}_2}\times \frac{\delta g}{\delta
\boldsymbol{\mu}}-\frac{\delta g}{\delta \mathbf{n}_2}\times
\frac{\delta f}{\delta \boldsymbol{\mu}}\right).
\end{multline}
The next section shows how the above bracket can be written in terms of the so called alignment tensors, analogously to our previous discussion concerning ordinary nematic molecules.

\begin{remark}[$\boldsymbol{V}$-shaped molecules]\normalfont
The orthogonality condition ${\bf n}_1\cdot{\bf n}_2=0$ can be weakened by considering
more general configurations of $V$-shaped molecules carrying two directors
 spanning a fixed angle $\vartheta$. In this case, the isotropy subgroup
is not the dihedral group $D_2$. Rather, it is formed of
$\pi$-rotations only, and thus it can be identified with the cyclic
group $\mathbb{Z}_2$. In this case, the order parameter manifold is
$M=SO(3)/\mathbb{Z}_2$ and the equations \eqref{biaxial} are still valid,
thereby describing the dynamics of a single $V$-shaped molecule with
directors ${\bf n}_1$ and ${\bf n}_2$, such that ${\bf n}_1\cdot{\bf
n}_2=\cos\vartheta$ at all times.
\end{remark}

\subsection{Alignment tensor dynamics for biaxial nematics} In analogy with ordinary nematic molecules, we now show how one can write the Lie-Poisson equations for biaxial molecules in terms of the alignment tensor which is usually given by a linear combination of the two symmetric matrices
\begin{align}
{\sf A}&=\frac12\left(\mathbf{n}_1\mathbf{n}_1^T-\frac13\,\mathbf{I}\right)
\,,\qquad
{\sf B}=\frac12\left(\mathbf{n}_2\mathbf{n}_2^T-\left(\mathbf{n}_1\times\mathbf{n}_2\right)\left(\mathbf{n}_1\times\mathbf{n}_2\right)^T\right)
.
\end{align}
Following \cite{SoViDu2003}, we shall begin by considering the above matrices as two separate variables, so that the equality $\delta f(\boldsymbol\sigma,\mathbf{n}_1,\mathbf{n}_2)=\delta f(\boldsymbol\sigma,{\sf A},{\sf B})$ yields the relations
\begin{align*}
\frac{\delta f}{\delta {\bf n}_1}&=\mathbf{n}_1^T\frac{\delta
f}{\delta
{\sf A}}-\mathbf{n}_2\times\left(\mathbf{n}_1\times\mathbf{n}_2\right)^T\frac{\delta
f}{\delta {\sf B}}
\\
\frac{\delta f}{\delta {\bf n}_2}&=\mathbf{n}_2^T\frac{\delta
f}{\delta
{\sf B}}+\mathbf{n}_1\times\left(\mathbf{n}_1\times\mathbf{n}_2\right)^T\frac{\delta
f}{\delta {\sf B}}
\end{align*}
In analogy to the procedure that we have followed for the case of ordinary nematic particles, we now express the following quantity in terms of $({\sf A},{\sf B})$:
\begin{align*}
\mathbf{n}_1\cdot\frac{\delta f}{\delta
\boldsymbol{\mu}}\times\frac{\delta g}{\delta
\mathbf{n}_1}+\mathbf{n}_2\cdot\frac{\delta f}{\delta
\boldsymbol{\mu}}\times\frac{\delta g}{\delta \mathbf{n}_2}
=
&\ -\frac{\delta f}{\delta
\boldsymbol{\mu}}\cdot\mathbf{n}_1\times\left(\mathbf{n}_1^T\frac{\delta
f}{\delta
{\sf A}}-\mathbf{n}_2\times\left(\mathbf{n}_1\times\mathbf{n}_2\right)^T\frac{\delta
f}{\delta {\sf B}}\right)
\\
&\
 -\frac{\delta f}{\delta
\boldsymbol{\mu}}\cdot\mathbf{n}_2\times\left(\mathbf{n}_2^T\frac{\delta
f}{\delta
{\sf B}}+\mathbf{n}_1\times\left(\mathbf{n}_1\times\mathbf{n}_2\right)^T\frac{\delta
f}{\delta {\sf B}}\right)
\\
=&\ - \frac{\delta f}{\delta
\boldsymbol{\mu}}\cdot\left(\mathbf{n}_1\times\mathbf{n}_1^T\frac{\delta
f}{\delta {\sf A}}+\mathbf{n}_2\times\mathbf{n}_2^T\frac{\delta f}{\delta
{\sf B}}\right.
\\
&\ \hspace{2.5cm}
-\left.\left(\mathbf{n}_1\times\mathbf{n}_2\right)\times\left(\mathbf{n}_1\times\mathbf{n}_2\right)^T\frac{\delta
f}{\delta {\sf B}}\right)
\\
=&\ \operatorname{Tr}\!\left({\sf A}\left[\frac{\delta f}{\delta
\hat{\mu}}\,,\frac{\delta g}{\delta
{\sf A}}\right]\right)-\operatorname{Tr}\!\left(\mathbf{n}_2\mathbf{n}_2^T\frac{\delta
g}{\delta {\sf B}}\frac{\delta f}{\delta \hat{\boldsymbol{\mu}}}\right)
\\
&\
+\operatorname{Tr}\!\left(\left(\mathbf{n}_1\times\mathbf{n}_2\right)\left(\mathbf{n}_1\times\mathbf{n}_2\right)^T\frac{\delta
g}{\delta {\sf B}}\frac{\delta f}{\delta \hat{\boldsymbol{\mu}}}\right)
\\
=&\
 \operatorname{Tr}\!\left({\sf A}\left[\frac{\delta f}{\delta
\hat{\boldsymbol{\mu}}}\,,\frac{\delta g}{\delta {\sf A}}\right]\right)+
 \operatorname{Tr}\!\left({\sf B}\left[\frac{\delta f}{\delta
\hat{\boldsymbol{\mu}}}\,,\frac{\delta g}{\delta {\sf B}}\right]\right) \,,
\end{align*}
where we have used the same formulas that emerged in the case of
nematic particles.
Therefore, we can write the Poisson bracket \eqref{Poisson_biaxial_particle} for biaxial particles in
terms of the tensor order parameters as
\begin{multline}
\{f,g\}(\boldsymbol{\mu},{\sf A},{\sf B})=\boldsymbol{\mu}\!\cdot\!\left(\frac{\delta
f}{\delta \boldsymbol{\mu}}\times \frac{\delta g}{\delta
\boldsymbol{\mu}}\right)+\operatorname{Tr}\!\left(\!{\sf A}\left(\left[\frac{\delta
f}{\delta \hat{\boldsymbol{\mu}}}\,,\frac{\delta g}{\delta
{\sf A}}\right]-\left[\frac{\delta g}{\delta \hat{\boldsymbol{\mu}}}\,,\frac{\delta
f}{\delta {\sf A}}\right]\right)\right)
\\+\operatorname{Tr}\!\left(\!{\sf B}\left(\left[\frac{\delta f}{\delta
\hat{\boldsymbol{\mu}}}\,,\frac{\delta g}{\delta {\sf B}}\right]-\left[\frac{\delta
g}{\delta \hat{\boldsymbol{\mu}}}\,,\frac{\delta f}{\delta
{\sf B}}\right]\right)\right).
\label{AlTensor-Biaxials}
\end{multline} 
At this point, it is easy to see that any linear combination of the form ${\sf Q}=\alpha{\sf A}+\beta{\sf B}$ leads precisely to the bracket \eqref{Poisson_ordpartensor}. Indeed, this follows upon noticing that
\[
\frac{\delta f}{\delta {\sf A}}=\alpha\frac{\delta f}{\delta {\sf Q}}\,,\qquad\frac{\delta f}{\delta {\sf B}}=\beta\frac{\delta f}{\delta {\sf Q}}
\]
and by replacing the above functional derivatives in the bracket \eqref{AlTensor-Biaxials}.

\begin{remark}[Other order parameter spaces]\normalfont The present
Euler-Poincar\'e approach applies to any sort of order parameter
manifold. For example, in superfluid dynamics one is faced to more
complicated coset structures such as $\left(SO(3)\times
SO(3)\right)/\left(SO(2)\times\Bbb{Z}_2\right)$, which is the order
parameter space for the $B$-phase of superfluid Helium-3
\cite{Mi1980}. Another interesting example is provided by
$^3$He-$A$, i.e. the $A$-phase of $^3$He, whose order parameter
space is $\left(SO(3)\times SO(3)\right)/SO(3)=SO(3)$.
In the context of liquid crystals one of the most complicated examples is
provided by smectics, whose order parameter space involves the
special Euclidean group \cite{Mi1980}. Other possibilities still
include more general complex fluids, such as micromorphic or
micropolar fluids \cite{Eringen1987}. See \cite{GBRa2008} for a
geometric treatment similar to the present one.
\end{remark}

\begin{remark}[A fully symmetric case: magnetic moment dynamics]\normalfont
 All the above examples are indicative of how often symmetry breaking
 appears in physics. However, one should
not forget that fully symmetric Lie-Poisson systems also emerge in
condensed matter applications. A famous example is provided by the
magnetic moments of ferromagnetic media, whose evolution takes place
on coadjoint orbits of the rotation group $SO(3)$. In this sense,
the dynamics of the electron magnetic moment
$\dot{\boldsymbol\mu}=\gamma\boldsymbol\mu\times{\bf H}$ (where $\bf
H$ is the external magnetic field and $\gamma$ is a physical
constant) is a Lie-Poisson system on $\mathfrak{so}^*(3)$ with
Hamiltonian $h(\boldsymbol\mu)=\gamma\boldsymbol\mu\cdot{\bf H}$.
Thus, the magnetic moment dynamics possesses the same geometric
interpretation as the rigid body dynamics, with the only difference
that the latter also determines a geodesic flow on $SO(3)$. A
posteriori, one selects a special coadjoint orbit (the unit sphere
$S^2$) to be consistent with the definition of spin as a unit
vector. Indeed, magnetic moments possess precisely the same symmetry properties of the usual spin variable, whose dynamics governs the theory of spin glasses \cite{HoKu1988}.
\end{remark}

\section{Lagrange-Poincar\'e approach to symmetry breaking}
\label{sec_Lagr_red}

While the last two sections presented the Euler-Poincar\'e and Lie-Poisson formulations for systems with symmetry breaking, the present section will show how the same systems allow for another geometric description, whose underlying general theory is known under the name of Lagrangian reduction \cite{CeHoMaRa1998,CeMaRa2001}. While the Euler-Poincar\'e theory applies to systems possessing a Lie group as the configuration space, Lagrangian reduction may be used to approach any  system on a tangent bundle possessing a continuous symmetry. As we shall see, the resulting Lagrange-Poincar\'e equations produce a special variable possessing a purely geometric
character, and whose physical nature is analogue to the color charge in Yang-Mills theories. In these theories, the motion of
a colored particle in a Yang-Mills field is a trajectory on a
principal bundle $B$ determined by an invariant Hamiltonian on
$T^*B$, where the Poisson bracket is canonical. In the particular
case when the Hamiltonian is quadratic, the resulting Wong's
geodesic equations produce the well known coadjoint dynamics for the
color charge. It is clear that for Abelian Yang-Mills theories, the
color is constant, like in the case of electromagnetism (cf.
\cite{MaRa1999}). We shall see how a very similar situation also
appears for the dynamics of nematic particles, whose celebrated
Ericksen-Leslie equations are equivalent to the Euler-Lagrange
equations on $T\Bbb{R}P^2$.

\subsection{General Lagrangian reduction: a brief review}\label{Sec:LR} Consider a
\textit{right\/} action $\Phi: G\times Q\rightarrow Q$ of a Lie
group $G$ on a manifold $Q$. Let $L:TQ\rightarrow\mathbb{R}$ be a
$G$-invariant Lagrangian under the cotangent-lifted action of $G$ on
$TQ$. Because of this invariance, we get a well defined reduced
Lagrangian $\ell : (TQ)/G \rightarrow\mathbb{R}$ satisfying
$\ell([v_q])=L(v_q)$.
Assuming the group action is free and proper, the quotient space $(TQ)/G$ is intrinsically a vector bundle over $T(Q/G)$ with a fiber modeled on the Lie algebra $\mathfrak{g}$. Using a connection $\mathcal{A}$ on the principal bundle $\pi:Q\rightarrow Q/G$ we have a vector bundle isomorphism
\[
\alpha_\mathcal{A}:(TQ)/G\longrightarrow T(Q/G)\oplus\tilde{\mathfrak{g}},\quad [v_q]\longmapsto \alpha_\mathcal{A}([v_q]):=\left(T\pi(v_q),[q,\mathcal{A}(v_q)]_G\right)
\]
over $Q/G$, where the associated bundle $\tilde{\mathfrak{g}}:=Q\times _G\mathfrak{g}$, is defined as the quotient space of $Q\times\mathfrak{g}$ relative to the right action $(q,\xi)\mapsto (\Phi_g(q),\operatorname{Ad}_{g^{-1}}\xi)$ of $G$. The elements of $\tilde{\mathfrak{g}}$ are denoted by $\bar v=[q,\xi]_G$. Using the isomorphism $\alpha_\mathcal{A}$, we can consider $\ell$ as a function defined on $T(Q/G)\oplus\tilde{\mathfrak{g}}$, and we write $\ell(x,\dot x,\bar v)$ to emphasize the dependence of $\ell$ on $(x,\dot x)\in T(Q/G)$ and $\bar v\in\tilde{\mathfrak{g}}$. However one should keep in mind that $x, \dot x$,
and $\bar v$ cannot be considered as being independent variables unless $T (Q/G)$ and $\tilde{\mathfrak{g}}$
are trivial bundles.

We now formulate the Lagrangian reduction theorem.

\begin{theorem}\label{Lagrangian_reduction} The following conditions are equivalent:
\begin{itemize}
\item[\bf{i}] Hamilton's variational principle
\[
\delta\int_{t_0}^{t_1}L(q,\dot q)dt=0,
\]
holds, for variations $\delta q(t)$ vanishing at the endpoints.
\item[\bf{ii}] The curve $q(t)$ satisfies the Euler-Lagrange equations for $L$ on $TQ$.
\item[\bf{iii}] The reduced variational principle
\[
\delta\int_{t_0}^{t_1}\ell(x,\dot x,\bar v)dt=0
\]
holds, for variations $\delta x\oplus\delta^\mathcal{A} \bar v$ of the curve $x(t)\oplus \bar v(t)$, where $\delta^\mathcal{A}\bar v$ has the form
\begin{equation}\label{FtCollins}
\delta^\mathcal{A}\bar v=\frac{D}{Dt}\bar\eta -[\bar v,\bar \eta]+\tilde{\mathcal{B}}(\delta x,\dot x),
\end{equation}
with the boundary conditions $\delta x(t_i)=0$ and $\bar \eta(t_i)= 0$, for $i = 0, 1$.
\item[\bf{iv}] The following \textbf{vertical} and \textbf{horizontal Lagrange-Poincar\'e equations}, hold:
\begin{equation}\label{reduced_E-L}
\left\lbrace
\begin{array}{l}
\displaystyle\vspace{0.2cm}\frac{D}{Dt}\frac{\partial \ell}{\partial \bar v}(x,\dot x,\bar v)=-\operatorname{ad}^*_{\bar v}\frac{\partial \ell}{\partial \bar v}(x,\dot x,\bar v)\\
\displaystyle\frac{\nabla \ell}{\partial x}(x,\dot x,\bar v)-\frac{\nabla}{dt}\frac{\partial \ell}{\partial \dot x}(x,\dot x,\bar v)=\left\langle\frac{\partial \ell}{\partial \bar v}(x,\dot x,\bar v),\mathbf{i}_{\dot x}\tilde{\mathcal{B}}(x)\right\rangle.
\end{array}\right.
\end{equation}
\end{itemize}
\end{theorem}

We now comment on the various expressions appearing in parts \textbf{ii} and \textbf{iii}. In the expression \eqref{FtCollins}, $D/Dt$ denotes the covariant time derivative of the curve $\bar\eta(t)\in\tilde{\mathfrak{g}}$ associated to the principal connection $\mathcal{A}$, that is, for $\bar\eta(t)=[q(t),\xi(t)]_G$, we have
\[
\frac{D}{Dt}\big[q(t),\xi(t) \big]_G= \Big[ q(t),\,\dot\xi(t)+[ \mathcal{A}(q(t),\dot q(t)),\xi(t) ] \Big]_G.
\]
The bracket $[\bar v,\bar \eta]$ denotes the Lie bracket induced by $\mathfrak{g}$ on each fiber of $\tilde{\mathfrak{g}}$. The two-form $\tilde{\mathcal{B}}\in \Omega^2(Q/G,\tilde{\mathfrak{g}})$ is the curvature on the base $Q/G$ induced by the curvature form $\mathcal{B}=\mathbf{d}\mathcal{A}+[\mathcal{A},\mathcal{A}]\in\Omega^2(Q,\mathfrak{g})$ of $\mathcal{A}$. Notice that for the formulation of the Lagrange-Poincar\'e equations, the introduction of an arbitrary connection $\nabla$ on the manifold $Q/G$ is needed. For simplicity a torsion free connection is chosen. The partial derivatives
\[
\frac{\partial \ell}{\partial \dot x}(x,\dot x,\bar v)\in T^*_x(Q/G)\quad\text{and}\quad \frac{\partial \ell}{\partial \bar v}(x,\dot x,\bar v)\in \tilde{\mathfrak{g}}^*_x
\]
are the usual fiber derivatives of $\ell$ in the vector bundles $T(Q/G)$ and $\tilde{\mathfrak{g}}$, and
\[
\frac{\nabla \ell}{\partial x}(x,\dot x,\bar v)\in T^*_x(Q/G)
\]
is the partial covariant derivative of $\ell$ relative to the given connection $\nabla$ on $Q/G$ and to the principal connection $\mathcal{A}$ on $Q$. We refer to \cite{CeMaRa2001} for details and proofs regarding the Lagrange-Poincar\'e equations.
Of course, there is an analogue result on the Hamiltonian side. Given a $G$ invariant Hamiltonian $H$ on $T^*Q$, Poisson reduction yields the  so called \textit{Hamilton-Poincar\'e equations} on the vector bundle $T^*(Q/G)\oplus \widetilde{\mathfrak{g}}^*$, see \cite{CeMaPeRa2003}. Given a Hamiltonian $\mathsf{h}=\mathsf{h}(x,\pi,\bar\mu): T^*(Q/G)\oplus\widetilde{\mathfrak{g}}^*$, the Hamilton-Poincar\'e equations are
\begin{equation}\label{Hamilton_Poincare}
\left\lbrace
\begin{array}{l}
\displaystyle\vspace{0.2cm}\frac{D}{Dt}\bar \mu=-\operatorname{ad}^*_{\frac{\partial \mathsf{h}}{\partial\bar\mu}}\bar\mu\\
\displaystyle\vspace{0.2cm}\dot x=\frac{\partial \mathsf{h}}{\partial\pi}\\
\displaystyle\frac{\nabla}{dt}\pi=-\frac{\nabla \mathsf{h}}{\partial x}-\left\langle\bar \mu,\mathbf{i}_{\frac{\partial \mathsf{h}}{\partial\pi}}\tilde{\mathcal{B}}(x)\right\rangle,
\end{array}\right.
\end{equation}
where $\partial \mathsf{h}/\partial \pi$ and $\partial \mathsf{h}/\partial \bar\mu$ are fiber derivatives, $\nabla \mathsf{h}/\partial x$ is the partial derivative defined in terms of the affine connection $\nabla$ on $Q/G$, see equations (16)-(19) in \cite{CeMaPeRa2003}.

\subsection{Lagrangian reduction on the coset bundle $G\to G/G_{n_0}$}\label{sec_LPA}

In this Section we specialize the situation described in
\S\ref{Sec:LR} to consider the case of a Lie group configuration
manifold $Q=G$, which is acted on by its isotropy subgroup
$G_{n_0}$, where $n_0$ is a fixed point on order parameter manifold
$M$. By the quotient $TG/G_{n_0}\simeq
T(G/G_{n_0})\oplus\widetilde{\mathfrak{g}_{n_0}}$ (or, on the
Hamiltonian side: $T^*G/G_{n_0}\simeq
T^*(G/G_{n_0})\oplus\widetilde{\mathfrak{g}_{n_0}}^*$), this
description provides the Lagrange-Poincar\'e (or
Hamilton-Poincar\'e) approach to symmetry breaking, thereby
extending the new phase space $T^*(G/G_{n_0})$ to include a
charge-like variable taking values in the dual isotropy subalgebra
$\mathfrak{g}_{n_0}^*$. This process provides a rigorous geometric
framework that naturally explains the emergence of the order
parameter space $G/G_{n_0}$ as the effective configuration manifold,
starting from an original system defined on the broken symmetry
group $G$. This is the situation always appearing in physical
applications (particularly in condensed matter physics) and its
geometric formulation apparently differs from the Euler-Poincar\'e
approach, which does not treat $G/G_{n_0}$ as the effective
configuration space.

In what follows, we suppose that the isotropy subgroup $G_{n_0}\subset G$ is a Lie group, with \makebox{$\operatorname{dim}G_{n_0}\geq 1$}. Since the $G_{n_0}$-action is free and proper, we have the right principal bundle
$\pi_{n_0}:G\rightarrow G/G_{n_0}\simeq
\operatorname{Orb}(n_0),\quad g\mapsto gG_{n_0}\simeq gn_0$. Given a
principal connection $\mathcal{A}$ we have the usual vector bundle
isomorphism
\begin{equation}\label{vector_bundle_isom}
TG/G_{n_0}\rightarrow T\operatorname{Orb}(n_0)\oplus \widetilde{\mathfrak{g}_{n_0}},\quad [v_g]\mapsto \left((v_gg^{-1})_M(n),[g,\mathcal{A}(v_g)]_{G_{n_0}}\right)
\end{equation}
over $\operatorname{Orb}(n_0)$, where $n=gn_0$ and
$\widetilde{\mathfrak{g}_{n_0}}$ is the adjoint bundle of the
isotropy subalgebra $\mathfrak{g}_{n_0}$.
Recall from Remark \ref{i_n_0} that we have the diffeomorphism
\begin{equation}\label{diffeo1}
\bar i_{n_0}:TG/G_{n_0}\rightarrow\mathfrak{g}\times\operatorname{Orb}(n_0)\subset \mathfrak{g}\times M,\quad [v_g]_{G_{n_0}}\mapsto (v_gg^{-1},gn_0).
\end{equation}
Therefore, by composing with \eqref{vector_bundle_isom} we get the vector bundle isomorphism
\begin{equation}\label{diffeo}
\mathfrak{g}\times\operatorname{Orb}(n_0)\rightarrow  T\operatorname{Orb}(n_0)\oplus \widetilde{\mathfrak{g}_{n_0}},\quad (\xi,n)\mapsto \left(\xi_M(n),[g,\mathcal{A}(\xi g)]_{G_{n_0}}\right)
\end{equation}
over $\operatorname{Orb}(n_0)$, where $g\in G$ is such that $gn_0=n$.

As a concrete example of a principal connection, we consider the mechanical connection associated to a Ad-invariant inner product $\gamma$ on $\mathfrak{g}$. The associated Riemannian metric on $G$ is bi-invariant and is denoted by $\left\langle\!\left\langle\cdot,\cdot \right\rangle\!\right\rangle$. Recall that we have
(see \cite{MaMiOrPeRa2007})
\[
\mathcal{A}(v_g)=\mathbb{I}(g)^{-1}(\mathbf{J}(v_g)).
\]
where the {\it locked inertia tensor} $\mathbb{I}(g):\mathfrak{g}_{n_0}\rightarrow\mathfrak{g}_{n_0}^*$ is given by
\[
\langle\mathbb{I}(g)\eta,\zeta\rangle:=\langle\!\langle\eta_G(g),\zeta_G(g)\rangle\!\rangle=\langle\!\langle g\eta,g\zeta\rangle\!\rangle=\left.\gamma\right|_{\mathfrak{g}_{n_0}\!\!}(\eta,\zeta),\quad\text{with}\quad \xi,\eta\in\mathfrak{g}_{n_0}.
\]
Therefore the explicit expression of the locked inertia tensor is
$\mathbb{I}(g)\eta=\left.\gamma\right|_{\mathfrak{g}_{n_0}\!\!}(\eta,\_\,)$,
where $\left.\gamma\right|_{\mathfrak{g}_{n_0}}$ denotes the restriction of the inner product $\gamma$ on $\mathfrak{g}$ to $\mathfrak{g}_{n_0}\subset \mathfrak{g}$.
Note that when $\xi\in\mathfrak{g}$ then $\mathbb{I}(g)^{-1}\left(\gamma(\xi,\_\,)|_{\mathfrak{g}_{n_0}}\right)=\mathbb{P}_{n_0}(\xi)$,
where $\mathbb{P}_{n_0}:\mathfrak{g}\rightarrow\mathfrak{g}_{n_0}$
is the orthogonal projector associated to $\gamma$ and $\gamma(\xi,\_\,)|_{\mathfrak{g}_{n_0}}$ denotes the restriction to $\mathfrak{g}_{n_0}$ of the linear form $\gamma(\xi,\_\,)\in\mathfrak{g}^*$. Indeed, one
has
\[
\mathbb{I}(g)\,\mathbb{P}_{n_0}(\xi)=\left.\gamma\right|_{\mathfrak{g}_{n_0}\!\!}(\mathbb{P}_{n_0}(\xi),\_\,)=\gamma(\xi,\_\,)|_{\mathfrak{g}_{n_0}}
\]
The map ${\bf J}:TG\to\mathfrak{g}_{n_0}^*$ is given by
\[
\langle\mathbf{J}(v_g),\zeta\rangle:=\langle\!\langle
v_g,\zeta_G(g)\rangle\!\rangle=\langle\!\langle
v_g,g\zeta\rangle\!\rangle=\left.\gamma (g^{-1}v_g,\zeta)\right|_{\mathfrak{g}_{n_0}},\quad\text{for all}\quad \zeta\in\mathfrak{g}_{n_0}.
\]
Therefore
$\mathbf{J}(v_g)=\left.\gamma(g^{-1}v_g,\_\,)\right|_{\mathfrak{g}_{n_0}}$
and
\[
\mathcal{A}(v_g)=\mathbb{I}(g)^{-1}(\mathbf{J}(v_g))=\mathbb{I}(g)^{-1\!}\left(\left.\gamma(g^{-1}v_g,\_\,)\right|_{\mathfrak{g}_{n_0}}\right)=\mathbb{P}_{n_0}(g^{-1}v_g)
\]
In this case, the vector bundle isomorphism
\eqref{vector_bundle_isom} reads
\begin{equation}\label{vector_bundle_isom_machanical}
TG/G_{n_0}\rightarrow T\operatorname{Orb}(n_0)\oplus \widetilde{\mathfrak{g}_{n_0}},\quad [v_g]\mapsto \left((v_gg^{-1})_M(n),[g,\mathbb{P}_{n_0}(g^{-1}v_g)]_{G_{n_0}}\right)
\end{equation}
and the diffeomorphism \eqref{diffeo} is
\begin{equation}\label{diffeo_mechanical}
\mathfrak{g}\times\operatorname{Orb}(n_0)\rightarrow  T\operatorname{Orb}(n_0)\oplus \widetilde{\mathfrak{g}_{n_0}},\quad (\xi,n)\mapsto \left(\xi_M(n),[g,\mathbb{P}_{n_0}(\operatorname{Ad}_{g^{-1}}\xi)]_{G_{n_0}}\right)
\end{equation}
In order to compute the curvature of the mechanical connection, we use the formula $\mathcal{B}=\mathbf{d}\mathcal{A}+[\mathcal{A},\mathcal{A}]$. We have
\[
\mathbf{d}\mathcal{A}(u_g,v_g)=\mathbf{d}(\mathcal{A}(Y))X-\mathbf{d}(\mathcal{A}(X))Y-\mathcal{A}([X,Y]),
\]
where $X,Y\in\mathfrak{X}(G)$ are two vector fields extending $u_g,v_g$. Using the left-invariant vector fields $X(g)=g\xi$ and $Y(g)=g\eta$, where $\xi,\eta\in\mathfrak{g}$, we have
\[
\mathbf{d}\mathcal{A}(u_g,v_g)=-\mathcal{A}([X,Y](g))=-\mathcal{A}(g[\xi,\eta])=-\mathbb{P}_{n_0}([\xi,\eta])
\]
thus, we get
\begin{equation}\label{mechanical_curvature}
\mathcal{B}(u_g,v_g)=[\mathbb{P}_{n_0}(\xi),\mathbb{P}_{n_0}(\eta)]-\mathbb{P}_{n_0}([\xi,\eta]),\quad \xi=g^{-1}u_g,\;\; \eta=g^{-1}v_g.
\end{equation}

Suppose that we have chosen a fixed reference point $n_0\in M$. Using Lagrangian reduction (Theorem \ref{Lagrangian_reduction}) for the $G_{n_0}$-invariant Lagrangian $L_{n_0}:TG\rightarrow\mathbb{R}$ and the reduced Lagrangian $\ell_{n_0}:T\operatorname{Orb}(n_0)\oplus\widetilde{\mathfrak{g}_{n_0}}$, we obtain the Lagrange-Poincar\'e equations \eqref{reduced_E-L} in terms of $(n,\dot{n})\in T\operatorname{Orb}(n_0)$'.

\begin{remark}
\normalfont If the Lagrangian $L_{n_0}:TG\rightarrow\mathbb{R}$ is
hyperregular, one can obtain the Hamiltonian description for the
corresponding Hamiltonian $H_{n_0}:T^*G\rightarrow\mathbb{R}$. The reduction
process is analogue and one ends up with the \textit{Hamilton-Poincar\'e equations} on the vector bundle $T^*\operatorname{Orb}(n_0)\oplus \widetilde{\mathfrak{g}_{n_0}}^*$, see \eqref{Hamilton_Poincare} with reduced Hamiltonian $\mathsf{h}_{n_0}$ for the general theory.
We will refer to $\ell_{n_0}$ as the \textit{Lagrange-Poincar\'e
(LP) Lagrangian} and to $\mathsf{h}_{n_0}$ as the \textit{Hamilton-Poincar\'e
(HP) Hamiltonian}.
\end{remark}

\subsection{Two equivalent approaches for symmetry breaking}

At this stage, it is useful to discuss the relation between the Euler-Poincar\'e and Lagrange-Poincar\'e descriptions associated to a $G$-invariant Lagrangian $L:TG\times M\rightarrow\mathbb{R}$. The equivalence of the two approaches arises as follows. Fix a reference point $n_0\in M$ and consider the induced Lagrangians $L_{n_0}:TG\rightarrow\mathbb{R}$, $l:\mathfrak{g}\times M\rightarrow\mathbb{R}$, and $\ell_{n_0}:T\operatorname{Orb}(n_0)\oplus_{\operatorname{Orb}(n_0)}\widetilde{\mathfrak{g}_{n_0}}\rightarrow\mathbb{R}$.
Then the following are equivalent:
\begin{itemize}
\item $g\in G$ is a solution of the \textit{Euler-Lagrange equations} for $L_{n_0}$.
\item $\xi:=\dot gg^{-1}\in\mathfrak{g}$ and $n:=gn_0\in M$ are solution of the \textit{Euler-Poincar\'e equations} \eqref{EP} for $l$.
\item $n:=gn_0\in \operatorname{Orb}(n_0)$ and $\bar\xi:=[g,\mathcal{A}(\dot g)]_{G_{n_0}}\in\widetilde{\mathfrak{g}_{n_0}}$ are solutions of the \textit{Lagrange-Poincar\'e equations} \eqref{reduced_E-L} for $\ell_{n_0}$.
\end{itemize}

\begin{remark}[The case of transitive actions]\normalfont In the case where $G=\mathcal{O}$ is an order parameter group acting transitively on the order parameter space $M$, we have $\operatorname{Orb}(n_0)=M$, thus the reduced Lagrangian is defined on the vector bundle $TM\oplus_M\widetilde{\mathfrak{p}}$, where $\mathfrak{p}$ is the Lie algebra of $\mathcal{P}$. In particular the diffeomorphisms \eqref{vector_bundle_isom}, \eqref{diffeo1}, \eqref{diffeo} become
\begin{align*}
&T\mathcal{O}/\mathcal{P}\rightarrow TM\oplus\widetilde{\mathfrak{p}},\quad [v_\chi]_\mathcal{P}\mapsto \left((v_\chi\chi^{-1})_M(n),[\chi,\mathcal{A}(v_\chi)]_\mathcal{P}\right)\\
&\bar i_{n_0}:T\mathcal{O}/\mathcal{P}\rightarrow\mathfrak{o}\times M,\quad [v_\chi]_\mathcal{P}\mapsto \left(v_\chi\chi^{-1},\chi n_0\right)\\
&\mathfrak{o}\times M\rightarrow TM\oplus\widetilde{\mathfrak{p}},\quad (\nu,n)\mapsto \left(\nu_M(n),[\chi,\mathcal{A}(\nu\chi)]_\mathcal{P}\right).
\end{align*}
\end{remark}

\subsection{Lagrange-Poincar\'e formulation of uniaxial nematics}

We now treat the particular case of \textit{nematics}.
Here the order parameter space is the projective plane
$M=\mathbb{R}P^2$. Recall that $\mathbb{R}P^2$ is
the non-orientable two-dimensional manifold given by the quotient of
the two-sphere by the antipodal relation. An element $\mathbf{n}\in
\mathbb{R}P^2$ is an equivalence class
$\mathbf{n}=[\mathbf{v}]$, where $\mathbf{v}\in S^2$ is a unit
vector.
The broken symmetry is the group of rotations $SO(3)$ acting on
the directors on the left by
$\mathbf{n}\mapsto \chi\mathbf{n}:=[\chi\mathbf{v}]$.
The isotropy group of a fixed direction $\mathbf{n}_0$ is the
infinite dihedral group $\mathcal{P}=D_\infty$ generated by the
rotations around the axis $\mathbf{n}_0$ and rotation by $\pi$
around an orthogonal axis. When the $z$-direction $[(0,0,1)]$ is
chosen for the reference axis, then the isotropy group $D_\infty$
consists of matrices of the form
\begin{equation}\label{D_infty}
\mathcal{M}=\left(\begin{array}{cc}
\M & 0\\
0&\operatorname{det}(\M)
\end{array}\right)\in D_{\infty\,},\quad\text{ with }\ \M\in O(2).
\end{equation}
The formula above shows how the group $D_\infty$ is actually isomorphic to
$O(2)$, which is the group usually appearing in the condensed matter literature.

\begin{remark}\normalfont
The group $D_\infty$ should not be confused with $D_{\infty,h}$
which contains also $-I_3$ (i.e. minus the identity matrix) and is not a subgroup of $SO(3)$ but of
$O(3)$. Note that $D_\infty$ is not Abelian. For example the
$\pi$-rotation $\operatorname{diag}(1,-1,-1)$ does not commute
with rotations around the vertical axis.
\end{remark}
The Lie algebra
$\mathfrak{d}_\infty$ of $D_\infty$ is given by matrices of the
form
\[
\left(\begin{array}{ccc}
0& -\sigma & 0\\
\sigma &0&0\\
0&0&0
\end{array}\right),\;\;\sigma\in\mathbb{R}.
\]
Using the hat map, it can be identified with the subspace
$\operatorname{Span}(0,0,1)$ of $\mathbb{R}^3$ on which the adjoint action
acts by matrix multiplication. By identifying $\operatorname{Span}(0,0,1)$
with the real line, and using the notations of \eqref{D_infty} we
obtain the adjoint action
\[
\operatorname{Ad}_\mathcal{M}\sigma=\operatorname{det}(\M)\,\sigma
\]
(recall that $\operatorname{det}(\M)=\pm 1$).
The identification $SO(3)/D_\infty\simeq \mathbb{R}P^2$ is given by
\[
[\chi]\in SO(3)/D_\infty\mapsto \mathbf{n}=[(\chi_{13},\chi_{23},\chi_{33})]\in \mathbb{R}P^2\simeq S^2/\Bbb{Z}_2.
\]
Giving a reference point $\mathbf{n}_0\in\mathbb{R}P^2$ and a Lagrangian $L_{\mathbf{n}_0}:TSO(3)\rightarrow\mathbb{R}$ describing the nematic particle, one can easily obtain the Euler-Poincar\'e  formulation associated to the Lagrangian
$l:\mathfrak{so}(3)\times \mathbb{R}P^2\to\Bbb{R}$.
In order to obtain the Lagrange-Poincar\'e equations for $\ell_{\mathbf{n}_0}:TSO(3)/D_\infty\rightarrow\mathbb{R}$, one needs to use the adjoint bundle. As we will see, it will be more comfortable to work with the sphere $S^2$ instead of the projective plane.
The adjoint bundle of the right principal bundle $SO(3)\rightarrow \mathbb{R}P^2$ is the quotient space
\[
\widetilde{\mathfrak{d}_\infty}=\left(SO(3)\times\mathfrak{d}_\infty\right)/D_\infty=\left(SO(3)\times\mathbb{R}\right)/D_\infty
\]
relative to the right action of $\mathcal{M}\in D_\infty$ on $(\chi,r)\in
SO(3)\times\mathbb{R}$ given by
\[
(\chi,r)\mapsto (\chi \mathcal{M},\operatorname{det}(\M)r)\,.
\] Note that an
element in the fiber
$\big(\widetilde{\mathfrak{d}_\infty}\big)_{\!\mathbf{n}}$ of the
vector bundle $\widetilde{\mathfrak{d}_\infty}\to\Bbb{R}P^2$ reads $[\chi,r]$,
where $\chi\in SO(3)$ is such that $\mathbf{n}=\chi\mathbf{n}_0$,
that is, $[\chi]=\mathbf{n}$. The fact that $\mathcal{M}$ also acts on $r$ is
due to the fact that $D_\infty$ is not Abelian.

In order to simplify
the approach and to work with more explicit formulas, we replace the
order parameter space $\mathbb{R}P^2$ with the two sphere $S^2$. The
breaking symmetry group $SO(2)$ is now Abelian. The reduced
Lagrangian is now defined on $\mathfrak{so}(3)\times S^2$ but one
has to recall that it is invariant under a change of sign for the
variable in $S^2$. In this case, the isotropy group is $SO(2)$, and
the adjoint bundle
$\widetilde{\mathfrak{so}(2)}=SO(3)\times_{SO(2)}\mathbb{R}$ is a
trivial bundle, since $SO(2)$ is Abelian and thus the adjoint action
on its Lie algebra $\mathfrak{so}(2)\simeq\mathbb{R}$ is trivial.


We now describe the mechanical connection associated to the $\operatorname{Ad}$-invariant inner product $\gamma(\xi,\eta)=\frac{1}{2}\operatorname{trace}(\xi^T\eta)$ on $\mathfrak{so}(3)$. Recall that the Lie algebra $\mathfrak{so}(3)$ and $\mathfrak{so}(2)$ are identified with $\mathbb{R}^3$ and $\mathbb{R}(0,0,1)$, via the hat map. On $\mathbb{R}^3$ the inner product $\gamma$ is the standard inner product and the projection $\mathbb{P}:\mathfrak{so}(3)\simeq\mathbb{R}^3\rightarrow \mathfrak{so}(2)\simeq\mathbb{R}$ is simply given by taking the third component. The mechanical connection is thus given by
\begin{equation}\label{mech_conn_nematics}
\mathcal{A}(v_\chi)=\mathbb{P}(\chi^{-1}v_\chi)=(\chi^{-1}v_\chi)_3.
\end{equation}
According to formula \eqref{mechanical_curvature}, the curvature
is
\[
\mathcal{B}(u_\chi,v_\chi)=0-(\boldsymbol{\nu}\times\boldsymbol{\kappa})_3=-{\nu}_1\,{\kappa}_2+{\nu}_2\,{\kappa}_1,\qquad\boldsymbol{\nu}=\chi^{-1}u_\chi,
\boldsymbol{\kappa}=\chi^{-1}v_\chi.
\]
In order to compute the reduced curvature on $S^2$ we first note
that the tangent map to the projection
$\pi_{\mathbf{n}_0}:SO(3)\rightarrow S^2,\; \chi\mapsto \chi
\mathbf{n_0}$ reads
\[
T\pi_{\mathbf{n}_0}:TSO(3)\rightarrow TS^2,\quad \hat{\boldsymbol{\nu}}\chi\mapsto \boldsymbol{\nu}\times\chi\mathbf{n}_0=\boldsymbol{\nu}\times\mathbf{n}.
\]
Note that if $\dot{\mathbf{n}}\in T_\mathbf{n}S^2$ is given, then we
have
$T\pi_{\mathbf{n}_0}\big(\widehat{\mathbf{n}\times\dot{\mathbf{n}}}\,\chi\big)=\dot{\mathbf{n}}$,
where the `hat' map denotes the usual isomorphism
$\Bbb{R}^3\simeq\mathfrak{so}(3)$. Therefore, the reduced curvature
is
\begin{align}\label{reduced_curvature}
\tilde{\mathcal{B}}_\mathbf{n}(\dot{\mathbf{n}},\dot{\mathbf{m}})&=
\mathcal{B}\left(\widehat{\mathbf{n}\times\dot{\mathbf{n}}}\,\chi\,,\,\widehat{\mathbf{n}\times\dot{\mathbf{m}}}\,\chi\right)=
-\left(\chi^{-1}(\mathbf{n}\times\dot{\mathbf{n}})\times \chi^{-1}(\mathbf{n}\times\dot{\mathbf{m}})\right)_3\nonumber\\
&=-\left((\mathbf{n}_0\times\chi^{-1}\dot{\mathbf{n}})\times (\mathbf{n}_0\times\chi^{-1}\dot{\mathbf{m}})\right)_3=
-\left(\chi^T\dot{\mathbf{n}}\times\chi^T\dot{\mathbf{m}}\right)_3\nonumber\\
&=-\left(\chi^T(\dot{\mathbf{n}}\times\dot{\mathbf{m}})\right)_3=
-\mathbf{n}\!\cdot\!(\dot{\mathbf{n}}\times\dot{\mathbf{m}}).
\end{align}
In the last equality we used the formula
$(\chi^T\boldsymbol{\nu})_3=\chi_{i3}\boldsymbol{\nu}_i=\mathbf{n}\cdot\boldsymbol{\nu}$,
valid since $\chi\in SO(3)$ is such that
$\chi\mathbf{n_0}=\mathbf{n}$, where $\mathbf{n}_0=(0,0,1)$. Up to a
sign the curvature is given by the volume of the polytope generated
by the vectors $\mathbf{n}$, $\dot{\mathbf{n}}$ and
$\dot{\mathbf{m}}$.

Using the mechanical connection, and the fact that the adjoint bundle is trivial, the diffeomorphism \eqref{diffeo_mechanical} reads
\begin{equation}\label{diffeo_nematics}
\mathfrak{so}(3)\times S^2\rightarrow TS^2\times\mathbb{R},\qquad (\boldsymbol{\nu},\mathbf{n})\mapsto (\mathbf{n},\boldsymbol{\nu}\times\mathbf{n},\mathbf{n}\!\cdot\!\boldsymbol{\nu})=(\mathbf{n},\dot{\mathbf{n}},r).
\end{equation}
Indeed, the infinitesimal generator associated to $\boldsymbol{\nu}\in \mathfrak{so}(3)\simeq\mathbb{R}^3$ reads as
\makebox{
$\boldsymbol{\nu}_{S^2}(\mathbf{n})
=\boldsymbol{\nu}\times\mathbf{n}\in T_\mathbf{n}S^2
$}
and the second component of \eqref{diffeo_mechanical} is
\[
\mathbb{P}(\operatorname{Ad}_{\chi^{-1}}\boldsymbol{\nu})=(\chi^{-1}\boldsymbol{\nu})_3=(\chi^T\boldsymbol{\nu})_3=\chi_{i3}\boldsymbol{\nu}_i=\mathbf{n}\!\cdot\!\boldsymbol{\nu},
\]
In terms of equivalence classes $[v_\chi]\in TSO(3)/SO(2)$, the variables $\mathbf{n}$ and $\boldsymbol{\nu}$ are given by
$\mathbf{n}=\chi\mathbf{n}_0$, and $\boldsymbol{\nu}=v_\chi\chi^{-1}$.

We now obtain the formula for the inverse of the vector bundle map \eqref{diffeo_nematics}. Denoting $r:=\mathbf{n}\!\cdot\!\boldsymbol{\nu}$ and $\dot{\mathbf{n}}:=\boldsymbol{\nu}\times \mathbf{n}$ we have $\mathbf{n}\times (\boldsymbol{\nu}\times \mathbf{n})=(\mathbf{n}\!\cdot\!\mathbf{n})\boldsymbol{\nu}-\mathbf{n}(\mathbf{n}\!\cdot\!\boldsymbol{\nu})=\boldsymbol{\nu}-r\mathbf{n}$, and
\begin{equation}\label{nu}
\boldsymbol{\nu}= \mathbf{n}\times (\boldsymbol{\nu}\times
\mathbf{n})+r\mathbf{n}=\mathbf{n}\times
\dot{\mathbf{n}}+r\mathbf{n}.
\end{equation}
This proves that the inverse of the vector bundle map
\eqref{diffeo_nematics} is
\[
(\mathbf{n},\dot{\mathbf{n}},r)\in TS^2\times\mathbb{R}\mapsto
(\mathbf{n}\times
\dot{\mathbf{n}}+r\mathbf{n},\mathbf{n})=(\boldsymbol{\nu},\mathbf{n})\in
\mathfrak{so}(3)\times S^2.
\]
Therefore, the Lagrangian $l$ in \eqref{Nem-EP-Lagrangian} and the LP Lagrangian $\ell_{\mathbf{n}_0}$ (now simply
denoted by $\ell$, recall that $\mathbf{n}_0=(0,0,1)$), are related
by
\begin{equation}\label{relation_lagrangians}
l(\boldsymbol{\nu},\mathbf{n})=\ell(\mathbf{n},\boldsymbol{\nu}\times\mathbf{n},\mathbf{n}\!\cdot\!\boldsymbol{\nu}),
\quad \ell(\mathbf{n},\dot{\mathbf{n}},r)=l(\mathbf{n}\times \dot{\mathbf{n}}+r\mathbf{n},\mathbf{n}).
\end{equation}

We now compute the Lagrange-Poincar\'e equations associated to the reduced Lagrangian
$\ell=\ell(\mathbf{n},\dot{\mathbf{n}},r):TS^2\times\mathbb{R}\rightarrow\mathbb{R}$.
Since the group $SO(2)$ is Abelian, the Lie bracket is zero, the
covariant derivative coincides with usual derivative. Using formula
\eqref{reduced_curvature} for the reduced curvature of the
mechanical connection, the Lagrange-Poincar\'e equations
\eqref{reduced_E-L} read
\begin{equation}\label{barcellona}
\frac{d}{dt}\frac{\partial \ell}{\partial r}=0,\;\;\frac{\nabla}{dt}\frac{\partial \ell}{\partial \dot{\mathbf{n}}}-\frac{\nabla \ell}{\partial \mathbf{n}}=\frac{\partial \ell}{\partial r}\mathbf{n}\times\dot{\mathbf{n}},
\end{equation}
where the curvature term is evidently given by ${\bf n\times
\dot{n}}=-\mathbf{i}_{\dot{\bf n}}\mathcal{B}_{\bf n}$, and $\nabla/dt$,
$\nabla\ell/\partial\mathbf{n}$ are associated to the Levi-Civita
connection on $S^2$ induced by the inner product on $\mathbb{R}^3$.
The explicit form of the Lagrangian $\ell_{{\bf n}_0}$
can be written immediately upon using \eqref{relation_lagrangians} and recalling
equation \eqref{nematic-EPLagrangian}. We obtain the LP Lagrangian
\begin{align}\label{nematic_LP_Lagrangian}
\ell({\bf n,\dot{n}},r)&=l(\mathbf{n}\times \dot{\mathbf{n}}+r\mathbf{n},\mathbf{n})=
\frac12\,j\,|{\bf n\times \dot{n}}+r{\bf n}|^2-\phi({\bf n})\nonumber\\
&=\frac12\,j\,|{\bf n\times \dot{n}}|^2+\frac12\,j\,r^2-\phi({\bf
n})
\nonumber\\
&=\frac12\,j\,|\dot{\bf n}|^2+\frac12\,j\,r^2-\phi({\bf n})\, .
\end{align}
where the potential term may be expressed again by the quadratic expression $\phi({\bf n})=\lambda\left|{\bf
n\cdot k}\right|^2$, or alternatively by $\phi({\bf n})=\lambda\left|{\bf
n\times k}\right|^2$, for a fixed number $\lambda$ and an unsigned unit vector
${\bf k}$.
Upon inserting the expression \eqref{nematic_LP_Lagrangian} in the Lagrange-Poincar\'e equations \eqref{barcellona} we get
\[
\dot r=0,\quad j\frac{\nabla}{dt}\dot{\bf n}+\frac{\nabla\phi}{\partial\mathbf{n}}=j\,r\,{\bf n\times \dot{n}}.
\]
Note that $\phi$ does not depend on $\dot{\mathbf{n}}$, so $\nabla\phi/\partial\mathbf{n}$ coincides with the gradient of the map
$\phi:S^2\rightarrow\mathbb{R}$ with respect to the Riemannian
metric induced by $\mathbb{R}^3$ on $S^2$. In order to recover the
usual form of the equations, we will interpret $\mathbf{n}$ as a
curve in $\mathbb{R}^3$ and $\phi$ as a map defined on
$\mathbb{R}^3$. In this case we get the equation
\[
j\,\ddot{\bf n}-2q{\bf n}+\nabla\phi(\mathbf{n})=j\,r\,\dot{\bf n}\times{\bf n}\,
\quad \text{ with } \quad 2q:={\bf n}\cdot(j\,\ddot{\bf n}+\nabla\phi(\mathbf{n}))
\]
and where $\nabla\phi$ denotes the gradient of $\phi$ view as a map on $\mathbb{R}^3$. This equation evidently reduces to the celebrated Ericksen-Leslie equation for the
case $r=0$ (cf. \cite{Chandra1992}).

\begin{remark}[Physical nature of the variable $\boldsymbol{r}$]\normalfont The conserved variable \makebox{$r=\boldsymbol\nu_0\cdot{\bf
n}_0$} is evidently the projection of the angular velocity of the
nematic molecule on its director. Thus, when this quantity is
non-zero, it encodes the effect of rotations of the molecule about
the director. Because of the particular form of the Lagrangian, this
effect is taken into account only by the curvature term on the right
hand side of the Euler-Lagrange equation. An analogous situation
holds for the heavy top dynamics (see remark \ref{HTcomp} below).
However, for nematic molecules, such rotations about the director
are irrelevant, due to the rod-like nature of nematics. Thus, the
convention $r=0$ producing Ericksen-Leslie dynamics is the most
natural in this case. On the other hand, this is not true for the
dynamics of the heavy top, which is a rigid body (with fixed point)
of arbitrary shape.
\end{remark}

For the Hamiltonian side, one needs the dual vector bundle map of
\eqref{relation_lagrangians} over $S^2$. We have
\begin{equation}\label{Poisson-isom}
(\boldsymbol{\mu},\mathbf{n})\in \mathfrak{so}(3)^*\times S^2\mapsto
(\mathbf{n},\boldsymbol{\pi},w)=(\mathbf{n},\boldsymbol{\mu}\times\mathbf{n},\boldsymbol{\mu}\!\cdot\!\mathbf{n})\in
T^*S^2\times \mathbb{R}^*,
\end{equation}
whose inverse is given by
\[
(\mathbf{n},\boldsymbol{\pi},w)\in T^*S^2\times \mathbb{R}^*\mapsto
(\mathbf{n}\times\boldsymbol{\pi}+w\mathbf{n},\mathbf{n})=(\boldsymbol{\mu},\mathbf{n})\in
\mathfrak{so}(3)^*\times S^2.
\]
Therefore, the Hamiltonian $h$ in \eqref{Nem-EP-Hailtonian} produces
the corresponding HP Hamiltonian by
$
\mathsf{h}(\mathbf{n},\boldsymbol{\pi},w)=h(\mathbf{n}\times\boldsymbol{\pi}+w\mathbf{n},\mathbf{n})$.
The  nematic particle Hamiltonian  reads as
\begin{equation}\label{HP_Hamiltonian_nematics}
\mathsf{h}(\mathbf{n},\boldsymbol{\pi},w)=\frac{1}{2j}|\boldsymbol{\pi}|^2+\frac1{2j}\,w^2+\phi({\bf n})
\end{equation}
which evidently differs from the Lie-Poisson Hamiltonian
\[
h(\boldsymbol{\mu},{\bf
n})=\frac{1}{2j}\,|\boldsymbol{\mu}|^2+\phi({\bf n})
\]
only by the constant factor $w$,  without changing the content in
physical information. In the case of \eqref{HP_Hamiltonian_nematics}, the quadratic nature of the Hamiltonian produces a Kaluza-Klein construction analogue to that underlying Wong's equations in Yang-Mills theory. In this setting, substitution of the Hamiltonian \eqref{HP_Hamiltonian_nematics} in the the Hamilton-Poincar\'e
equations \eqref{Hamilton_Poincare} yields
\[
\dot{w}=0,\quad j\dot{\bf n}=\boldsymbol\pi,\quad \frac{\nabla}{dt}\boldsymbol\pi=-\frac{\nabla\phi}{\partial \mathbf{n}}-\frac{w}{j}\boldsymbol{\pi}\times\mathbf{n},
\]
so that the usual Ericksen-Leslie equation is recovered for $w=0$.

\begin{remark}[Comparison with the heavy top]\label{HTcomp}\normalfont
It is evident that the above reduction can be performed equivalently
for the heavy top dynamics by expanding the quotient
$TSO(3)/SO(2)\simeq TS^2\oplus\widetilde{\mathbb{R}}$, which gives
analogous equations to those above. Although the Lagrangian (and the
Hamiltonian) keep the same form as the single nematic particle, the
main difference resides in the form of the potential, whose explicit
expression is now $\phi(\boldsymbol{\Gamma})=\lambda\,\boldsymbol{\Gamma}\cdot \mathbf{k}$, where we use the notation of \S\ref{subsec_rigid_body}. This
expression is evidently $SO(2)$-invariant, rather than
$D_\infty$-invariant, consistently with the broken symmetry of the
heavy top system. Indeed, in the heavy top case, the dynamical
variable $\bf n$ belongs to the sphere $S^2\simeq SO(3)/SO(2)$
rather than to the projective plane $\Bbb{R}P^2\simeq
SO(3)/D_\infty$, thereby reflecting the different nature of the two
breaking symmetry subgroups $SO(2)$ and $D_\infty$. Moreover, we remark that the Ericksen-Leslie equation also holds for the heavy top dynamics provided the potential is now $\phi(\boldsymbol{\Gamma})=\lambda\,\boldsymbol{\Gamma}\cdot \mathbf{k}$. However, in this case, setting $r=0$ has no physical motivation.
\end{remark}

\begin{remark}[Lagrange-Poincar\'e approach for moving nematic particles]\normalfont
From a physical point of view, one should also take into account the
translational motion of nematic particles in the physical space $Q$,
thereby extending the LP Lagrangian to $\ell:
TQ\times\left(TS^2\times\Bbb{R}\right)\to\Bbb{R}$. This process has
the only effect of producing an extra Euler-Lagrange equation on
physical space, without changing any of the present geometric
construction. The final equations are then
\begin{equation*}
\frac{d}{dt}\frac{\partial \ell}{\partial \dot{q}}-\frac{\partial
\ell}{\partial q}=0\,,\qquad\frac{\nabla}{dt}\frac{\partial
\ell}{\partial \dot{\mathbf{n}}}-\frac{\nabla \ell}{\partial
\mathbf{n}}=\frac{\partial \ell}{\partial
r}\mathbf{n}\times\dot{\mathbf{n}}\,,\qquad\frac{d}{dt}\frac{\partial
\ell}{\partial r}=0.
\end{equation*}
with the Lagrangian
\[
\ell(q,\dot{q},{\bf n,\dot{n}},r)=\frac12\,\|
\dot{q}\|^2+\frac12\,j\,|\dot{\bf n}|^2-\Phi({\bf
n})+\frac12\,j\,r^2
\]
\end{remark}

\begin{remark}[Biaxial nematics]\label{remark_biaxial}\normalfont
Recall that the case of biaxial nematic particles involves the order
parameter space $SO(3)/D_2$, where $D_2$ is the finite dihedral
group. This discrete symmetry cannot be considered under the preceding Lagrange-Poincar\'e approach, since the latter is defined only for order parameter spaces $\mathcal{O/P}$, involving an isotropy subgroup
$\mathcal{P}\subset\mathcal{O}$, such that \makebox{$\operatorname{dim}(\mathcal{P})\geq1$}.
When the isotropy subgroup $\mathcal{P}$ is discrete, then the situation
requires more care in order to take into account the trivial nature of its Lie algebra
$\mathfrak{p}=\{0\}$. The broken symmetry group $\mathcal{O}$ becomes a principal
bundle $\mathcal{O}\to\mathcal{O/P}$ with discrete fiber and the reduction process in this case remains unknown.
\end{remark}

\subsection{Summary for uniaxial nematics}

This section gives an overview that summarizes the different approaches that have been carried out so far for uniaxial nematic particles. The starting point is the the unreduced  $SO(3)$-invariant Lagrangian
 $L:TSO(3)\times S^2 \rightarrow\mathbb{R}$ that  describes the dynamics of a single molecule. Then, we fix a reference direction $\mathbf{n}_0$ and consider the induced Lagrangians
\begin{align*}
L_{\mathbf{n}_0}&:TSO(3)\longrightarrow\mathbb{R},\quad L_{\mathbf{n}_0}=L_{\mathbf{n}_0}(v_\chi)\\
l&:\mathfrak{so}(3)\times S^2\longrightarrow\mathbb{R},\quad l=l(\boldsymbol{\nu},\mathbf{n})\\
\ell_{\mathbf{n}_0}&:TS^2\times\mathbb{R}\longrightarrow\mathbb{R},\quad \ell_{\mathbf{n}_0}=\ell_{\mathbf{n}_0}(\mathbf{n},\dot{\mathbf{n}},r),
\end{align*}
where we choose the reference configuration $\mathbf{n}_0=(0,0,1)$ for simplicity. Given a curve $\chi\in SO(3)$ and the vectors $\mathbf{n}=\chi\mathbf{n}_0\in S^2$ and $\boldsymbol{\nu}=\dot\chi\chi^{-1}\in\mathfrak{so}(3)$, the following are equivalent:
\begin{itemize}
\item [\bf{i}] The curve $\chi$ is a solution of the \textit{Euler-Lagrange equations} for $L_{\mathbf{n}_0}$.
\item [\bf{ii}] The curves $\boldsymbol{\nu}$ and $\mathbf{n}$ are solutions of the \textit{Euler-Poincar\'e equations} for $l$:
\begin{equation}\label{EP_nematics2}
\frac{d}{dt}\frac{\delta l}{\delta\boldsymbol{\nu}}+\frac{\delta l}{\delta\boldsymbol{\nu}}\times \boldsymbol{\nu}=\mathbf{n}\times \frac{\delta l}{\delta\mathbf{n}}
\,\qquad
\dot{\mathbf{n}}=\boldsymbol\nu\times{\bf n}
.
\end{equation}
\item [\bf{iii}] The curves $\mathbf{n}$ and $r$ are solutions of the \textit{Lagrange-Poincar\'e equations} for $\ell_{\mathbf{n}_0}$:
\begin{equation}\label{2}
\frac{d}{dt}\frac{\partial \ell_{\mathbf{n}_0}}{\partial r}=0,\;\;\frac{d}{dt}\frac{\partial \ell_{\mathbf{n}_0}}{\partial \dot{\mathbf{n}}}-\frac{\partial \ell_{\mathbf{n}_0}}{\partial \mathbf{n}}=\frac{\partial \ell_{\mathbf{n}_0}}{\partial r}\mathbf{n}\times\dot{\mathbf{n}}.
\end{equation}
\end{itemize}
To obtain the equations \eqref{EP_nematics2} from the Euler-Poincar\'e equations \eqref{EP}, we used the formulas
$\operatorname{ad}^*_{\boldsymbol{\nu}}\boldsymbol{\kappa}=-\boldsymbol{\nu}\times
\boldsymbol{\kappa}$ and the expression of the momentum map
\[
\mathbf{J}:T^*S^2\rightarrow\mathfrak{so}(3)^*,\quad \mathbf{J}(\mathbf{n},\boldsymbol{\pi})=\mathbf{n}\times\boldsymbol{\pi}.
\]
for the cotangent lifted action of $SO(3)$ on $T^*S^2$. The conservation law \eqref{cons_law} for nematics reads
\[
\frac{\partial}{\partial t}\left(\chi^{-1}\frac{\delta l}{\delta\boldsymbol{\nu}}\right)=\mathbf{n}_0\times\left(\chi^{-1}\frac{\delta l}{\delta\mathbf{n}}\right).
\]
On can pass from the Euler-Poincar\'e \eqref{EP_nematics2} to the
Lagrange-Poincar\'e equations \eqref{2}, using the relations:
\[
\boldsymbol{\nu}=\mathbf{n}\times\dot{\mathbf{n}}+r\mathbf{n}
\]
\[
r=\mathbf{n}\!\cdot\!\boldsymbol{\nu},\quad\dot{\mathbf{n}}=\boldsymbol{\nu}\times\mathbf{n}.
\]
The link between the Lagrangian variables $(\chi,\dot\chi)$ and the reduced variables $(\nu,\mathbf{n})$ and $(\mathbf{n},\dot{\mathbf{n}},r)$ is illustrated in the following diagram.

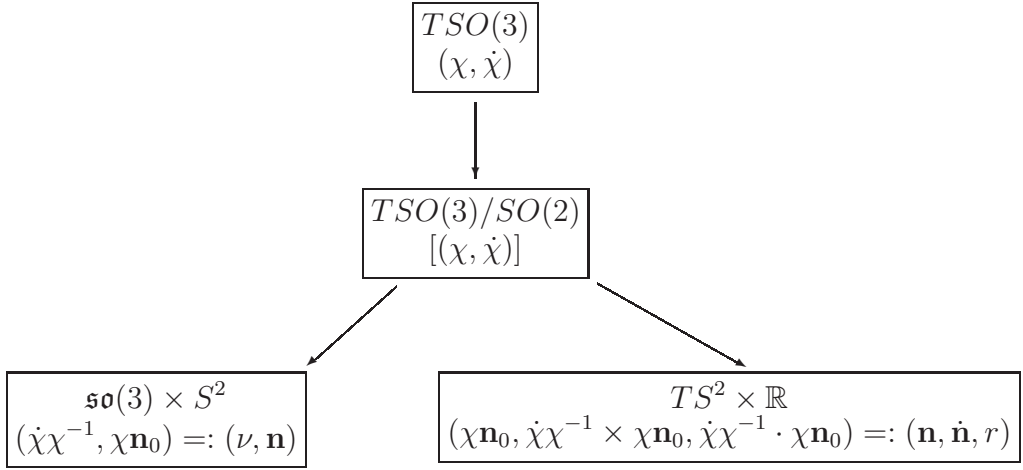
\begin{diagram}
\hspace{-3cm}   &        & \fbox{$\displaystyle\underset{\displaystyle(\chi,\dot\chi)}{TSO(3)}$}    &        & \\
\hspace{-3cm}   &        & \dTo &        & \\
   &        &   \fbox{$\displaystyle\underset{\displaystyle[(\chi,\dot\chi)]}{TSO(3)/SO(2)}$}  &        & \\
\hspace{-3cm}   & \ldTo  &      & \rdTo  & \\
\hspace{-3cm} \fbox{$\displaystyle\underset{\displaystyle(\dot\chi\chi^{-1},\chi\mathbf{n}_{0})=:(\nu,\mathbf{n})}{\mathfrak{so}(3)\times S^2}$} &        &      &        & \hspace{-2cm}\fbox{$\displaystyle\underset{\displaystyle(\chi\mathbf{n}_{0},\dot\chi\chi^{-1}\times\chi\mathbf{n}_{0},\dot\chi\chi^{-1}%
\cdot\chi\mathbf{n}_{0})=:({\bf n},\dot{\bf n},r)}{TS^2\times\Bbb{R}}$} \\
\end{diagram}

\begin{remark}[unit sphere vs. unit vectors]\normalfont
Instead of the sphere $M=S^2$, one can start with the vector space
$M=\mathbb{R}^3$ on which $SO(3)$ acts by matrix multiplication.
This is the point of view adopted in \cite{GBRa2008}. In this
case, if one chooses the initial condition $\mathbf{n}_0\in S^2$,
then we still have $\mathbf{n}(t)=\chi(t)\mathbf{n}_0\in S^2$, thus
these approaches are equivalent.
\end{remark}

\begin{remark}[Directors and the real projective plane]\normalfont Recall that the parameter space for nematics is the projective plane $\mathbb{R}P^2$ and not the sphere $S^2$. However, the results $\bf{i}-\bf{iv}$ are still true when one works with the projective plane, that is, when $\mathbf{n}$ is interpreted as a director instead as a unit vector. The only change concerns the Lagrange-Poincar\'e equations which are less explicit in the case of the projective plane. This is the reason why we use $S^2$ instead of $\mathbb{R}P^2$ above.
\end{remark}

\begin{remark}[Euler-Poincar\'e equations via Lagrangian reduction]\normalfont
From our discussions we realize that the Euler-Poincar\'e and Lagrange-Poincar\'e equations arise from two different approaches in reduction theory. However it is possible to obtain the Euler-Poincar\'e equations by standard Lagrangian reduction, provided one suitably enlarges the physical configuration space. We address the reader to \cite{CeHoMaRa1998,GBHoRa2009a} for the case of order parameter vector spaces. This picture produces the so called Clebsch constrained variational principle \cite{Ho2002}.
\end{remark}

\rem{ 
\begin{remark}[Biaxial nematics]\label{remark_biaxial}\normalfont
Recall that the case of biaxial nematic particles involves the order
parameter space $SO(3)/D_2$, where $D_2$ is the finite dihedral
group. Strictly speaking, such a discrete symmetry is not considered by the preceding Lagrange-Poincar\'e approach, since the latter is defined only for order parameter spaces $\mathcal{O/P}$, involving a Lie subgroup
$\mathcal{P}$ of the broken symmetry $\mathcal{O}$, such that $\operatorname{dim}(\mathcal{P})\geq1$.
When the isotropy subgroup $\mathcal{P}$ is discrete, then the situation
requires more care in order to take into account the trivial nature of its Lie algebra
$\mathfrak{p}=\{0\}$. The broken symmetry group $\mathcal{O}$ becomes a principal
bundle $\mathcal{O}\to\mathcal{O/P}$ with discrete fiber and the reduction proceeds analogously, with the exception that the adjoint bundle is absent.
Indeed, one has the following
\begin{proposition} Given a discrete subgroup $\mathcal{P}$ of a Lie group
$\mathcal{O}$, the map
\begin{eqnarray*}
(T\mathcal{O})/\mathcal{P} & \to & T(\mathcal{O}/\mathcal{P}) \\
\left[v_g\right] & \mapsto & T_g\pi(v_g)
\end{eqnarray*}
is a diffeomorphism.
\end{proposition}
\paragraph{Proof.}
 Since this
map is clearly surjective, it remains to show that $T_g\pi(v_g)=0$ implies $v_g=0$. To see this, we consider a curve $g(t)\in G$ such that $\dot g(0)=v_g$.
\todo{To be completed}
$\blacksquare$
\end{remark}
}    

\section{Hydrodynamics of nematic liquid crystals}\label{continuum_OP}

While the previous discussions have focused on the geometric dynamics of a single particle with broken symmetry, this section develops the same ideas in the more physical situation of a fluid system of particles with micro-structure,
such as nematic particles. Thus, this section provides the  link between the geometric treatment previously applied at the microscopic
 single-particle  level and the macroscopic models usually adopted for liquid
 crystal dynamics. We shall see how the same ideas apply to continuum media
 without substantial modifications. The extension to nematic fluid
 dynamics requires considering the diffeomorphism group, i.e. the  particle relabeling group  well known in the Lagrangian picture of fluid dynamics.

We begin by recalling our notation: $\mathcal{O}$ be the
order parameter group of a certain particle with broken symmetry,
and denote by $M$ the order parameter space, on which $\mathcal{O}$
acts transitively, with isotropy group $\mathcal{P}$. As we have
seen before, passing from one particle to a system of $N$ particles
simply consists in replacing the group $\mathcal{O}$ by the direct
product $\mathcal{O}^N$ acting on the cartesian product $M^N$. When
a continuum of particles with broken symmetry is considered, one
needs to consider as symmetry group the group
$G:=\mathcal{F}(\mathcal{D},\mathcal{O})$ of smooth maps form the
physical space $\mathcal{D}$ to the order parameter group
$\mathcal{O}$ of the single particle. We still denote by
$\chi:\mathcal{D}\rightarrow \mathcal{O}$ these maps. The order
parameter space is the manifold of maps
$\mathcal{M}:=\mathcal{F}(\mathcal{D},M)$ on which $G$ acts by the
naturally induced pointwise action.

 In order to
describe the hydrodynamic of such systems of particles (called
complex fluids), we need to include the group of diffeomorphisms in the symmetry group. For simplicity, we suppose that the fluid is incompressible, since the compressible case would require only a slight modification. Thus we need to consider the group $\operatorname{Diff}_{\rm vol}(\mathcal{D})$ of volume preserving diffeomorphisms of $\mathcal{D}$, relative to a fixed volume form $\mu$ on $\mathcal{D}$. It will be convenient to fix a Riemannian metric $g$ on $\mathcal{D}$ and to choose $\mu$ as the volume form associated to the metric.

\subsection{Euler-Poincar\'e formulation}

In the Euler-Poincar\'e framework, the dynamics of complex fluids is obtained by
considering the semidirect product
$G=\operatorname{Diff}_{\rm vol}(\mathcal{D})\,\circledS\,\mathcal{F}(\mathcal{D},\mathcal{O})$
acting on advected variables by linear or affine representation as
shown in \cite{GBRa2008}. Here the advected variable is the order
parameter function $n\in\mathcal{F}(\mathcal{D},M)$, on which $G$ acts by the left action
\begin{equation}\label{action_DF_on_n}
n\mapsto (\eta,\chi)n:=(\chi n)\circ\eta^{-1},\quad (\eta,\chi)\in G.
\end{equation}
Recall that $\chi n$ denotes the action of $\mathcal{F}(\mathcal{D},\mathcal{O})$ on $\mathcal{F}(\mathcal{D},M)$ naturally induced by the action of $\mathcal{O}$ on $M$. One easily checks that \eqref{action_DF_on_n} defines an action of the semidirect product $G=\operatorname{Diff}_{\rm vol}(\mathcal{D})\,\circledS\,\mathcal{F}(\mathcal{D},\mathcal{O})$.
Using the expression
\[
\operatorname{ad}^*_{(u,\nu)}(m,\kappa)=\left(\pounds_um+\kappa\!\cdot\!\nabla\nu,\pounds_u\kappa+\operatorname{ad}^*_\nu\kappa\right),
\]
for the infinitesimal coadjoint action, we obtain from \eqref{EP} the equations
\begin{equation}\label{liquid}
\left\{
\begin{array}{l}
\vspace{0.2cm}\displaystyle \frac{\partial }{\partial t}\frac{\delta l}{\delta u}+\pounds_u\frac{\delta l}{\delta u}+\frac{\delta l}{\delta \nu}\!\cdot\!\nabla\nu=-\frac{\delta l}{\delta n}\!\cdot\!\nabla n-\nabla p,\qquad\operatorname{div}u=0\\
\displaystyle \frac{\partial }{\partial t}\frac{\delta l}{\delta\nu}+\pounds_u\frac{\delta l}{\delta \nu}+\operatorname{ad}^*_\nu\frac{\delta l}{\delta \nu}=\mathbf{J}\circ\frac{\delta l}{\delta n},
\end{array}\right.
\end{equation}
together with the kinematic equation
\[
\dot n+u \!\cdot\! \nabla n=\nu_M\circ n.
\]
The Lagrangian $L_{n_0}:T \big(\operatorname{Diff}_{\rm vol}(\mathcal{D})\,\circledS\,\mathcal{F}(\mathcal{D},\mathcal{O})\big)\rightarrow\mathbb{R}$ is usually of the form
\[
L_{n_0}(\eta,\dot\eta,\chi,\dot\chi)=\frac{1}{2}\int_\mathcal{D}\|\dot\eta\|^2\mu+\frac{1}{2}\int_\mathcal{D}j|\dot\chi|^2\mu-\int_\mathcal{D}F\left((\chi n_0)\circ\eta^{-1},\nabla ((\chi n_0)\circ\eta^{-1})\right)\mu,
\]
where the norm of $\dot\eta$ is given by the Riemannian metric $g$,
the norm of $\dot\chi$ is associated to a right-invariant Riemannian
metric on $\mathcal{O}$, and $F$ is the free energy. One easily
checks that this expression is invariant under the right action of
$(\eta,\chi)\in G=\operatorname{Diff}_{\rm
vol}(\mathcal{D})\,\circledS\,\mathcal{F}(\mathcal{D},\mathcal{O})$
given by cotangent lift on $TG$ and acting on $n_0$ as
$n_0\mapsto \chi^{-1}(n_0\circ\eta)$.
The reduced Euler-Poincar\'e Lagrangian recovers the expression
\begin{equation}\label{typicalLagrangian}
l(u,\nu,n)=\frac{1}{2}\int_\mathcal{D}\|u\|^2\mu+\frac{1}{2}\int_\mathcal{D}j|\nu|^2\mu-\int_\mathcal{D}F(n,\nabla n)\mu,
\end{equation}
where $F$ is the free energy.

\subsubsection{Euler-Poincar\'e fluid equations for nematic molecules}

The case of nematic liquid crystals can be treated as above, choosing $\mathcal{O}=SO(3)$, $M=\mathbb{R}P^2$. In this case the free energy $F$ is usually given by the Oseen-Z\"ocher-Frank expression
\begin{equation}\label{standard_energy}
F(\mathbf{n},\nabla\mathbf{n})=\frac{1}{2}K_1(\operatorname{div}\mathbf{n})^2+\frac{1}{2}K_2(\mathbf{n}\cdot\operatorname{curl}\mathbf{n})^2+\frac{1}{2}K_3|\mathbf{n}\times\operatorname{curl}\mathbf{n}|^2,
\end{equation}
where the constants $K_1, K_2, K_3$ are respectively associated to the three
principal distinct director axis deformations in nematics,
namely, splay, twist, and bend. At this point, the system \eqref{liquid} produces the equations
\begin{equation}\label{liquid2}
\left\{
\begin{array}{l}
\vspace{0.2cm}\displaystyle \partial_t u+u\!\cdot\!\nabla u =-\partial_i\left(\nabla\mathbf{n}^T\!\cdot\!\frac{\partial F}{\partial\mathbf{n}_{,i}}\right)-\nabla p,\qquad\operatorname{div}u=0\\
\displaystyle j\left(\dot{\boldsymbol{\nu}}+u\!\cdot\!\nabla\boldsymbol{\nu}\right)=\mathbf{h}\times\mathbf{n},\qquad \mathbf{h}:=-\frac{\delta l}{\delta \mathbf{n}}=\frac{\partial F}{\partial\mathbf{n}}-\partial_i\left(\frac{\partial F}{\partial\mathbf{n}_{,i}}\right)
\end{array}\right.
\end{equation}
together with the kinematic equation
\[
\dot {\mathbf{n}}+u\!\cdot\!\nabla\mathbf{n}=\boldsymbol{\nu}\times\mathbf{n}.
\]
The Ericksen-Leslie fluid equations follow easily from the above relations, under the assumption that the initial condition verifies $\boldsymbol{\nu}_0\!\cdot\!\mathbf{n}_0=0$ \cite{GBRa2008}.
A direct computation using \eqref{red_Poisson} shows that the associated Poisson bracket reads
\begin{align}\label{liquid_crystals_bracket}
\{f,g\}(m,\boldsymbol{\mu},\mathbf{n})&=\left\langle m,\left[\frac{\delta
f}{\delta m},\frac{\delta g}{\delta m}\right]\right\rangle\nonumber+\left\langle \boldsymbol{\mu},\left(\frac{\delta
f}{\delta\boldsymbol{\mu}}\times\frac{\delta
g}{\delta\boldsymbol{\mu}}+\nabla\frac{\delta
f}{\delta\boldsymbol{\mu}}\cdot\frac{\delta
g}{\delta m}-\nabla\frac{\delta
g}{\delta\boldsymbol{\mu}}\cdot\frac{\delta
f}{\delta m}\right)\right\rangle\nonumber\\
&\quad+\left\langle \mathbf{n}\times\frac{\delta
f}{\delta\boldsymbol{\mu}}+\nabla\mathbf{n}\cdot\frac{\delta
f}{\delta m},\frac{\delta
g}{\delta\mathbf{n}}\right\rangle-\left\langle\mathbf{n}\times\frac{\delta
g}{\delta\boldsymbol{\mu}}+\nabla\mathbf{n}\cdot\frac{\delta
g}{\delta{m}},\frac{\delta
f}{\delta\mathbf{n}}\right\rangle,
\end{align}
where $m=\delta l/\delta u$ is the fluid momentum and the brackets
$\langle\,,\rangle$ denote $L^2$ duality. The first two terms is the
Lie-Poisson bracket, associated to the semidirect product
$\operatorname{Diff}_{\rm
vol}(\mathcal{D})\,\circledS\,\mathcal{F}(\mathcal{D},SO(3))$. Of
course, this bracket is consistent with that of the single particle,
see \eqref{Poisson_nematic_particle}.

From the discussion above, it is easy to generalize the Euler-Poincar\'e fluid equations for uniaxial nematics to the case of biaxial liquid crystals. Indeed, upon choosing $\mathcal{O}=SO(3)$ and $M=SO(3)/D_2$, these equations can be written down directly as follows:
\begin{equation*}
\left\{
\begin{array}{l}
\vspace{0.2cm}\displaystyle \frac{\partial }{\partial t}\frac{\delta l}{\delta u}+\pounds_u\frac{\delta l}{\delta u}+\frac{\delta l}{\delta \boldsymbol\nu}\cdot \nabla\boldsymbol\nu=-\sum_{l=1}^2\, \nabla \mathbf{n}_l\cdot\frac{\delta l}{\delta \mathbf{n}_l}-\nabla p,\qquad\operatorname{div}u=0\\
\displaystyle \frac{\partial }{\partial t}\frac{\delta l}{\delta\boldsymbol\nu}+\operatorname{div}\!\left(\frac{\delta l}{\delta \boldsymbol\nu}\,u\right)-\boldsymbol\nu\times\frac{\delta l}{\delta \boldsymbol\nu}=\sum_{l=1}^2\, \mathbf{n}_l\times\frac{\delta l}{\delta \mathbf{n}_l}\,,
\end{array}\right.
\end{equation*}
together with the kinematic equation
\[
\dot{\mathbf{n}}_l+u \!\cdot\! \nabla \mathbf{n}_l= \mathbf{n}_l \times\boldsymbol\nu\,,\qquad
l=1,2.
\]
On the Hamiltonian side, the corresponding Poisson bracket reads
\begin{align*}
\{f,g\}(m,\boldsymbol{\mu},\mathbf{n}_1,\mathbf{n}_2)=&\left\langle m,\left[\frac{\delta
f}{\delta m},\frac{\delta g}{\delta m}\right]\right\rangle\nonumber+\left\langle \boldsymbol{\mu},\left(\frac{\delta
f}{\delta\boldsymbol{\mu}}\times\frac{\delta
g}{\delta\boldsymbol{\mu}}+\nabla\frac{\delta
f}{\delta\boldsymbol{\mu}}\cdot\frac{\delta
g}{\delta m}-\nabla\frac{\delta
g}{\delta\boldsymbol{\mu}}\cdot\frac{\delta
f}{\delta m}\right)\right\rangle\nonumber
\\
&
+\sum_{l=1}^2
\left\langle \mathbf{n}_l\times\frac{\delta
f}{\delta\boldsymbol{\mu}}+\nabla\mathbf{n}_l\cdot\frac{\delta
f}{\delta m},\frac{\delta
g}{\delta\mathbf{n}_l}\right\rangle-\left\langle\mathbf{n}_l\times\frac{\delta
g}{\delta\boldsymbol{\mu}}+\nabla\mathbf{n}_l\cdot\frac{\delta
g}{\delta{m}},\frac{\delta
f}{\delta\mathbf{n}_l}\right\rangle.
\end{align*}
The next section extends the Euler-Poincar\'e approach for the alignment tensor dynamics of a single nematic molecule to the case of nematic liquid crystals.

\subsubsection{Alignment tensor dynamics in nematic liquid crystals}
 It is well known that fluid equations are obtained from the single particle dynamics by taking statistical averages with respect to some probability distribution function. The case of liquid crystals is not an exception to this procedure and one is usually interested in the dynamics of the averaged alignment tensor $\bar{\mathsf{Q}}$, where the `bar' symbol denotes an appropriate averaging of the single particle alignment tensor $\sf Q$. In order to formulate the Euler-Poincar\'e equations for liquid crystals in terms of the averaged quantity $\bar{\mathsf{Q}}$, one proceeds by simply replacing the order parameter space $M=\Bbb{R}P^2$ with the space of symmetric matrices $\operatorname{Sym}(3)$. The Euler-Poincar\'e reduction process is performed on the unreduced Lagrangian $L_{\mathsf{Q}_0}:T \big(\!\operatorname{Diff}_{\rm vol}(\mathcal{D})\,\circledS\,\mathcal{F}(\mathcal{D},SO(3))\big)\rightarrow\mathbb{R}$ with respect to the action
\[
(\eta,\chi){\sf Q}=\left(\chi{\sf Q}\chi^{-1}\right)\circ\eta^{-1},
\]
where evidently  ${\sf Q}\in\mathcal{F(D},\operatorname{Sym}(3))$ and we have suppressed the averaging notation. Thus, if one specializes the general equations \eqref{liquid} to the case under consideration, one obtains
\begin{equation}
\left\{
\begin{array}{l}
\vspace{0.2cm}\displaystyle \frac{\partial }{\partial t}\frac{\delta l}{\delta u}+\pounds_u\frac{\delta l}{\delta u}+\frac{\delta l}{\delta \boldsymbol\nu}\cdot \nabla\boldsymbol\nu=- \nabla {\sf Q}^{ij}\frac{\delta l}{\delta {\sf Q}^{ij}}-\nabla p,\qquad\qquad\qquad\ \operatorname{div}u=0\\
\displaystyle \frac{\partial }{\partial t}\frac{\delta l}{\delta\boldsymbol\nu}+\operatorname{div}\!\left(\frac{\delta l}{\delta \boldsymbol\nu}\,u\right)-\boldsymbol\nu\times\frac{\delta l}{\delta \boldsymbol\nu}=-\,
\!\overrightarrow{\,\left[{\sf Q},\frac{\delta
l}{\delta {\sf Q}}\right]\,}
,\qquad\quad\,
\dot{\sf Q}+u\!\cdot\!\nabla{\sf Q}=[\hat{\boldsymbol{\nu}},{\sf Q}]
\,,
\end{array}
\right.
\end{equation}
or more explicitly, for a Lagrangian of the form \eqref{typicalLagrangian} with $n$ replaced by $\sf Q$,
\begin{equation}
\left\{
\begin{array}{l}
\vspace{0.2cm}\displaystyle \partial_t u+u\!\cdot\!\nabla u =- \partial_k\left(\frac{\partial F}{\partial {\sf Q}^{ij}_{\;,k}}\right)\nabla{\sf Q}^{ij}-\nabla p,\qquad\qquad\quad\operatorname{div}u=0\\
\displaystyle j\left(\dot{\boldsymbol{\nu}}+u\!\cdot\!\nabla\boldsymbol{\nu}\right)=\,
\!\overrightarrow{\,\left[{\sf Q},{\sf H}\right]\,}
,\hspace{3.5cm}
\dot{\sf Q}+u\!\cdot\!\nabla{\sf Q}=[\hat{\boldsymbol{\nu}},{\sf Q}]
\,,
\end{array}
\right.
\end{equation}
where the molecular field $\sf H$ is given by
\[
{\sf H}=-\frac{\delta
l}{\delta {\sf Q}}=\frac{\partial F}{\partial {\sf Q}}-\partial_k\left(\frac{\partial F}{\partial {\sf Q}_{,k}}\right)
\]
with the notation
$\overrightarrow{A\,}_{\!i}=\varepsilon_{ijk}A_{jk}$ and the summation convention over repeated tensor indexes. The free energy $F({\sf Q,\nabla Q})$ appearing in the reduced Lagrangian $l(u,\boldsymbol\nu,\mathsf{Q})$ is usually given by the Landau-de Gennes free energy in the form \cite{deGennes1971}
\[
F({\sf Q,\nabla Q})=
\kappa_1\left\|\nabla {\sf Q}\right\|^2+\kappa_2\left\|\nabla\cdot
{\sf Q}\right\|^2+\kappa_3\operatorname{Tr}({\sf Q}\,\nabla\!\times\!{\sf Q})+
a_{21}\operatorname{Tr}({\sf Q}^2)+a_{31}\operatorname{Tr}({\sf Q}^3)
\]
where the tensor norms are given by total contraction of the indexes, i.e. $\left\|\nabla {\sf Q}\right\|^2=\sum\left(\partial_{k}{\sf Q}^{ij}\right)^2$ and $\left\|\nabla\cdot
{\sf Q}\right\|^2=\sum\left(\partial_i{\sf Q}^{ij}\partial_k{\sf Q}^{kj}\right)$, while $\left(\nabla\times\right)_{ij}=\epsilon_{ijk}\partial_k$ is considered as a matrix operator.
Higher order expansions in the alignment tensor are also possible.

On the Hamiltonian side, the corresponding Poisson bracket reads
\begin{align*}
\{f,g\}(m,\boldsymbol{\mu},\mathsf{Q})=&\left\langle m,\left[\frac{\delta
f}{\delta m},\frac{\delta g}{\delta m}\right]\right\rangle\nonumber+\left\langle \boldsymbol{\mu},\left(\frac{\delta
f}{\delta\boldsymbol{\mu}}\times\frac{\delta
g}{\delta\boldsymbol{\mu}}+\nabla\frac{\delta
f}{\delta\boldsymbol{\mu}}\cdot\frac{\delta
g}{\delta m}-\nabla\frac{\delta
g}{\delta\boldsymbol{\mu}}\cdot\frac{\delta
f}{\delta m}\right)\right\rangle\nonumber
\\
&
+
\operatorname{Tr}\left( \left[\mathsf{Q},\frac{\delta
f}{\delta\hat{\boldsymbol{\mu}}}\right]\frac{\delta
g}{\delta\mathsf{Q}}+\left(\frac{\delta
f}{\delta{m}}\!\cdot\!\nabla\!\right)\!\mathsf{Q}\,\frac{\delta
g}{\delta\mathsf{Q}}\right)-\operatorname{Tr}\left(\left[\mathsf{Q},\frac{\delta
g}{\delta\hat{\boldsymbol{\mu}}}\right]\frac{\delta
f}{\delta\mathsf{Q}}+\left(\frac{\delta
g}{\delta{m}}\!\cdot\!\nabla\!\right)\!\mathsf{Q}\,\frac{\delta
f}{\delta\mathsf{Q}}\right)
\end{align*}
and it is easy to recognize that the same relations also hold for biaxial liquid crystals.

\begin{remark}[Compressible fluid flows]\label{remark1_compressible_flow}\normalfont
In the Euler-Poincar\'e setting, the generalization to compressible
fluid flows is straightforward. Indeed, it is sufficient to consider
the whole diffeomorphism group $\operatorname{Diff}(\mathcal{D})$
and to enlarge the order parameter space by replacing
$\mathcal{F(D},M)$ with the product $\mathcal{F(D},M)\times{\rm
Den}(\mathcal{D})$, where ${\rm Den}(\mathcal{D})$ denotes the space
of densities on $\mathcal{D}$. Then one repeats the construction
above, upon considering the following  action of
$G=\operatorname{Diff}(\mathcal{D})\,\circledS\,\mathcal{F}(\mathcal{D},\mathcal{O})$
on \makebox{$\mathcal{F(D},M)\times{\rm Den}(\mathcal{D})$}
\[
(\eta,\chi)\cdot(\rho,{\bf n})=\left((\rho\circ\eta)J(\eta),\chi^{-1}({\bf
n}\circ\eta)\right)\,.
\]
where $J(\eta)$ is the Jacobian of $\eta$ with respect to a volume
form on $\mathcal{D}$. In this sense, compressibility is by itself
another example of symmetry breaking in infinite dimensions (cf.
\cite{HoMaRa1998}). Indeed, even in the case of ordinary isotropic
compressible fluids, the full ${\rm Diff}$-symmetry of the unreduced
Lagrangian $L_{\rho_0}:T{\rm Diff}(\mathcal{D})\to\Bbb{R}$ is broken
by the presence of the density variable $\rho_0\in{\rm
Den}(\mathcal{D})$, so that the only symmetry which is left is given
by the infinite-dimensional isotropy subgroup ${\rm
Diff}_{\!\rho_0}(\mathcal{D})$. Upon applying the Euler-Poincar\'e
approach, one performs the reduction in \eqref{lisboa}: $T{\rm
Diff}(\mathcal{D})/{\rm
Diff}_{\rho_0}(\mathcal{D})\simeq\mathfrak{X}(\mathcal{D})\times{\rm
Den}(\mathcal{D})$ because of the transitivity of the action (see
e.g. \cite{Khesin}). In principle, it is also possible to apply
Lagrangian reduction and write the Lagrange-Poincar\'e equations on
the reduced space $T{\rm Diff}(\mathcal{D})/{\rm
Diff}_{\rho_0}(\mathcal{D})\simeq T{\rm
Den}(\mathcal{D})\oplus\widetilde{\mathfrak{X}_{\rm
vol}(\mathcal{D})}$. Although the next discussion is devoted to
Lagrangian reduction, we shall consider again the case of
incompressible nematic liquid crystals.
\end{remark}

\subsection{Lagrange-Poincar\'e formulation of uniaxial nematic liquid crystals}

The Lagrange-Poincar\'e approach to continuum nematic
media produces another variant of the fluid equations for liquid
crystals. One starts with the same Lagrangian as before, namely
\[
L_{\mathbf{n}_0}(\eta,\dot\eta,\chi,\dot\chi)=\frac12\int_{\mathcal{D}}
\|\dot\eta\|^2\mu+\frac{1}{2}
\int_\mathcal{D}j\left|\dot{\chi}\right|^2\mu-\int_\mathcal{D}F\left((\chi \mathbf{n}_0)\circ\eta^{-1},\nabla((\chi \mathbf{n}_0)\circ\eta^{-1})\right)\mu.
\]
However we will now reduce the dynamics in two steps. The first step is a Lagrange-Poincar\'e reduction involving the internal variables. The second step involves the diffeomorphism group and is done by using the metamorphosis approach.

\subsubsection{First reduction stage: Lagrange-Poincar\'e approach}

As a first step of reduction, one applies the
Lagrange-Poincar\'e approach for $L_{\mathbf{n}_0}$ according to the vector bundle isomorphism
\[
T{\rm Diff}_{\rm vol}(\mathcal{D})\times
T\mathcal{F}(\mathcal{D},SO(3))/\mathcal{F}(\mathcal{D},SO(2))\ \to\
T{\rm Diff}_{\rm vol}(\mathcal{D})\times \left(
T\mathcal{F}(\mathcal{D},S^2)\oplus\mathcal{F(D)}\right)
\]
over ${\rm Diff}_{\rm
vol}(\mathcal{D})\times\mathcal{F}(\mathcal{D},S^2)$.
By a slight abuse of notation, we replaced $\operatorname{Orb}(n_0)\subset\mathcal{F}(\mathcal{D},S^2)$
 with $\mathcal{F}(\mathcal{D},S^2)$ itself. Also,  we have replaced the isotropy group $\mathcal{F}(\mathcal{D},D_\infty)$
 with the simpler choice
 $\mathcal{F}(\mathcal{D},SO(2))$, in analogy with the
preceding finite-dimensional treatment.
In this case, upon fixing $\mathbf{n}_0=(0,0,1)$, the Lagrange-Poincar\'e
reduction with respect to the mechanical connection yields the LP
Lagrangian $\ell(\eta,\dot\eta,\bar{\bf n},{\dot{\bar{\bf
n}}},\bar{r})$ on the reduced space
\[
T\operatorname{Diff}_{\rm vol}(\mathcal{D})\times\left(
T\mathcal{F}(\mathcal{D},S^2)\oplus
\mathcal{F}(\mathcal{D})\right)\ni
(\eta,\dot\eta,\bar{\mathbf{n}},\dot{\bar{\mathbf{n}}},\bar r)
\,.
\]
The LP Lagrangian is
\[
\ell(\eta,\dot\eta,\bar{\mathbf{n}},\dot{\bar{\mathbf{n}}},\bar
r)=\frac12\int_{\mathcal{D}} \|\dot\eta\|^2\mu+
\frac{j}{2}\int_\mathcal{D}|\dot{\bar{\mathbf{n}}}|^2\mu+\frac{j}{2}\int_\mathcal{D}\bar
r^2\mu-\int_\mathcal{D}F(\bar{\mathbf{n}}\circ\eta^{-1},\nabla(\bar{\mathbf{n}}\circ\eta^{-1})\mu.
\]

At this point, one recalls that the quantity
$\bar{r}\in\mathcal{F(D})$ in the above Lagrangian
is a constant parameter, due to the form of the Lagrangian. Thus, it
is allowed to set $\bar{r}\equiv0$ for convenience, so that the
rotational nematic dynamics is regulated by the ordinary
Euler-Lagrange equations on $\mathcal{F}(\mathcal{D},S^2)$.
Therefore, one obtains the Lagrangian
$\ell':T{\rm Diff}_{\rm vol}(\mathcal{D})\times
T\mathcal{F}(\mathcal{D},S^2)\to\mathbb{R}$
given by
\begin{align*}
\ell'(\eta,\dot\eta,\bar{\bf n},{\dot{\bar{\bf
n}}}):&=\ell(\eta,\dot\eta,\bar{\bf n},{\dot{\bar{\bf
n}}},0)\\
&=\frac12\int_{\mathcal{D}} \|\dot\eta\|^2\mu+
\frac{j}{2}\int_\mathcal{D}|\dot{\bar{\mathbf{n}}}|^2\mu-\int_\mathcal{D}F(\bar{\mathbf{n}}\circ\eta^{-1},\nabla(\bar{\mathbf{n}}\circ\eta^{-1}))\mu
\end{align*}
where one can verify directly the invariance of the free energy $F$
under the right action
\[
(\eta,\dot\eta,\bar{\mathbf{n}},\dot{\bar{\mathbf{n}}})\cdot\varphi
=
(\eta\circ\varphi,\dot\eta\circ\varphi,\bar{\mathbf{n}}\circ\varphi,\dot{\bar{\mathbf{n}}}\circ\varphi)
\]
by simply observing that
\[
F\!\left((\bar{\mathbf{n}}\circ\varphi)\circ(\eta\circ\varphi)^{-1},\nabla\!\left((\bar{\mathbf{n}}\circ\varphi)\circ(\eta\circ\varphi)^{-1}\right)\right)=
F(\bar{\mathbf{n}}\circ\eta^{-1},\nabla(\bar{\mathbf{n}}\circ\eta^{-1})).
\]
We now endow the manifold $\mathcal{F}(\mathcal{D},S^2)$ with the
Levi-Civita connection associated to the Riemannian metric given by
integration over $\mathcal{D}$ of the natural Riemannian on
$S^2\subset\mathbb{R}^3$.
Upon denoting by $\nabla/dt$ the covariant derivative with respect to the Levi-Civita connection of $S^2$, one writes the explicit form of the Euler-Lagrange equations
\[
\frac{d}{dt}\frac{\partial \ell'}{\partial\dot\eta}-\frac{\partial
\ell'}{\partial\eta}=0,\quad \frac{\nabla}{dt}\frac{\partial
\ell'}{\partial\dot{\bar{\mathbf{n}}}}-\frac{\nabla
\ell'}{\partial\bar{\mathbf{n}}}=0
\]
on the tangent bundle of $\operatorname{Diff}_{\rm
vol}(\mathcal{D})\times\mathcal{F}(\mathcal{D},S^2)$. The above equations are found
for example in \cite{Ho2002}, where they are shown to be equivalent
to the Ericksen-Leslie fluid's equations.  Of course, one can obtain these equations from \eqref{liquid2}, when the initial condition $\mathbf{n}_0$ and $\boldsymbol{\nu}_0$ are orthogonal.

\subsubsection{Second reduction stage: metamorphosis approach}

The Eulerian form of the fluid equations
is found by noting that the Lagrangian $\ell'$ is invariant under the cotangent-lift of the right action
$(\eta,\bar{\mathbf{n}})\mapsto
(\eta\circ\varphi,\bar{\mathbf{n}}\circ\varphi)$
of the diffeomorphism $\varphi\in\operatorname{Diff}_{\rm
vol}(\mathcal{D})$. Thus, one performs the reduction process
according to the quotient map
\[
T{\rm Diff}_{\rm vol}(\mathcal{D})\times
T\mathcal{F}(\mathcal{D},S^2)\rightarrow \mathfrak{X}_{\rm
vol}(\mathcal{D})\times T\mathcal{F}(\mathcal{D},S^2)
\]
given by $(\eta,\dot\eta,\bar{\mathbf{n}},\dot{\bar{\mathbf{n}}})\mapsto (u,\mathbf{n},D_t\mathbf{n}):=(\dot\eta\circ\eta^{-1},\bar{\mathbf{n}}\circ\eta^{-1},\dot{\bar{\mathbf{n}}}\circ\eta^{-1})$
so that the reduced Lagrangian is
\[
\ell(u,{\bf n},D_t{\bf n})
= \frac12\int_{\mathcal{D}}\|u\|^2\mu +\frac{j}{2}\int_\mathcal{D}|D_t{\bf n}|^2\mu-\int_\mathcal{D}F({\bf
n},\nabla{\bf n})\mu.
\]
This particular form of Lagrangian reduction is used in
\cite{HoTrYo2008} to formulate the metamorphosis equations in
imaging science. See also \cite{GBHoRa2009b} for a geometric description of metamorphosis reduction. Note that we have chosen to denote by $D_t{\bf n}$
the variable in $T_\mathbf{n\,}\mathcal{F}(\mathcal{D},S^2)$ since
the dynamics will yield the relation $D_t{\bf
n}=(\partial_t+u\!\cdot\!\nabla){\bf n}$. A direct computation shows
that the reduced equations are given by
\begin{equation}\label{2nd_red}
\left\{
\begin{array}{l}
\vspace{0.2cm}\displaystyle \dot u+u\!\cdot\!\nabla u =-\partial_i\left(\nabla\mathbf{n}^T\!\cdot\!\frac{\partial F}{\partial\mathbf{n}_{,i}}\right)-\nabla p,\qquad\operatorname{div}u=0\\
\displaystyle j\left(\frac{\nabla}{dt}+u\!\cdot\!\nabla\right)D_t\mathbf{n}=\mathbf{h},\qquad \mathbf{h}:=-\frac{\delta l}{\delta \mathbf{n}}=\frac{\partial F}{\partial\mathbf{n}}-\partial_i\left(\frac{\partial F}{\partial\mathbf{n}_{,i}}\right).
\end{array}\right.
\end{equation}
These
equations recover the Ericksen-Leslie fluid equations.

\rem{  
 \todo{Ok, so now this
$\bar{\boldsymbol{n}}$ is the full reduced variable, and we should
use the notation $\mathbf{n},\dot{\mathbf{n}}$ for consistence with
the Euler-Poincar\'e approach, see Remark
\ref{important_remark}.\\
Important remark for the reader: the second reduction is a standard
Lagrangian reduction and not an Euler-Poincar\'e reduction as
described in Section \ref{subsec_lagr}, since $T\operatorname{Orb}(\mathbf{n}_0)$
is not interpreted as a parameter space, but is part of the tangent
bundle.}
}      
\begin{remark}\normalfont
This second reduction step is a standard Lagrangian reduction and
not an Euler-Poincar\'e reduction as described in Section
\ref{subsec_lagr}, since $T\mathcal{F}(\mathcal{D},S^2)$ is part of
the tangent bundle and may not be interpreted as a parameter space.
\end{remark}

\begin{remark}[Metamorphosis for parameter-dependent Lagrangians]\normalfont
In order to account for compressibility in a natural way, one needs
to extend the Lagrangian to depend also on the parameter
$\rho_0\in{\rm Den}(\mathcal{D})$, so that
\[
L_{(\rho_0,{\bf n}_0)}:T{\rm Diff}(\mathcal{D})\times T\mathcal{F(D},SO(3))\to\Bbb{R}.
\]
Then, after the first Lagrange-Poincar\'e reduction stage on ${\bf n}_0=(0,0,1)$, one obtains the reduced Lagrangian
\[
\ell'_{\rho_0}:T{\rm Diff}(\mathcal{D})\times \big(T\mathcal{F(D},S^2)\oplus\mathcal{F(D)}\big)\to\Bbb{R}.
\]
At this point, after setting $\bar{r}=0$, one performs a metamorphosis reduction stage on the above Lagrangian according to the invariance property
\[
\ell'(\eta,\dot\eta,\rho_0,\bar{\mathbf{n}},\dot{\bar{\mathbf{n}}})=\ell'(\dot\eta\circ{\eta^{-1}},(\rho_0\circ{\eta^{-1}})J(\eta^{-1}),\bar{\mathbf{n}}\circ{\eta^{-1}},\dot{\bar{\mathbf{n}}}\circ{\eta^{-1}})
=\ell(u,\rho,\mathbf{n},D_t\mathbf{n})
\]
where we have used the same notation as in Remark \ref{remark1_compressible_flow}, although the above reduction process differs from that described in \ref{remark1_compressible_flow}. Therefore, the reduced Lagrangian $\ell$ is defined such as
\[
\ell:\mathfrak{X}(\mathcal{D})\times {\rm Den}(\mathcal{D})\times T\mathcal{F(D},S^2)\to \Bbb{R}.
\]
The reduction process just described is an example of a metamorphosis reduction for parameter-dependent Lagrangians. Notice that this process can be extended to any type of $G$-invariant Lagrangian
\[
L_{m_0}:TG\times TN\to\Bbb{R}.
\]
where $N$ is a manifold, while $m_0$ is a parameter belonging to another order parameter space $M$. This method extends the Euler-Poincar\'e approach  presented in \S\ref{subsec_lagr}.
\end{remark}

\subsubsection{Hamilton-Poincar\'e formulation and its Poisson bracket}

Notice that one can Legendre-transform the equations thereby
obtaining a Hamiltonian $\mathsf{h}(m,{\bf n},\boldsymbol{\pi})$,
on $\mathfrak{X}^*_{\rm vol}(\mathcal{D})\times
T^*\mathcal{F}(\mathcal{D},S^2)$.
Indeed, upon introducing the fluid momentum and the conjugate director variables
\[
m=\frac{\delta \ell}{\delta u}\in\mathfrak{X}^*_{\rm vol}(\mathcal{D})\quad\text{and}\quad
\left({\bf n},\boldsymbol\pi\right)=\left({\bf n},\frac{\delta \ell}{\delta (D_t{\bf n})}\right)\in T^*\mathcal{F}(\mathcal{D},S^2)\,,
\]
the Hamiltonian functional
\[
\mathsf{h}(m,{\bf n},\boldsymbol\pi)=\left\langle m\,,u\right\rangle+\left\langle \boldsymbol\pi\,,D_t{\bf n}\right\rangle-\ell(u,{\bf n},D_t{\bf n})
\]
produces the following Poisson bracket via Legendre transformation
\begin{align}\label{nemLC-PB}
\{f,g\}(m,{\bf n},\boldsymbol\pi)=&\left\langle m,\left[\frac{\delta f}{\delta
m},\frac{\delta g}{\delta m}\right]\right\rangle
+
\big\{f,\,g\big\}_{\,T^{*\!}\mathcal{F}(\mathcal{D},S^2)}
\nonumber
\\
&
+
\left\langle \frac{\delta
f}{\delta ({\bf n},\boldsymbol\pi)}\,,\pounds_{\frac{\delta g}{\delta m}} \left({\bf n},\boldsymbol\pi\right) \right\rangle
-
\left\langle\frac{\delta
g}{\delta ({\bf n},\boldsymbol\pi)}\,, \pounds_{\frac{\delta f}{\delta m}}\left({\bf n},\boldsymbol\pi\right) \right\rangle,
\end{align}
where
$\left\{\cdot,\,\cdot\right\}_{\,T^{*\!}\mathcal{F}(\mathcal{D},S^2)}$
denotes the canonical Poisson bracket on
$T^{*\!}\mathcal{F}(\mathcal{D},S^2)$,  $\pounds$ denotes Lie
derivative and the Lie bracket $[\cdot,\,\cdot]$ stands for minus
the Jacobi-Lie bracket on vector fields, as usual in fluid
mechanics.
The same Hamiltonian structure can also be obtained by a two steps
reduction from the Hamiltonian
$H_{n_0}:T^*\big(\operatorname{Diff}_{\rm vol}(\mathcal{D})\times
\mathcal{F}(\mathcal{D},SO(3))\big)\rightarrow\mathbb{R}$ associated
to $L_{n_0}$. In this process, one proceeds analogously by ignoring
the conjugate variable $\bar{w}=\delta\ell/\delta\bar{r}$ (momentum
associated to $\bar{r}$), which is possible because of the special
form of the Hamiltonian.

The
Poisson bracket formulation of condensed matter systems (especially
liquid crystals) is a rather relevant topic in the physics
literature; cf e.g. \cite{Volovick1980,Lubensky2003}.
 Here we emphasize that
the Poisson structures \eqref{liquid_crystals_bracket} and
\eqref{nemLC-PB} arise from {\it two  different reductions} of the
canonical Hamiltonian structure on $T^*\big(\operatorname{Diff}_{\rm
vol}(\mathcal{D})\times \mathcal{F}(\mathcal{D},SO(3))\big)$ and
they both produce {\it the same Ericksen-Leslie equations}.

The Hamiltonian system
described by \eqref{nemLC-PB} is of the general form
$h:\mathfrak{g}^*\times P\to\Bbb{R}$, upon choosing
$\mathfrak{g}=\mathfrak{X}_{\rm vol}(\mathcal{D})$ and
$P=T^*\mathcal{F}(\mathcal{D},S^2)$. However, this construction
differs from that treated in the first part of this paper and one
cannot simply transfer the Lie-Poisson setting discussed previously
to this infinite-dimensional case. Indeed, the difference resides in
the fact that $P$ carries its own Poisson structure and thus it is a
Poisson manifold by itself. Such a construction appears quite often
in condensed matter systems, where $P=T^*\mathcal{F(D},M)$
\cite{HoKu1982}, and it also emerges in electromagnetic fluid
dynamics, where $P=T^*\Omega^1(\mathcal{D})$ is the phase space of
Maxwell equations \cite{Ho1987}. This type of Hamiltonian systems
arising from an unreduced Hamiltonian on $T^*G\times P$ has been
extensively studied in \cite{KrMa1987}, where many interesting
properties are presented. For example, an interesting consequence of
the special type of the bracket \eqref{nemLC-PB} is that it allows
for a Poisson isomorphism that eliminates all the terms in the
second line of \eqref{nemLC-PB}. More precisely, given an
equivariant momentum map ${\bf J}:P\to\mathfrak{g}^*$,  there exists
a Poisson diffeomorphism $(\mu,p)\mapsto (\mu+\mathbf{J}(p),p)$
sending the reduced Poisson structure on $(T^*G\times
P)/G=\mathfrak{g}^*\times P$ to the product Poisson structure on
$\left(T^*G/G\right)\times P=\mathfrak{g}^*\times P$ (cf.
Proposition 2.2 in \cite{KrMa1987}). In the case of uniaxial
nematics, this corresponds to introducing a new variable \[ {\sf
m}:=m+{\bf J}(\mathbf{n},\boldsymbol\pi)\,,
\]
so that
the new set of variables $\left({\sf m},\mathbf{n},\boldsymbol\pi\right)$ carries the following `untangled' Poisson bracket:
\begin{align}\label{untPB}
\{f,g\}({\sf m},{\bf n},\boldsymbol\pi)=&\left\langle {\sf m},\left[\frac{\delta f}{\delta
{\sf m}},\frac{\delta g}{\delta {\sf m}}\right]\right\rangle
+
\big\{f,\,g\big\}_{\,T^{*\!}\mathcal{F}(\mathcal{D},S^2)}
\end{align}
This bracket produces ordinary Hamilton's equations on $T^{*}\mathcal{F}(\mathcal{D},S^2)$ as well as a Lie-Poisson equation on $\mathfrak{X}^*_{\rm vol}(\mathcal{D})$.
In the special case of the bracket \eqref{nemLC-PB} for liquid crystals, one can introduce the following momentum map, associated to the cotangent lifted action of $G=\operatorname{Diff}_{\rm vol}(\mathcal{D})$ on $P=T^*\mathcal{F}(\mathcal{D},S^2)$:
\begin{equation}\label{MomapForNematicLiqXals}
{\bf J}(\mathbf{n},\boldsymbol\pi)=\nabla\mathbf{n}^T\cdot\boldsymbol\pi+\nabla\varphi\,\in\mathfrak{X}_{\rm vol}(\mathcal{D})^*
\end{equation}
where $(\mathbf{n},\boldsymbol\pi)\in T^*\mathcal{F(D},S^2)$ and $\varphi$ is a variable such that $\operatorname{div}({\bf J}(\mathbf{n},\boldsymbol\pi))=0$.
Then, the new Hamiltonian reads as
\[
h\left({\sf m},\mathbf{n},\boldsymbol\pi\right)=\frac12\int_{\mathcal{D}}\left\|{\sf m}-\nabla\mathbf{n}^T\cdot\boldsymbol\pi-\nabla\varphi\right\|^2\mu +\frac{1}{2j}\int_\mathcal{D}|\boldsymbol\pi|^2\mu-\int_\mathcal{D}F({\bf
n},\nabla{\bf n})\mu.
\]
This approach has been sometimes referred to as `untangling' and the Poisson bracket \eqref{untPB} is called `untangled Poisson bracket'. See  \cite{Ho1986,Ho1987,HoKu1982} for examples of how the untangling and entangling processes are used in the physics of charged fluids and superfluids.

\subsubsection{The helicity invariant}
Another important property of Hamiltonian systems of the general form
$h:\mathfrak{g}^*\times P\to\Bbb{R}$ involves Casimir functions. Indeed, from the results in \cite{KrMa1987}, it follows immediately  that the Poisson bracket \eqref{nemLC-PB} allows for an interesting class of Casimir functions. Indeed, given an equivariant momentum map ${\bf J}:P\to\mathfrak{g}^*$ and any Casimir function $C(\mu)$ for the Lie-Poisson bracket on $\mathfrak{g}^*$, the function
\begin{equation}\label{formula_casimir}
C(\mu,p)=C(\mu+{\bf J}(p))
\end{equation}
is a Casimir for the reduced Poisson bracket on $(T^*G\times P)/G=\mathfrak{g}^*\times P$ (cf. Corollary 2.3 in \cite{KrMa1987}). In the case of liquid crystals, the momentum map \eqref{MomapForNematicLiqXals} can be used to produce an explicit expression for the helicity of nematic liquid crystals. Indeed, since it is well known that the helicity $\mathscr{H}(m)=\left\langle\operatorname{curl}m,m\right\rangle$ is a Casimir for the Lie-Poisson bracket on $\mathfrak{X}_{\rm vol}(\mathcal{D})^*$, then direct substitution of the momentum map ${\bf J}(\mathbf{n},\boldsymbol\pi)$ in formula \eqref{formula_casimir} yields a Casimir for the Poisson bracket \eqref{nemLC-PB}. This Casimir is explicitly written as the following helicity functional for nematic liquid crystals
\begin{equation}\label{Casimir_nematics}
\mathscr{H}(m,\mathbf{n},\boldsymbol\pi)=\int_\mathcal{D} \left(m+\nabla\mathbf{n}^T\cdot\boldsymbol\pi\right)\cdot\operatorname{curl}\left(m+\nabla\mathbf{n}^T\cdot\boldsymbol\pi\right).
\end{equation}
This helicity invariant can also be written as a Casimir for the Lie-Poisson bracket \eqref{liquid_crystals_bracket}. Indeed, one can pull back $\mathscr{H}$ using the mapping $(m,\boldsymbol{\mu},\mathbf{n})\mapsto (m,\mathbf{n},\boldsymbol{\mu}\times\mathbf{n})$, which is suggested by the isomorphism \eqref{Poisson-isom} (with $w=0$) holding for the dynamics of the single nematic molecule.
The resulting expression for the helicity is then
\begin{equation}\label{Casimir_nematics_LP}
\mathscr{H}(m,\boldsymbol{\mu},\mathbf{n})=\int_\mathcal{D} \left(m+\boldsymbol{\mu}\cdot(\mathbf{n}\times\nabla\mathbf{n})\right)\cdot\operatorname{curl}\left(m+\boldsymbol{\mu}\cdot(\mathbf{n}\times\nabla\mathbf{n})\right).
\end{equation}
More rigorously, the Casimir property can be verified by restricting the Lie-Poisson construction in \eqref{liquid_crystals_bracket}
to the submanifold $\mathscr{P}=\{(m,\boldsymbol{\mu},\mathbf{n})\mid \boldsymbol{\mu}\cdot\mathbf{n}=0\}$, which is analogous to setting $w=0$ in \eqref{Poisson-isom} for a single nematic molecule. Then, the following arguments show that $\mathscr{P}$ can be endowed with the Poisson structure \eqref{liquid_crystals_bracket}, thereby making it into a Poisson submanifold of $\mathcal{W}=\mathfrak{X}_{\rm vol}(\mathcal{D})^*\times\mathcal{F}(\mathcal{D},\mathfrak{so}(3))^*\times\mathcal{F}(\mathcal{D},S^2)$. Let $(m(t),\boldsymbol{\mu}(t),\mathbf{n}(t))$ be a solution of Hamilton's equation $\dot f=\{f,h\}$ on $\mathcal{W}$ relative to an arbitrary Hamiltonian $h$. Since $\partial_t(\boldsymbol{\mu}\cdot\mathbf{n})+(u\cdot\nabla)(\boldsymbol{\mu}\cdot\mathbf{n})=0$, then we obtain $(\boldsymbol{\mu}(t)\cdot\mathbf{n}(t))=(\boldsymbol{\mu}(0)\cdot\mathbf{n}(0))\circ\eta_t^{-1}$, where we have denoted by $\eta_t$ the flow of $u$.  Consequently, any Hamiltonian vector field on $\mathcal{W}$ restricted to $\mathscr{P}$ is tangent to $\mathscr{P}$ and this shows that $\mathscr{P}$ is a quasi Poisson submanifold. Thus, there is a unique Poisson structure on $\mathscr{P}$ making it into a Poisson submanifold, see Prop 4.1.23 in \cite{OrRa2004}. It is readily seen that this Poisson structure has the same expression \eqref{liquid_crystals_bracket}.
\begin{theorem}
The expression
\begin{equation}
\mathscr{H}(m,\boldsymbol{\mu},\mathbf{n})=\int_\mathcal{D} \big(m+\boldsymbol{\mu}\cdot(\mathbf{n}\times\nabla\mathbf{n})\big)\cdot\operatorname{curl}\big(m+\boldsymbol{\mu}\cdot(\mathbf{n}\times\nabla\mathbf{n})\big).
\end{equation}
is a Casimir function on the Poisson manifold $\mathscr{P}=\{(m,\boldsymbol{\mu},\mathbf{n})\mid \boldsymbol{\mu}\cdot\mathbf{n}=0\}$, endowed with the bracket \eqref{liquid_crystals_bracket}.
\end{theorem}
\textbf{Proof.} We shall show that for a solution $(m(t),\boldsymbol{\mu}(t),\mathbf{n}(t))$ of an arbitrary Hamiltonian system on $\mathscr{P}$, we have
\[
\left(\frac{\partial\,}{\partial t}+\pounds_{\!\textstyle\frac{\delta h}{\delta m}}\right)\mathcal{C}=-\mathbf{d} p.
\]
where we have defined the differential one form $\mathcal{C}:=m+\boldsymbol{\mu}\cdot(\mathbf{n}\times\nabla\mathbf{n})$.
A direct calculation shows that
\begin{align*}
\left(\frac{\partial\,}{\partial t}+\pounds_{\!\textstyle\frac{\delta h}{\delta m}}\right)\mathcal{C}
=&
-\boldsymbol\mu\cdot\frac{\delta h}{\delta \boldsymbol\mu}+ \nabla \mathbf{n}\cdot\frac{\delta h}{\delta \mathbf{n}}-\operatorname{grad}p
+
\left(\frac{\delta h}{\delta
\boldsymbol{\mu}}\times\boldsymbol{\mu}+\frac{\delta h}{\delta \bf
n}\times{\bf n}\right)\cdot(\mathbf{n}\times\nabla\mathbf{n})
\\
&+
\boldsymbol{\mu}\cdot\left(\left(\frac{\delta
h}{\delta \boldsymbol{\mu}}\times{\bf n}\right)\times\nabla\mathbf{n}\right)
+
\boldsymbol{\mu}\cdot\left(\mathbf{n}\times\nabla\!\left(\frac{\delta
h}{\delta \boldsymbol{\mu}}\times{\bf n}\right)\right)
\\
=&
-\boldsymbol\mu\cdot\frac{\delta h}{\delta \boldsymbol\mu}+ \nabla \mathbf{n}\cdot\frac{\delta h}{\delta \mathbf{n}}-\operatorname{grad}p
+
\boldsymbol{\mu}\cdot\left(\mathbf{n}\times\mathbf{n}\times\nabla\frac{\delta h}{\delta \boldsymbol\mu}\right)
-
{\bf n}\cdot\left(\frac{\delta h}{\delta \bf n}\times\mathbf{n}\times\nabla\mathbf{n}\right)
\\
=&
-\boldsymbol\mu\cdot\frac{\delta h}{\delta \boldsymbol\mu}+ \nabla \mathbf{n}\cdot\frac{\delta h}{\delta \mathbf{n}}-\operatorname{grad}p
-
\boldsymbol\mu\cdot\left(\left(\nabla\frac{\delta h}{\delta \boldsymbol\mu}\cdot\mathbf{n}\right)\mathbf{n}-\nabla\frac{\delta h}{\delta \boldsymbol\mu}\right)
\\
&
-
\mathbf{n}\cdot\left(\left(\nabla\mathbf{n}\cdot\frac{\delta h}{\delta \mathbf{n}}\right)\mathbf{n}-\left(\frac{\delta h}{\delta \mathbf{n}}\cdot\mathbf{n}\right)\nabla\mathbf{n}\right)
\\
=&
-\mathbf{d} p
\end{align*}
where we have used the Jacobi identity for double cross products, and the properties $|\mathbf{n}|^2=1$ and $\boldsymbol\mu\cdot\mathbf{n}=\mathbf{n}\cdot\nabla\mathbf{n}=0$, respectively. Thus, upon writing the helicity as $\mathscr{H}=\int_\mathcal{D}\mathcal{C}\wedge{\bf d}\mathcal{C}$,
one obtains
\[
\partial_t\left(\mathcal{C}\wedge{\bf d}\mathcal{C}\right)=-{\bf d}p\wedge{\bf d}\mathcal{C}-\operatorname{div}\!\left(\left(\mathcal{C}\wedge{\bf d}\mathcal{C}\right)\frac{\delta h}{\delta m}\right) {\rm d}^3x
\]
so that $\mathscr{H}=\int_{\mathcal{D}}\mathcal{C}\wedge{\bf d}\mathcal{C}=\operatorname{const}$ along any Hamiltonian flow on $\mathscr{P}$.
$\blacksquare$

\bigskip\noindent
We notice that a simple consequence of the above theorem is the
\begin{corollary}
With the initial condition $\boldsymbol{\mu}_0\cdot\mathbf{n}_0=0$, the circulation theorem for the equations of liquid crystals reads as
\[
\frac{d}{dt}\oint_{\gamma(t)} \big(m+\boldsymbol{\mu}\cdot(\mathbf{n}\times\nabla\mathbf{n})\big)=0\,,
\]
\end{corollary}
which follows directly from the property $\left(\partial _t+\pounds_{\delta h/\delta m}\right)\mathcal{C}=-\mathbf{d}p$. Similar arguments on the relation between the Kelvin-Noether circulation and the helicity invariant also hold for superfluid dynamics \cite{HoKu1987}.

\begin{remark}[Two dimensional flows]\normalfont
It is important to notice that formula \eqref{formula_casimir} provides a whole class of Casimir functions for 2D flows. Indeed, it is well known that, in two dimensions, any function $\Phi(\omega)$ of the vorticity $\omega=\mathbf{\hat{z}}\cdot\operatorname{curl}\!\left(m\right)$ yields a Casimir $C=\int \Phi(\omega)$ and thus, by \eqref{formula_casimir}, any function $\Phi\!\left(\omega+\mathbf{\hat{z}}\cdot\operatorname{curl}\!\left(\nabla{\bf n}\cdot\boldsymbol\pi\right)\right)$ produces a new Casimir
\[
C=\int_{\mathcal{D}} \Phi\!\left(\omega+\left\{\pi_a,n^a\right\}\right)=\int_{\mathcal{D}} \Phi\!\left(\omega+\varepsilon_{abc}\!\left\{\sigma^b n^c,n^a\right\}\right)
\] 
for 2D flows of uniaxial nematics. Here, the operation $\left\{\cdot,\cdot\right\}$ is the canonical Poisson bracket in the planar coordinates $(x,y)$. This observation may have important consequences in terms of the stability properties of liquid crystals, following the approach in \cite{HoMaRaWe1985}.
\end{remark}

\subsection{Reduction schemes for uniaxial nematic liquid crystals}

Starting from the same Lagrangian $L_{n_0}$ on the tangent bundle of $\operatorname{Diff}_{\rm vol}(\mathcal{D})\times\mathcal{F}(\mathcal{D},SO(3))$ we have obtained the Ericksen-Leslie fluid equations in two different ways. For completeness, this section considers nonvanishing $r$.

First, by considering as symmetry group the semidirect product
$\operatorname{Diff}_{\rm
vol}(\mathcal{D})\,\circledS\,\mathcal{F}(\mathcal{D},SO(3))$, we
obtain the equations of motion by Euler-Poincar\'e reduction,
following the general approach described in \S\ref{subsec_lagr}. The
reduced space is given by $\big(\mathfrak{X}_{\rm
vol}(\mathcal{D})\,\circledS\,
\mathcal{F}(\mathcal{D},\mathfrak{so}(3))\big)\times\mathcal{F}(\mathcal{D},S^2)$
and the reduction map reads
\begin{align*}
(\eta,\dot\eta,\chi,\dot\chi,\mathbf{n}_0)\longmapsto &\left((\eta,\dot\eta,\chi,\dot\chi)(\eta,\chi)^{-1},(\eta,\chi)\mathbf{n}_0\right)\\
&=\left(\dot\eta\circ\eta^{-1},(\dot\chi\chi^{-1})\circ\eta^{-1},(\chi \mathbf{n}_0)\circ\eta^{-1}\right)\\
&=:(u,\boldsymbol{\nu},\mathbf{n})
\end{align*}
according to the general formula \eqref{diffeo_EP}. This approach is consistent with that given in \cite{GBRa2008}.

Second, we use Lagrangian reduction by stages (see
\cite{CeMaRa2001}), and obtain the motion equations on the reduced
space $\mathfrak{X}_{\rm vol}(\mathcal{D})\times
T\mathcal{F}(\mathcal{D},S^2)\times\mathcal{F}(\mathcal{D})$. First
we apply Lagrange-Poincar\'e reduction for the internal structure, with respect to
the mechanical connection (here we choose $\mathbf{n}_0=(0,0,1)$).
We get
\begin{align*}
(\eta,\dot\eta,\chi,\dot\chi,\mathbf{n}_0)\longmapsto &\left(\eta,\dot\eta,\chi\mathbf{n}_0,\left(\dot\chi\chi^{-1}\right)\times (\chi\mathbf{n}_0),(\chi\mathbf{n}_0)\!\cdot\!\dot\chi\chi^{-1}\right)\\
&=:(\eta,\dot\eta,\bar{\bf n},\bar{\boldsymbol{\nu}}\times\bar{\mathbf{n}},\bar{\boldsymbol{\nu}}\!\cdot\!\bar{\mathbf{n}})=:(\eta,\dot\eta,\bar{\mathbf{n}},\dot{\bar{\mathbf{n}}},\bar r).
\end{align*}
Then, we use Lagrangian reduction with respect to the fluid variables and have
\[
(\eta,\dot\eta,\bar{\bf n},\dot{\bar{\mathbf{n}}},\bar r)\longmapsto (\dot\eta\circ\eta^{-1},\bar{\bf
n}\circ\eta^{-1},\dot{\bar{\mathbf{n}}}\circ\eta^{-1},\bar r\circ\eta^{-1})=:\left(u,\mathbf{n},D_t{\bf
n},r\right).
\]
This approach is consistent with that given in \cite{Ho2002}.
Note the following important relation between the intermediate and fully reduced variables. We have
\begin{align*}
\bar{\mathbf{n}}&=\chi\mathbf{n}_0, \quad\mathbf{n}=(\chi\mathbf{n}_0)\circ\eta^{-1}=\bar{\mathbf{n}}\circ\eta^{-1}\\
\bar{\boldsymbol{\nu}}&=\dot\chi\chi^{-1},\quad\boldsymbol{\nu}=(\dot\chi\chi^{-1})\circ\eta^{-1}=\bar{\boldsymbol{\nu}}\circ\eta^{-1}\\
\bar r&=(\chi\mathbf{n}_0)\!\cdot\!\dot\chi\chi^{-1}=\bar{\boldsymbol{\nu}}\!\cdot\!\bar{\mathbf{n}},\quad r=\boldsymbol{\nu}\!\cdot\!\mathbf{n}=\bar r\circ\eta^{-1}.
\end{align*}

One can pass from one approach to the other by using the same transformations \eqref{diffeo_nematics} and \eqref{nu} as for the single particle case, except that now they are applied to vector valued functions instead of vectors. This clarifies the link between the approaches given in \cite{Ho2002} and \cite{GBRa2008}. The diagram below clarifies the various reduction processes involved in the hydrodynamics of
nematic liquid crystals.
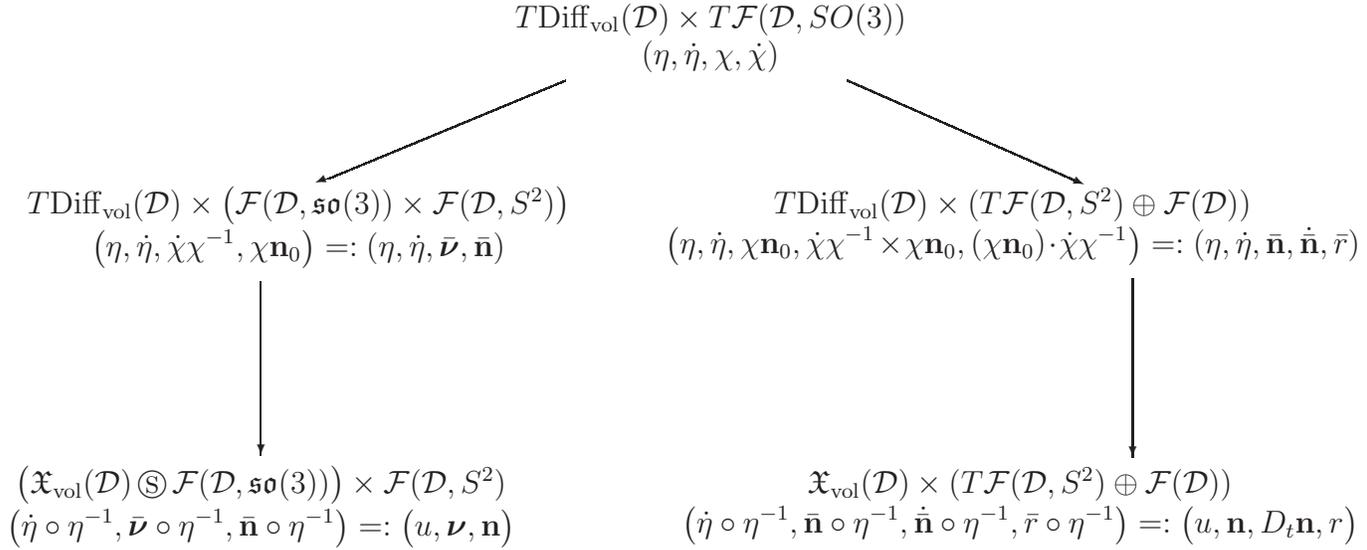
\begin{diagram}
&  & \underset{\displaystyle\left(\eta,\dot{\eta},\chi,\dot\chi\right)}{T{\rm Diff}_{\rm vol}(\mathcal{D}) \times T\mathcal{F(D},SO(3))} &  &
\\
& \ldTo  &      & \rdTo  &
\\
\underset{\displaystyle\left(\eta,\dot{\eta},\dot\chi\chi^{-1},\chi{\bf n}_0\right)
=:\left(\eta,\dot{\eta},\bar{\boldsymbol{\nu}},\bar{\bf n}\right)}{T{\rm Diff}_{\rm vol}(\mathcal{D}) \times \big(\mathcal{F(D},\mathfrak{so}(3))\times \mathcal{F(D},S^2)\big)}\hspace{-1cm} &   &      &   & \hspace{-3.2cm}\underset{\displaystyle\left(\eta,\dot\eta,\chi{\bf n}_0,\dot\chi\chi^{-1\!}\times \!\chi\mathbf{n}_0,(\chi\mathbf{n}_0)\!\cdot\!\dot\chi\chi^{-1}\right)
=:(\eta,\dot\eta,\bar{\mathbf{n}},\dot{\bar{\mathbf{n}}},\bar{r})}{T{\rm Diff}_{\rm vol}(\mathcal{D}) \times \left(T\mathcal{F(D},S^2)\oplus \mathcal{F(D})\right)}
\\
&&&&
\\
\dTo & & & & \dTo
\\
\underset{\displaystyle\big(\dot\eta\circ\eta^{-1},\bar{\boldsymbol{\nu}}\circ\eta^{-1},\bar{{\bf n}}\circ\eta^{-1}\big)
=:\big(u,{\boldsymbol{\nu}},{\bf n}\big)}{\big( \mathfrak{X}_{\rm vol}(\mathcal{D})\,\circledS \,\mathcal{F(D},\mathfrak{so}(3))\big)\times \mathcal{F(D},S^2)} & & & & \hspace{-3cm}\underset{\displaystyle\left(\dot\eta\circ\eta^{-1},\bar{\mathbf{n}}\circ\eta^{-1},\dot{\bar{\mathbf{n}}}\circ\eta^{-1},\bar{r}\circ\eta^{-1}\big)=:\big(u,{\bf n},D_t{\mathbf{n}},r\right)}{\mathfrak{X}_{\rm vol}(\mathcal{D})\times \left(T\mathcal{F(D},S^2)\oplus \mathcal{F(D})\right)}
\end{diagram}

\section{Conclusions}
This paper has developed symmetry-reduction methods for systems with broken
symmetry, whose main example consists of nematic particles in liquid crystals
(ordinary or biaxial). After extending the Euler-Poincar\'e theory, this has been applied to the case of a transitive group action on an order parameter manifold, thereby showing how the latter carries its natural coset structure
typically appearing in symmetry breaking. This setting has been applied to nematic particles ($SO(3)/O(2)$), biaxial nematics ($SO(3)/D_2$) and $V$-shaped molecules ($SO(3)/\Bbb{Z}_2$), and its general validity can be used to reproduce dynamics on any order parameter space. Moreover, the Euler-Poincar\'e and Lie-Poisson dynamics for nematic and biaxial systems have been formulated also in terms of the alignment tensors, which are widely used in the Landau-de Gennes theory.

As a further step, the Lagrangian reduction technique
(Lagrange-Poincar\'e reduction) has been applied directly to the
space $(TG)/G_0$, where $G_0\subset G$ is the isotropy group
determining the ``breaking symmetry'', which is left in the system
when the full $G$-symmetry is broken. This technique has showed how
the reduced configuration space may be identified with the coset
$G/G_0$ (order parameter space), provided the adjoint bundle term in
$\widetilde{\mathfrak{g}_0}$ can be set to zero. This is indeed the
case for nematic particles, so that the resulting dynamics is given
by the Ericksen-Leslie equation on $SO(3)/O(2)$ for a single
director. The choice of the mechanical connection enabled us to
explain why Ericksen-Leslie dynamics can be considered as ordinary
Hamiltonian dynamics on the projective plane, upon setting
$\mathbf{n}\cdot\boldsymbol{\nu}=0$ for consistency with the
rod-like nature of nematic particles. Moreover, the mechanical
connection produced explicit transformations that relate directly
the Euler-Poincar\'e and the Lagrange-Poincar\'e approaches. The
Lagrange-Poincar\'e approach restricts to the case when the isotropy
subgroup $G_0\subset G$ is a Lie group with ${\rm dim}\,G_0\geq1$,
which excludes the case when $G_0$ is discrete. It is then an
interesting open question whether it is possible to extend the
Lagrange-Poincar\'e reduction process $(TG)/G_0$ to account for a
discrete symmetry group $G_0$.

All the above techniques have been extended at the continuum level, to produce
also hydrodynamic models, thereby recovering the well known Ericksen-Leslie
equation for the dynamics of the director field. As a result of this extension, all the geometric features of the microscopic particle dynamics are naturally transfered at the macroscopic fluid level without substantial modifications. Interestingly enough, the
application of Lagrangian reduction to the hydrodynamics of nematic liquid
crystals produces a ``generalized Lie-Poisson'' Hamiltonian structure on
$\mathfrak{g}^*\times P$, involving a Lie algebra $\mathfrak{g}$ and a Poisson manifold $P$. This type of Hamiltonian structure
was discovered in \cite{KrMa1987} and it appears very frequently in hydrodynamic physical models as well as in other contexts, such as imaging sciences.
The last part of this paper applied the theory on this generalized Lie-Poisson construction to produce new Poisson bracket structures, as well as new Casimir functions for liquid crystal dynamics. In particular, two explicit expressions for the helicity of uniaxial nematic liquid crystals have been presented in \eqref{Casimir_nematics} and \eqref{Casimir_nematics_LP}.

\subsubsection*{Acknowledgments}
We are indebted to Darryl Holm and Tudor Ratiu for many helpful discussions.

{\footnotesize

\bibliographystyle{new}

\begin{thebibliography}{300}



\bibitem{AbMa1978}
Abraham, R. and J.~E. Marsden [1978], {\em Foundations of
Mechanics}. \newblock Benjamin-Cummings Publ. Co, Updated
1985 version, reprinted by Perseus Publishing, second
edition.


\bibitem{BePr1987}
Ben-Mizrachi, A. and Procaccia, I. [1983], Microscopic derivation of
nonlinear hydrodynamics in ordered systems with applications to
nematic liquid crystals, {\it Phys. Rev. A} {\bf 27}(4),  2126--2139.



\bibitem{CeHoMaRa1998}
Cendra, H., D.~D. Holm, J.~E. Marsden and T.~S. Ratiu [1998], Lagrangian Reduction, the Euler-Poincar\'e Equations and Semidirect Products, \textit{Amer. Math. Soc. Transl.}, \textbf{186}, 1--25.

\bibitem{CeMaPeRa2003}
Cendra, H., J.~E. Marsden, S. Pekarsky, and T.~S. Ratiu [2003],
Variational principles for Lie-Poisson and Hamilton-Poincar\'e
equations. \textit{Mosc. Math. J.} \textbf{3}(3), 833--867.


\bibitem{CeMaRa2001}
Cendra, H., J.~E. Marsden, and T.S. Ratiu [2001b], Lagrangian Reduction by Stages. \textit{Mem. Amer. Math. Soc.}, \textbf{152}, no. 722.

\bibitem{Chandra1992}
Chandrasekhar, S. [1992], {\it Liquid Crystals}, Cambridge
University Press, Cambridge.


\bibitem{deGennes1971}
de Gennes, P.G. [1971], Short range order effects in the isotropic phase of nematics and cholesterics. {\it Mol. Cryst. Liq. Cryst.}, {\bf 12}, 193--214

\bibitem{deGennes1993}
de Gennes, P. G. and J. Prost [1993], {\it The Physics of Liquid Crystals}, Second Edition. Oxford University Press, Oxford.

\bibitem{Volovick1980}
Dzyaloshinskii, I.E., Volovick, G.E. [1980], Poisson brackets in
condensed matter systems, {\it Ann. Phys.} {\bf 125}, 67--97.


\bibitem{Eringen1987}
Eringen, A.~C. [1987], A unified continuum theory of liquid
crystals, {\it ARI} {\bf 50}, 73--84.



\bibitem{GBHoRa2009a}
Gay-Balmaz, F., D.~D. Holm, and T.~S. Ratiu [2009a], Variational principles for spin systems and the Kirchhoff rod, \textit{J. Geom. Mech.}, \textbf{1} (4), 417--444.

\bibitem{GBHoRa2009b}
Gay-Balmaz, F., D.~D. Holm, and T.~S. Ratiu [2009b], \emph{Geometric dynamics of optimization}, preprint, {\tt arXiv:0912.2989}

\bibitem{GBRa2008}
Gay-Balmaz, F. and T.~S. Ratiu [2008], The geometric structure of complex fluids, \textit{Adv. Appl. Math.}, \textbf{42} (2), 176--275.

\bibitem{GiHoKu1983}
Gibbons, J., D.~D. Holm and B.~A. Kupershmidt [1983], The Hamiltonian
structure of classical chromohydrodynamics, \textit{Physica D},
\textbf{6}, 179--194.


\bibitem{Ho2002}
Holm, D.~D. [2002], Euler-Poincar\'e dynamics of perfect complex fluids, in \textit{Geometry, Dynamics and Mechanics: 60th Birthday Volume for J.E.
Marsden.} P. Holmes, P. Newton, and A. Weinstein, eds., Springer-Verlag.

\bibitem{Ho1986}
Holm, D.~D. [1986], Hamiltonian dynamics of a charged fluid, including electro- and magnetohydrodynamics, \textit{Phys. Lett. A}, {\bf 114}, 137--141.

\bibitem{Ho1987}
Holm, D.~D. [1987], Hamiltonian dynamics and stability analysis of
neutral electromagnetic fluids with induction, \textit{Physica D},
{\bf 25}, 261--287.

\bibitem{HoKu1982}
Holm, D.~D. and B.~A. Kupershmidt [1982], Poisson structures of
superfluids, \textit{Phys. Lett. A}, \textbf{91}(9), 425--430.

\bibitem{HoKu1984}
Holm, D.~D. and B.~A. Kupershmidt [1984], Yang-Mills magnetohydrodynamics:
Nonrelativistic theory, \textit{Phys. Rev. D}, \textbf{30}, 2557--2560.

\bibitem{HoKu1987}
Holm, D.~D. and B.~A. Kupershmidt [1987], Superfluid plasmas: Multivelocity
nonlinear hydrodynamics of superfluid solutions with charged condensates
coupled
electromagnetically, \textit{Phys. Rev. A}, \textbf{36}, 3947--3956.

\bibitem{HoKu1988}
Holm, D.~D. and B.~A. Kupershmidt [1988], The analogy between spin glasses and
Yang-Mills fluids, \textit{J. Math. Phys.} \textbf{29}, 21--30.

\bibitem{HoMaRa1998}
Holm D.~D., J.~E. Marsden and T.~S. Ratiu [1998], The Euler-Poincar\'e
equations
and semidirect products with applications to continuum theories, \textit{Adv.
in Math.}, \textbf{137}, 1--81.

\bibitem{HoMaRaWe1985}
Holm, D.D., J.~E. Marsden and T.~S. Ratiu, A. Weinstein  [1985],  Nonlinear stability of fluid and plasma equilibria, {\it Phys. Rep.}, {\bf 123}, 1--116


\bibitem{HoTrYo2008}
Holm, D.~D., A. Trouv\'e, and L. Younes [2009], The Euler-Poincar\'e theory of metamorphosis, \textit{Quart. Appl. Math.}  {\bf 67}, 661--685




\bibitem{Khesin}
Khesin, B. and R. Wendt [2009], {\it Geometry of infinite-dimensional groups}.  Springer-Verlag.

\bibitem{KrMa1987}
Krishnaprasad, P.~S. and J.~E. Marsden [1987],
Hamiltonian structure and stability for rigid bodies with flexible attachments,
\textit{Arch. Rational Mech. Anal.}, \textbf{98}, 71--93.


\bibitem{Le1979}
Leslie, F.~M. [1979], {\it Theory of flow phenomena in liquid
crystals}, in Advances in Liquid Crystals, vol. 4, (G. H. Brown,
ed.) Academic, New York pp. 181.

\bibitem{LeTo1977}
Lev, B.~I., Tomchuk, P.~M. [1977], Connection between the
phenomenological and microscopic approaches in the theory of liquid
crystals, {\it Theor. Math. Phys.} {\bf 32}(1), 101--113.

\bibitem{Levy1985}
Levy, D. [1985], A spin model with non-abelian $\Pi_1$, {\it Phys.
Lett. B} {\bf 154}(1), 57--62


\bibitem{MaRa1999}
Marsden, J.~E., Ratiu, T.~S. [1994] {\it Introduction to mechanics and
symmetry.} Springer-Verlag.

\bibitem{MaMiOrPeRa2007}
Marsden, J.~E., G. Misio\l ek, J.-P. Ortega, M. Perlmutter, and T.~S. Ratiu
[2007], \textit{Hamiltonian Reduction by Stages}, Springer Lecture Notes in Mathematics, \textbf{1913}, Springer-Verlag 2007.

\bibitem{MaRaWe1984}
Marsden, J.~E., Ratiu, T.~S., Weinstein, A. [1984], Semidirect
products and reduction in mechanics.  {\it Trans. Amer. Math. Soc.}
{\bf 281}(1), 147--177.

\bibitem{Mermin1979}
Mermin, N.~D. [1979], The topological theory of defects in ordered
media, {\it Rev. Mod. Phys.} {\bf 51}, 591-648.

\bibitem{Mi1980}
Michel, L. [1980], Symmetry defects and broken symmetry. Configurations. Hidden symmetry. \textit{Rev. Modern Phys.} \textbf{52}(3), 617--651.


\bibitem{Nemtsov1975}
Nemtsov, V.~B. [1975], Statistical theory of hydrodynamic and
kinetic processes in liquid crystals, {\it Theor. Math. Phys.} {\bf
25}(1), 1019--1028.


\bibitem{OrRa2004}
Ortega J.-P. and T.S. Ratiu [2004], \textit{Momentum maps and Hamiltonian reduction}, Progress  in Mathematics, 222, Birkh\"auser, Boston,
2004.




\bibitem{SoViDu2003}
Sonnet, A.~M., Virga, E.~G, Durand, G.E. [2003],
Dielectric shape dispersion and biaxial transitions in nematic liquid crystals, {\it Phys Rev E} {\bf 67}, 061701

\bibitem{Lubensky2003}
Stark, H., Lubensky, T.C. [2003], Poisson-bracket approach to the
dynamics of nematic liquid crystals, {\it Phys. Rev. E} {\bf 67},
061709.






\end{thebibliography}

}

\end{document}